\documentclass[prc,aps,amsfonts,twocolumn,floatfix,superscriptaddress,nofootinbib,preprintnumbers]{revtex4-1}

\usepackage{lineno,hyperref}
\modulolinenumbers[5]
\usepackage{graphicx,amssymb,color,amsmath,soul}
\usepackage{natbib}
\usepackage[normalem]{ulem}
\usepackage{braket}
\usepackage{cases}
\usepackage{mathtools}
\usepackage{caption}
\usepackage{subcaption}
\usepackage{multirow}

\definecolor{orange}{rgb}{1,0.5,0}

\begin{document}

\title{
Entanglement Rearrangement in Self-Consistent Nuclear Structure Calculations}

\author{Caroline Robin}
\altaffiliation{Present affiliation : Fakult\"at f\"ur Physik, Universit\"at Bielefeld, D-33615, Bielefeld, Germany.}
\email{crobin@physik.uni-bielefeld.de}
\affiliation{Institute for Nuclear Theory, University of Washington, Seattle, WA 98195, USA.}

\author{Martin J. Savage}
\email{mjs5@uw.edu}
\affiliation{Institute for Nuclear Theory, University of Washington, Seattle, WA 98195, USA.}

\author{Nathalie Pillet}
\affiliation{CEA, DAM, DIF, F-91297 Arpajon, France.}
\affiliation{Universit\'e Paris-Saclay, CEA, LMCE, 91680 Bruy\`eres-le-Ch\^atel, France}

\date{\today}


\begin{abstract}
\begin{description}
 \item[Background] 
 Entanglement plays a central role in a diverse  array of increasingly important research areas, including quantum computation, simulation, measurement, sensing, and communication. 
Extensive suites of investigations have been performed to better understand entanglement in atomic and molecular quantum many-body systems, 
while the exploration of entanglement in the structure of nuclei and their reactions is presently in its infancy.
 \item[Purpose] The goal of this work is to begin investigating the entanglement properties of nuclei from first principles nuclear many-body calculations. 
 We attempt to  identify common features and emergent structures of entanglement that could ultimately lead to new and natural many-body schemes.
 With an eye toward quantum accelerators in future hybrid-supercomputers, criteria for partitioning nuclear many-body calculations into quantum and classical components may provide advantages in future large-scale computations.
 Along the way we look for explanations of  the relative success of phenomenological models such as the nuclear shell model, and for better ways to match to low-energy nuclear effective field theories and lattice QCD calculations 
 to nuclear many-body techniques
 that are based upon entanglement.
 \item[Method] 
 We explore the entanglement between single-particle states in  $^{4}$He and $^{6}$He. 
 The patterns of entanglement emerging from different single-particle bases are compared, and possible links with the convergence of observables are explored, in particular, ground-state energies. 
The nuclear wavefunctions are obtained by performing active-space no-core configuration-interaction calculations using a two-body nucleon-nucleon interaction derived from chiral effective field theory. 
Entanglement measures within single-particle bases exhibiting different degrees of complexity
are determined,
 in particular, harmonic oscillator (HO), Hartree-Fock (HF), natural (NAT) and variational natural (VNAT) bases. 
 Specifically, single-orbital entanglement entropy, two-orbital mutual information and negativity are studied.
 \item[Results] 
 The entanglement structures in 
 $^4$He and $^6$He are found to be more localized within NAT and VNAT bases than within a HO basis for the optimal HO parameters we have worked with.
  In particular a core-valence structure clearly emerges from the full no-core calculation of $^{6}$He. 
  The two-nucleon mutual information shows that the VNAT basis, which typically exhibits good convergence properties, effectively decouples the active and inactive spaces. 
 \item[Conclusions] 
 Measures of one- and two-nucleon entanglement are found to 
be useful in analyzing the structure of nuclear wave functions,
in particular the efficacy of basis states,
and may provide useful metrics toward developing more efficient 
schemes for ab initio computations of the structure and reactions of nuclei, and
quantum  many-body systems more generally.
\end{description}
\end{abstract}

\maketitle
\section{Introduction}
\noindent
Developing quantitative first-principles predictive capabilities for computing the structure and reactions of nuclei remains a grand challenge in nuclear physics research.  
From a fundamental standpoint, nuclei emerge from  quantum chromodynamics (QCD)~\cite{Politzer:1973fx,Gross:1973id} and the standard model of electroweak interactions~\cite{Glashow:1961tr,Weinberg:1967tq,Salam:1968rm} at  low-energies,
and display a delicate balance between classical and quantum physics. 
A compelling explanation remains to be uncovered for why nuclei can be approximately described by collections of nucleons with a hierarchy of two-, three- and higher-body forces, rather than a single composite of quarks and gluons.  Lattice QCD~\cite{Wilson:1974sk,Creutz:1979kf,Balian:1974ts} calculations have shown that this emergence persists  over a significant range of standard model parameters beyond the physical light quark masses~\cite{Beane:2012vq,Beane:2013br,Yamazaki:2015asa}.

Nuclear structure calculations have advanced dramatically since the 
1970's by building interactions around the approximate global chiral symmetries of QCD~\cite{Weinberg:1978kz,Weinberg:1990rz,Weinberg:1991um}, 
by utilizing renormalization group techniques~\cite{Kaplan:1998tg,Kaplan:1998we,Szpigel:1999gf,Bogner:2003wn} to effectively smooth the short-range nature of the nuclear interactions 
(and electroweak or beyond standard model operators) in a way that is consistent with flowed nuclear many-body wavefunctions, 
by major advances in high-performance computing,
and advances in algorithms for computing quantities related to nuclear many-body systems.
Before these advances, it was thought that reliable calculations would not be achievable because, 
for example, 
the repulsive-core of NN interactions coupled  large numbers of many-body states, rendering a converged  diagonalization of the many-body Hamiltonian impractical.
These new capabilities are enabling precision calculations of properties of light and medium nuclei, see for example Refs.~\cite{Carlson:2014vla,Carlson:2017ebk,Gandolfi:2017arm,Contessi:2017rww,Bansal:2017pwn,Lonardoni:2018nob,Johnson:2018hrx,King:2020wmp,BARRETT2013131,PhysRevLett.107.072501,Vary:2015dda,HERGERT2016165,PhysRevLett.118.032502,Hagen:2013nca,Lahde:2013uqa,PhysRevC.89.024323,PhysRevLett.111.062501}.

In ways closely resembling effective field theory (EFT) constructions in perturbative quantum field theories (QFTs),  the  development of low-energy effective theories of nuclei that faithfully reproduce low-energy observables and formulated in terms of effective Hamiltonians is conceptually well understood.
Such constructions are complicated by the nonperturbative nature of the nuclear systems, and the role of induced multi-nucleon forces, with a power-counting that is more complicated and less obvious than EFTs, and invariance under systematic changes to the model space highlight the evolution of relevant operator structures, see for example, Refs.~\cite{Haxton:2002kb,Bogner:2003wn,Signoracci:2010bz}.  
The faithful reproduction of results requires including all of the relevant low-energy degrees of freedom in the active model space, and short-distance operators alone cannot substitute.  
Therefore, using the appropriate effective single nucleon states, or states that are perturbatively close,
that is to say that choosing a "good" single nucleon basis, significantly impacts the cost of numerical computation (see, for example, Ref.~\cite{Lietz:2016qfb}) and accuracy of results.

Efforts to better understand 
the role of quantum information and entanglement 
in quantum many-body systems and quantum field theories have begun.
These include investigations of information (Shannon) and von Neumann entropies to study the complexity of nuclear states and chaotic behavior in nuclei \cite{ZELEVINSKY1995141,ZELEVINSKY199685,PhysRevE.58.56,VOLYA200327,PhysRevC.88.044325}.
In  higher energy processes, the role of entanglement in dynamical processes related to QCD, such as fragmentation~\cite{Berges:2017zws,Berges:2017hne,Berges:2018cny}, 
heavy-ion collisions~\cite{Ho:2015rga,Kovner:2015hga,Kovner:2018rbf,Armesto:2019mna}, 
and deep inelastic scattering~\cite{Kharzeev:2017qzs,Tu:2019ouv} is being examined, 
and  suggestive hints have been found in the results of experiment~\cite{Baker:2017wtt}.
Recently, it has been shown that chiral symmetry and entanglement are 
interconnected in describing the  decomposition of the nucleon spin~\cite{Beane:2019loz}.
Further, there are indications that entanglement may play an important role in the power counting hierarchy of effective theories of nuclear forces~\cite{Beane:2018oxh}.  
In particular, it is found that 
entanglement preserving low-energy strong interactions lead to enhanced global emergent spin-flavor symmetries~\cite{Beane:2018oxh}, such as Wigner's SU(4) symmetry for two light quarks~\cite{Wigner:1936dx} and SU(16) symmetry 
for three light quarks~\cite{Wagman:2017tmp},
consistent with t-channel exchanges of the $\sigma$-field with $I=J=0$. 
These are symmetries beyond those present in the QCD Lagrange density, and also beyond those predicted in the large-$N_c$ limit of QCD~\cite{Kaplan:1995yg,Kaplan:1996rk}.~\footnote{It is interesting to note that the hierarchy of nuclear forces appears to be somewhat insensitive to $N_c$ also~\cite{Detmold:2014kba}.}
Exploiting such emergent symmetries can be used to mitigate the sign problem in Monte Carlo studies of light nuclei, through a two-step algorithm, with and without the symmetry-violating interactions, using  adiabatic projection techniques~\cite{Pine:2013zja,Elhatisari:2016hby}.
  The connections between entanglement, symmetries and sign problems are manifest in these computations, and remain to be better understood.
The appearance of entanglement hierarchy in nuclear effective field theories that could be more fundamental than the expansion parameters that have so far been identified, i.e. momentum and quark masses,
motivates us, in part, to explore the entanglement structure of nuclear many-body systems.
This is to begin to establish a phenomenological features of the entanglement structure of nuclei, to attempt to identify a better organizational many-body scheme, and to possibly uncover a connection between the nuclear EFT and many-body entanglement structures.

There has been remarkable progress during the last twenty years in understanding fundamental aspects of entanglement in quantum many-body systems and quantum field theories.  The concepts of bound- and distillable-entanglement, of importance for quantum communication and also for understanding the nature of quantum systems, are two such developments. 
Useful measures of bi-partite and multi-partite entanglement have been developed that have different sensitivities to these forms of entanglement, for example, 
entanglement entropy, mutual information, negativity, log-negativity, tangle and concurrence, see, for example, Refs.~\cite{Hill_1997,yczkowski_1998,Belavkin_2002,eisert2006entanglement,RevModPhys.81.865,PhysRevA.65.032314,Plenio_2005}. 
\\
In atomic nuclei, various partitions can be applied to the wave function, providing information on the nature of entanglement between different components of the nucleus.
For example, the nuclear wavefunction can be  written as a superposition of tensor products of proton and neutron configurations, leading to a natural bi-partitioning of the nuclear state to investigate entanglement between proton and neutron subsystems.
A first study of this type of entanglement 
was recently undertaken in inspiring work by Gorton and Johnson~\cite{GortonThesis,GortonJohnson2019a}.
They explored the behavior of the corresponding von Neumann entropy in an effort to identify the Slater determinants that dominate the entanglement between the two sectors. In the context of phenomenological shell-model calculations, they found that the
entanglement entropy decreases with growing isospin asymmetry, and increases with excitation energy. They are moving toward identifying and using a "weak entanglement approximation" for computational purposes. 
\\
It is interesting to also investigate the entanglement between single nucleon states.
Such entanglement is conceptually more challenging due to indistinguishability in collections of protons and collections of neutrons. This is a well-known and still debated issue~\cite{Shi2003,Benatti:2014gaa,RevModPhys.81.865,LoFranco2016}. It lies in the fact, that, when dealing with identical particles, the Hilbert space formulation of the many-body state does not have a tensor-product structure, which prevents partitioning of the system. One solution is to work in the Fock space formulation (occupation number representation) and define entanglement between single-particle states, rather than between single particles~\cite{Shi2003}. 
In this formulation, entanglement naturally depends on the single-particle basis used to define the system, and excludes
entanglement due to the antisymmetry of the wave function.
This form of "orbital entanglement" or "mode entanglement" has been investigated in atomic and molecular systems (see, for example, Refs.\cite{RISSLER2006519,Boguslawski2015}). 
The entanglement between single-particle states in $^{64}$Ge in the framework of density matrix renormalization group (DMRG) using a phenomenological shell model interaction~\cite{PhysRevC.92.051303}
is one of the few studies that have been performed on this topic in nuclei. 
\\
In the present work, we investigate orbital entanglement in the context of configuration-interaction calculations of light Helium nuclei, $^4$He and $^6$He,
using an interaction derived from chiral effective field theory ($\chi$EFT).
Since practical nuclear structure calculations typically require truncations of the many-body wavefunction, 
it is known that the nature of the underlying single-particle basis is important as it can potentially accelerate the convergence of observables, such as energies and radii, with respect to the size of the model space \cite{Robin:2015,Robin:2016,Robin:2020,Constantinou:2016,Tichai:2018}. In this context, it is therefore interesting to explore the connection between the quality of a single-nucleon basis and its entanglement properties.
In this work, we establish underlying patterns of entanglement between single-particle states in bases used for nuclear structure calculations, specifically the harmonic oscillator (HO), Hartree-Fock (HF), "Natural" (NAT) and "Variational Natural" (VNAT) bases. 
In particular, we explore relations between the convergence of the ground-state energy and the containment of entanglement within the active model space.
We focus on the distribution of single-nucleon-state entanglement entropy, two-nucleon-state mutual information and negativity to reflect bound and distillable entanglement.

\section{Measures of Entanglement in $^4$He} \label{sec:4He}
\noindent
The eigenstates of nuclei, denoted as $\ket{\Psi}$, can be written as linear combinations of Slater determinants $\ket{\phi_\alpha}$ of nucleon wavefunctions, which can be decomposed into neutron ($\nu$) and proton ($\pi$) components:
\begin{eqnarray}
\ket{\Psi} &=& \sum_\alpha \mathcal{A}_\alpha \ket{\phi_\alpha}  \label{eq:wf1} \\
           &\equiv& \sum_{\alpha_\pi \alpha_\nu} \mathcal{A}_{\alpha_\pi \alpha_\nu} \ket{\phi_{\alpha_\pi}} \otimes \ket{\phi_{\alpha_\nu}} \; . 
\end{eqnarray}
Each Slater determinant $\ket{\phi_\alpha}$ represents a configuration of nucleons in a basis of single-particle states $\{ i \}$:
\begin{equation}
\ket{\phi_\alpha} = \prod_{i \in \alpha} a^\dagger_i \ket{0} \; ,
\end{equation}
where $\ket{0}$ denotes the true particle vacuum.
The basis states are denoted by their quantum numbers $\{ i \}= \{n_i,l_i,j_i,m_i,\tau_i\}$ (principal quantum number, orbital angular momentum (AM), total AM, total AM projection, and isospin projection), and can be, for example, states of a HO, or states associated with a self-consistent potential.
We will refer to these basis states alternatively as single-particle states, or orbitals\footnote{Note that the terminology used here differs from the one used in quantum chemistry studies, such as Refs.~\cite{RISSLER2006519,Boguslawski2015}, in which "orbitals" can usually be doubly occupied.}.
\noindent To easily access the measures of entanglement we are focusing on, 
the expansion in Eq.~(\ref{eq:wf1}) is rewritten in terms of the single-particle occupation states.
In this occupation number formalism the Slater determinants read
\begin{eqnarray}
\ket{\phi_\alpha} = | n^\alpha_1  n^\alpha_2 ... n^\alpha_N \rangle \equiv \bigotimes_{i=1}^N |n_i^\alpha\rangle\; ,
\end{eqnarray}
where $N$ is the total number of single-particle states, and $n^\alpha_i$ is the occupation number of state $i$ in the configuration $\ket{\phi_\alpha}$, \textit{i.e.}
\begin{numcases}{}
n^\alpha_i = 0 \mbox{ if $i$ is empty in configuration} \ket{\phi_\alpha}  \nonumber \\
n^\alpha_i = 1 \mbox{ if $i$ is occupied in configuration} \ket{\phi_\alpha} \; .
\nonumber
\end{numcases}
The sum of the occupation numbers is equal to the total number of nucleons,
$n_1 + .... n_N = A$.
The many-body wavefunction can then be written as
\begin{eqnarray}
\ket{\Psi} &=& \sum_{\alpha} \mathcal{A}_{\alpha} \ | n^\alpha_1  n^\alpha_2 ... n^\alpha_N \rangle \; , \nonumber \\
                &\equiv& \sum_{n_1 ... n_N} \mathcal{A}_{n_1 ... n_N}\  | n_1  n_2 ... n_N \rangle \; . \label{eq:new_wf}
\end{eqnarray}
In defining the density matrix, it is convenient to consider 
the Hilbert space of a nucleus in terms of three spaces, $A$, $B$ and $C$, that span the entire space.
For the purposes of this work, while more general assignments can be made, we assign $A$ and $B$ to be single-nucleon basis states, while $C$ includes the remaining states.  The nuclear wavefunction yields a density operator 
$\hat\rho_{ABC}$.
To determine two-nucleon measures of entanglement, the states in $C$ are traced over, to give 
$\hat\rho_{AB} = {\rm Tr}_C\left[ \hat\rho_{ABC} \right]$.
Similarly, to determine single-nucleon measures of entanglement, the $B$ space is traced over,
$\hat\rho_{A} = {\rm Tr}_B\left[ \hat\rho_{AB} \right]$.
The von Neumann entanglement entropy associated with a density matrix $\hat\rho_A$ is defined as $S(\hat\rho_A) = -{\rm Tr} \hat\rho_A \log \hat\rho_A$. 
The mutual information (MI) between states $A$ and $B$
is defined as $I(A:B) = S(\hat\rho_A) +S(\hat\rho_B) -S(\hat\rho_{AB})$, and 
represents a measure of classical and quantum correlations, bound entanglement and distillable entanglement.
An upper limit to the distillable entanglement in $A$ and $B$ is defined by the negativity 
${\cal N}(\hat\rho) = (||\hat\rho_{AB}^{\Gamma_A}||_1-1)/2$,
and the related logarithmic-negativity, $E_{\cal N}(\hat\rho_{AB}) = 
\log ||\hat\rho_{AB}^{\Gamma_A}||_1$, where $||\hat\rho_{AB}^{\Gamma_A}||_1$ is the 1-norm of the partial transpose of the density matrix with respect to  $A$. 
\subsection{Methods} \label{sec:method}
Many-body configurations $\ket{\phi_\alpha}$ are selected in terms of a truncation in the number of shells of the single-particle basis. That is, we include all possible configurations in an active model space containing a given number of major shells $N_{{\rm tot}}$, that is varied.
The calculations are performed using a two-body interaction derived from $\chi$EFT. In particular, the bare NNLO$_{{\rm opt}}$ interaction~\cite{NNLO_OPT_2013}, 
with counterterms that have been fit in order to 
minimize explicit three-body forces is used.
Starting from a HO basis, the goal of our study is to investigate how entanglement evolves and rearranges while optimizing and modifying  the single-particle states.
In particular, measures of entanglement are investigated when the nuclear state $\ket{\Psi}$ is expanded in 
\begin{enumerate}
    \item a HO single-particle basis;
    
    \item a HF single-particle basis obtained by performing an {\it a priori} Hartree-Fock calculation using the NNLO$_{{\rm opt}}$ interaction;
    
    \item a "natural" (NAT) basis that diagonalizes the one-body density matrix $\gamma_{ij}=\braket{\Psi|a^{\dagger}_j a_i|\Psi}$. This basis is obtained by performing a  diagonalization of the two-body Hamiltonian matrix in the many-body configuration space spanned by a HO basis (with $N_{{\rm tot}}$ shells). Once the expansion coefficients $\{\mathcal{A}_\alpha\}$ in Eq.~(\ref{eq:new_wf}) are obtained, the one-body density is computed and diagonalized to obtain the "natural" single-particle states;
    
    \item what we will refer to as "variational natural" (VNAT) basis. This basis is obtained by applying a variational principle to the energy of the correlated state $\ket{\Psi}$, with respect to the single-particle orbitals. This leads to a non-linear equation where the one-nucleon states incorporate the effect of two-body correlations. Specifically,
\begin{equation}
\left[ \hat{h}(\gamma) , \hat\gamma \right] = \hat{G}(\sigma) \; ,
\label{e:VNAT}
\end{equation}
is solved.
In Eq. (\ref{e:VNAT}), $\hat{h}(\gamma) $ is a general mean-field Hamiltonian:
\begin{eqnarray}
{h}_{ij}(\gamma)  &=& K_{ij} + \sum_{kl} \braket{ik|\widetilde{V}^{NN}|jl} \gamma_{lk} \; ,
\label{e:mean-field}
\end{eqnarray}
where $K$ denotes the intrinsic kinetic energy and $\widetilde{V}^{NN}$ the anti-symmetrized two-body interaction.
$\sigma$ denotes the two-body correlation matrix of the state $\ket{\Psi}$:
\begin{eqnarray}
\sigma_{il,jk} &=& \braket{\Psi|a^{\dagger}_i a^{\dagger}_j a_{k} a_{l}|\Psi} - \gamma_{li} \gamma_{kj} + \gamma_{lj} \gamma_{ki} \; ,
\end{eqnarray} 
and $G(\sigma)$ is the source term containing the effect of two-body correlations beyond the mean-field $h(\gamma)$:
\begin{eqnarray}
G(\sigma)_{ij} &=&  \frac{1}{2} \sum_{klm} \sigma_{ki,lm} \braket{kl|\widetilde{V}^{NN}| {jm}} \nonumber \\
               &&  - \frac{1}{2}  \sum_{klm} \braket{ik|\widetilde{V}^{NN}| {lm}}  \sigma_{jl,km}  \; .
\label{e:G}                        
\end{eqnarray}
The single-particle states are taken as eigenfunctions of the one-body density $\gamma$ which satisfies the variational equation given in Eq.~(\ref{e:VNAT}). As Eq.~(\ref{e:VNAT}) is coupled to Eq.~(\ref{eq:new_wf}),  both equations are solved iteratively until convergence of the system is achieved. More details on the practical procedure can be found in Ref.~\cite{Robin:2015}.
We note that this approach is usually called 
multi-configurational self-consistent field (MCSCF) or
multi-configurational Hartree-Fock (MCHF) in quantum chemistry. \\
\end{enumerate}
When the model space involves a truncation of the single-particle basis, the ordering of the orbitals matters. In the calculations, the HO states are ordered by their quantum numbers by increasing values of $N=2n+l$ and decreasing angular momentum $j$. The HF states are ordered by increasing single-particle energies, while the NAT and VNAT orbitals are ordered by decreasing occupation numbers.
\\
For our calculations, all single-particle bases are expanded on a set of 7 HO shells with frequency $\hbar\Omega = 30$ MeV. This value was found to be the optimal frequency in terms of energy minimization when expanding the wavefunction on the HO basis.


\subsection{Entanglement entropy of single-particle states}
It is interesting to start by  evaluating the entanglement of one single-nucleon state, or orbital, ($A \equiv i$) with the rest of the basis ($B \cup C$), within the nuclear ground state. This is achieved by calculating the single-orbital reduced density matrix $\rho^{(i)}$,
which can be obtained by performing permutations in Eq. (\ref{eq:new_wf}):
\begin{eqnarray}
\ket{\Psi} &=& \sum_{n_1 ...n_i... n_N} \mathcal{A}_{n_1 ...n_i... n_N} | n_1  n_2 ... n_i ... n_N \rangle \nonumber \\
               &=& \sum_{n_1 ...n_i... n_N} \mathcal{A}_{n_1 ...n_i... n_N}  \times \varphi_i \nonumber \\
               && \hspace{1.5cm} \times | n_1  n_2 ... n_{i-1} n_{i+1} ... n_N \rangle  \otimes  \ket{n_i} \nonumber \\
               & \equiv & \sum_{n_i, BC} \mathcal{A}_{ BC n_i}  \times \varphi_i \, \ket{BC} \otimes \ket{n_i} \; ,
\end{eqnarray}
where ${BC} \equiv   (n_1  n_2 ... n_{i-1} n_{i+1} ... n_N )$ and $\varphi_i$ is the phase resulting from the permutation. 
\\
The one-orbital reduced density matrix $\rho^{(i)}_{n_i,n_i'}$ (i.e. the matrix elements of $\hat\rho_A$) becomes
\begin{eqnarray}
\rho^{(i)}_{n_i,n_i'}  = \sum_{ BC } \bra{BC } \braket{n_i | \Psi}  \braket{\Psi | n_i'} \ket{BC} \; .
\end{eqnarray}
$\rho^{(i)}$ 
is then simply a 2x2 matrix with elements that can be written in terms of the diagonal elements of one-nucleon density matrix $\gamma_{ii} = \braket{\Psi | a^\dagger_i a_i  |\Psi} $ as
\begin{eqnarray}
\rho^{(i)} = 
\begin{pmatrix}
1 - \gamma_{ii}  & 0 \\
0 & \gamma_{ii}
\end{pmatrix}
\; ,
\label{e:rho_1}
\end{eqnarray}
in the occupation number basis,
$|n_i\rangle 
= \{ |0\rangle , |1\rangle \}$. The derivation of Eq.~(\ref{e:rho_1}) is given in Appendix~\ref{app1}.
The single-orbital entanglement entropy, $S_{i}^{(1)}$, 
characterizing the entanglement between single-particle state $i$ and the other orbitals in the nucleus, is 
\begin{eqnarray}
S_{i}^{(1)} = - 
{\rm Tr}\left[ \rho^{(i)} \ln \rho^{(i)} \right] 
= 
- \sum_{k = 1}^2 \omega_k^{(i)} \ln \omega_k^{(i)} \; ,
\end{eqnarray}
where $\omega_k^{(i)}$ are the eigenvalues of  $\rho^{(i)}$. 
\\ \\
The entanglement entropy acquires its maximum value, $\ln(2)$, 
when the single-particle state $i$ has an occupation number of $\frac{1}{2}$,
and vanishes when the state is fully occupied or empty.
Therefore, if the nuclear state 
$|\Psi \rangle$ reduces to a single Slater determinant (in the single-particle basis $\{i\}$), the entanglement entropy $S^{(1)}$ is zero in that basis. \\ \\
\begin{figure}
\centering{\includegraphics[width=\columnwidth]{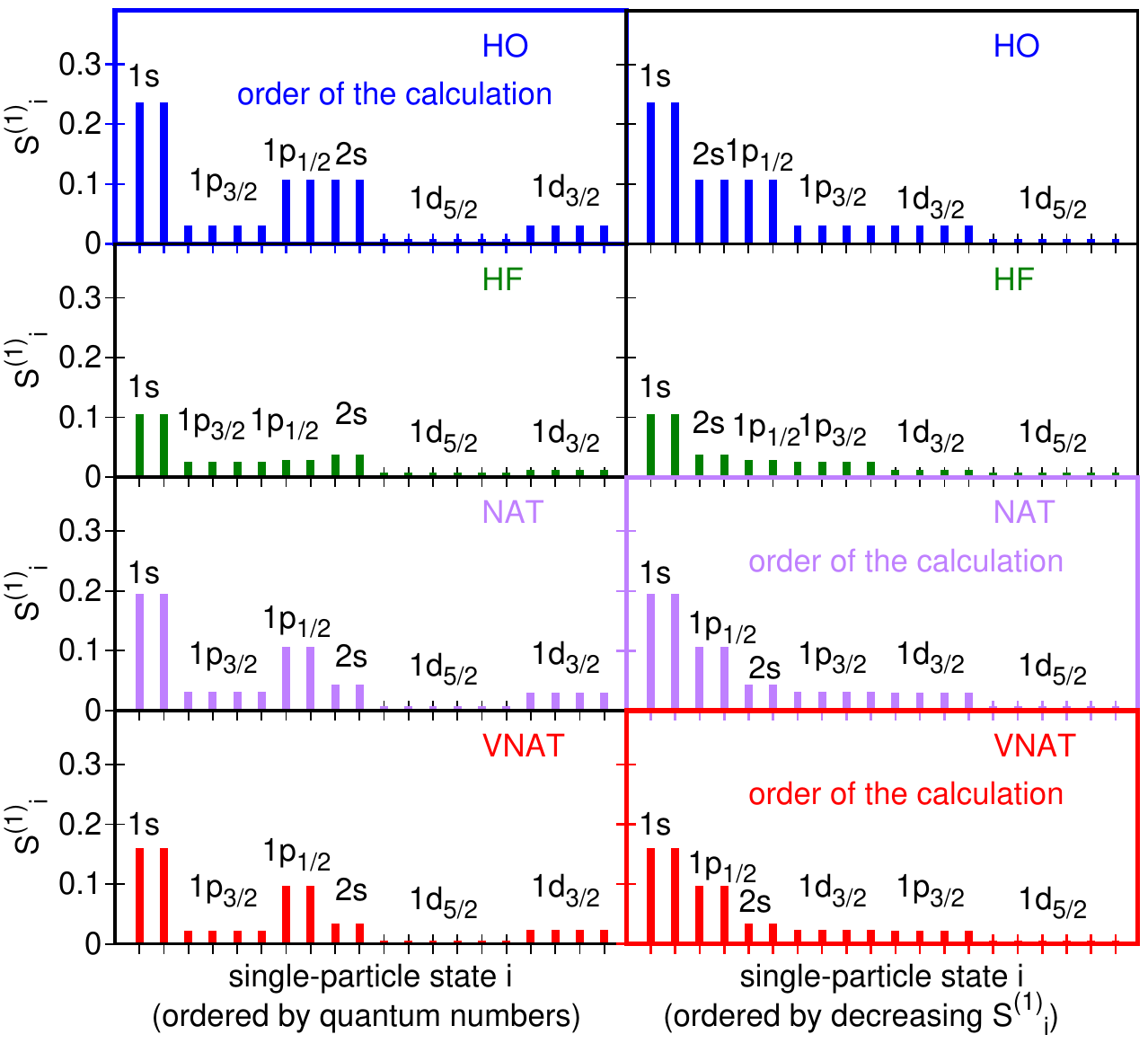} }
\caption{
(Color Online) Single-orbital entanglement entropy $S^{(1)}_i$ of the HO, HF, NAT and VNAT single-neutron states $i=(n_i,l_i,j_i,m_i,\tau_i=-1/2)$ in $^{4}$He, obtained with a model space of 3 shells.
The left panels shows $S^{(1)}$
ordered by the states in the HO basis, while the right panels correspond to ordering by decreasing values of $S^{(1)}$, which coincides with the ordering of the calculation (by occupation number) for the NAT and VNAT bases.}
\label{f:S1_4He_3shells}
\end{figure}
\begin{figure*}
\centering{\includegraphics[width=\textwidth]{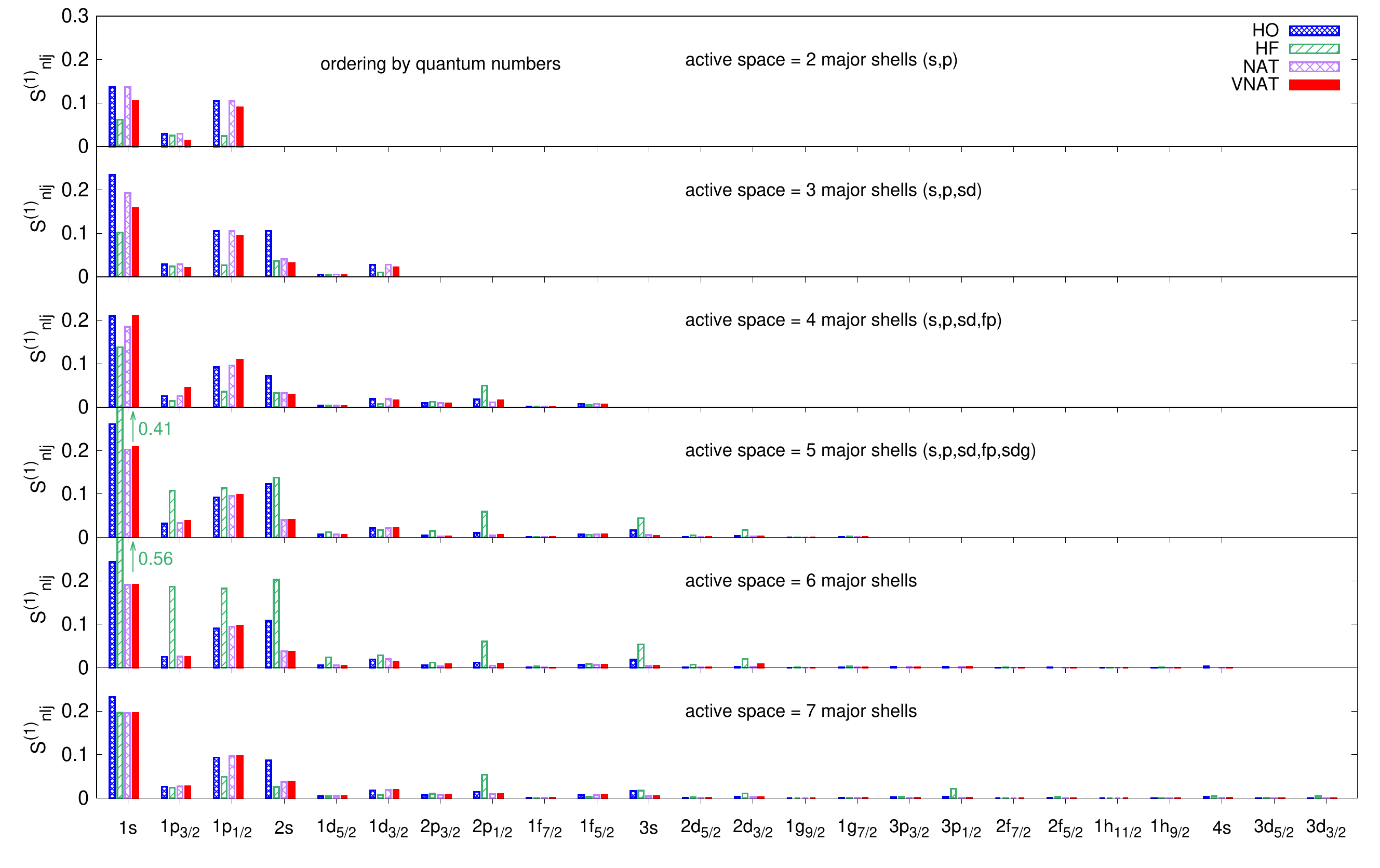} }
\caption{(Color Online) Single-orbital entanglement entropy $S^{(1)}_{nljm} = S^{(1)}_{nlj}$ of the HO, HF, NAT and VNAT single-neutron states in $^{4}$He, for different sizes of the active model space.}
\label{f:S1_4He}
\end{figure*}
Figure~\ref{f:S1_4He_3shells} shows the entanglement entropies $S^{(1)}$ of the active neutron single-particle states in the $^4$He ground state wavefunction obtained with the HO, HF, NAT and VNAT bases, with a model space of $N_{tot}=3$ active shells. Due to the spherical symmetry, the entanglement entropy of states with same $nlj$ and different AM projections $m_i$ are equal: $S^{(1)}_{nlj,m_i} \equiv S^{(1)}_{nlj}$.
The entropies of the proton orbitals are very similar to the neutron ones and  are not shown. 
To allow for a direct comparison, the states on the left panels have been ordered by quantum numbers (decreasing $N=2l+1$ and increasing $j$, corresponding to the ordering of the HO states).
On the right panels the states are ordered by decreasing values of $S^{(1)}$. We note that in the practical calculation with the NAT and VNAT bases, the states are ordered by decreasing occupation numbers, which naturally coincides with the ordering in decreasing $S^{(1)}$. 
Typically, the entanglement 
entropy of a state with given quantum numbers is smaller in the NAT and VNAT bases than in the HO basis, and, in particular, the entropy of the VNAT $1s$ and $2s$ orbitals are importantly decreased compared to the HO states. 
The HF states exhibit small entanglement entropy for this truncation of the model space. We will comment more extensively on the HF basis in the next paragraph.
\\ \\
Figure~\ref{f:S1_4He} shows the evolution of the entanglement entropy $S^{(1)}_{nlj}$ as  the size of the model space is varied from two to seven major shells. In this figure the states are ordered by quantum numbers. We note that due to mixing of high-lying single-particle states during the self-consistent procedure, and in order to have a consistent truncation between the different bases the model space with "6 major shells" includes the first 114 single-particle states. This means that in the HO basis the $4s$ subshell is included in that model space.
Examining Fig.~\ref{f:S1_4He}, in an active space comprising only $N_{tot}=2$ major shells, the single orbital entanglement entropy is underestimated. This is because such a small model space cannot accommodate sufficient correlation, and the wave function resembles a Slater determinant. 
When increasing the number of active shells, the conclusions drawn for Fig.~\ref{f:S1_4He_3shells} also apply. The entanglement entropy appears to stabilize somewhat more rapidly with increasing size of the active space in the NAT and VNAT bases, compared to the HO basis.
This can be seen more clearly in Table~\ref{t:S1_tot}, which shows the sum of one-orbital entanglement entropy over the active single-particle states
$S^{(1)}_{tot}=\sum_i S^{(1)}_{i}$. 
In the HO basis, $S^{(1)}_{tot}$ is found to fluctuate somewhat with increasing  model space, even for modestly large numbers of shells. In the NAT and VNAT basis the total entropy starts to stabilize 
with a model space of 4 shells around a value of 1.0, and is found to be 
systematically smaller compared with other bases. 
In fact, it was shown in Ref.~\cite{PhysRevA.92.042326} that  $S^{(1)}_{tot}$ is minimized in the eigenbasis of the one-body density $\gamma$.
Keep in mind that, in the present set of calculations, the full configuration space is exhausted for an active space of 7 shells. In that case Eq.~(\ref{e:VNAT}) is automatically fulfilled, and the NAT and VNAT bases coincide.
\\
Overall the HF basis exhibits non-convergent behavior. The entanglement entropy of the lowest HF single-particle states (in terms of ordering by quantum numbers) is consistently underestimated compared to the other bases for model spaces of 2,3, and 4 shells. However a jump occurs when including a fifth shell. This can be somewhat understood from the composition of the nuclear state, which exhibits a large 0p-0h component ($\approx 98-94 \%$) for small model spaces, while  decreasing to $\approx 70 \%$ when the 
5$^{\rm th}$  shell is included, and to $\approx 53 \%$ when including the 6$^{\rm th}$ shell. 
The $3p$ and $4s$ shells are pushed out to the end of the basis during the HF calculations, and are not present in the model space with 6 shells, contrarily to the other bases. This is also found when using softer interactions. 
The 0p-0h component then increases to $~91 \%$ when the model space exhausts the full configuration space (7 shells), and the entanglement entropy of the lowest orbitals reduces dramatically. 
These results are consistent with pathologies in the convergence of the ground-state energy in the HF basis~\cite{Tichai:2018}. 
Of course, $^4$He is a light nucleus where a mean field cannot be firmly established. Therefore, we would have to perform calculations in heavier systems in order to see if these pathologies persist.
For this reason, in the following discussions of entanglement, we will focus on the HO, VNAT (and NAT) bases, and not consider further the HF basis. \\ \\
\begin{table}
 \centering
\begin{tabular}{ccccc} 
\hline   
\hline
$N_{tot}$ & HO        & HF   & NAT    & VNAT\\
\hline
2 shells & 0.596     & 0.270 & 0.596 & 0.441 \\
3 shells & 1.143     & 0.487 & 0.929 & 0.746\\
4 shells & 1.065     & 0.686 & 0.928 & 1.063 \\
5 shells & 1.348     & 2.327 & 1.036 & 1.042\\
6 shells & 1.264     & 3.434 & 0.972 & 0.963\\ 
7 shells & 1.217     & 1.069 & 1.006 & 1.006 \\ 
\hline
\hline
\end{tabular}
\caption{Sum of single-orbital entanglement entropies $S^{(1)}_{tot}=\sum_i S^{(1)}_{i}$ for the different bases and with different number of shells $N_{tot}$ in the active model space.}
\label{t:S1_tot}
\end{table}
\begin{table}
 \centering
\begin{tabular}{ccccc} 
\hline   
\hline
$N_{tot}$ & HO           & HF   & NAT    & VNAT\\
\hline
2 shells & -19.15     & -16.58 & -19.15 &  -21.40 \\
3 shells & -23.29     & -17.35 & -23.29 &  -24.89 \\
4 shells & -25.72     & -20.48 & -25.72 &  -26.61 \\
5 shells & -26.88     & -23.37 & -26.88 &  -27.27 \\
6 shells & -27.44     & -23.81 & -27.44 &  -27.47 \\ 
7 shells & -27.50     & -27.50 & -27.50 &  -27.50 \\ 
\hline
\hline
\end{tabular}
\caption{Ground-state energy of $^4$He (in MeV) obtained with the different bases and with different number of shells $N_{tot}$ in the active space. All bases being expanded on 7 HO shells, they all lead to the same energy when the active space comprises 7 shells and exhausts the full configuration space. }
\label{t:E_gs}
\end{table}

By looking at the one-orbital entanglement entropies it is difficult to distinguish the NAT and VNAT bases as they show very similar profiles (see {\it e.g.} Fig.~\ref{f:S1_4He_3shells}). However the convergence of the ground-state energy in Table~\ref{t:E_gs} shows that the VNAT states are marginally "better" than the NAT ones when comparing with the target value \footnote{In the present calculations, we have used an underlying harmonic oscillator frequency of $\hbar \Omega = 30$ MeV, which is found to be optimal in terms of energy minimization using the HO basis~\cite{Robin:2020}. The ground-state energy obtained with the HO and NAT bases importantly depends on this frequency, and the improvement obtained using the VNAT basis is found to be greater for other oscillator frequencies.}. With the present truncation scheme, the NAT basis leads to the same energy as the HO basis. This is understood as the natural orbitals only mix HO states that are partially occupied. In other words they only mix HO states that are within the active model space. Since the ground-state wavefunction (\ref{eq:new_wf}) includes all configurations in the active space, the ground-state energy is invariant. This is different with the VNAT basis, which, due to Eq.~(\ref{e:VNAT}), mixes both active and inactive HO orbitals. 
At this point, it is not obvious how measures of
entanglement could be used to distinguish the 
NAT and VNAT bases.  As entanglement is derived from reduced densities, it does not give information on the coupling between the active and inactive orbital spaces. We will attempt to address this issue  in Section~\ref{sec:NATvsVNAT} by considering measures of two-orbital entanglement.

\subsection{Two-orbital entanglement entropy and Mutual information}

The mutual information (MI) within a pair of single-particle states $(A,B)\equiv(i,j)$, 
can be determined from the two-orbital reduced density matrix obtained by tracing over the rest of the basis $C$,
\begin{eqnarray}
\rho^{(ij)}_{n_i n_j ,n_i' n_j'}  = \sum_{ C } \bra{C} \braket{n_j n_i | \Psi}  \braket{\Psi | n_i' n_j '} \ket{C} \; ,
\end{eqnarray}
(i.e., the matrix elements of $\hat\rho_{AB}$ discussed earlier)
where the nuclear wavefunction is structured as
\begin{eqnarray}
\ket{\Psi} 
               &=& \sum_{n_1 ...n_i...n_j... n_N} \mathcal{A}_{n_1 ...n_i...n_j... n_N}  \times \varphi_i  \varphi_j \nonumber \\
               && \times | n_1  n_2 ... n_{i-1} n_{i+1} ... n_{j-1} n_{j+1} ... n_N \rangle  \otimes  \ket{n_i n_j} \nonumber \\
               & \equiv & \sum_{n_i, n_j, C} \mathcal{A}_{ C n_i n_j}  \times \varphi_i  \varphi_j \, \ket{C} \otimes \ket{n_i n_j} \; .
\end{eqnarray}

As derived in Appendix~\ref{app2}, in the basis ${\ket{n_i n_j}}=\{\ket{00},\ket{01},\ket{10},\ket{11} \}$ 
the two-orbital reduced density matrix becomes
\begin{eqnarray}
 \resizebox{0.95\hsize}{!}{%
        $
\rho^{(ij)} = 
\begin{pmatrix}
1 - \gamma_{ii} - \gamma_{jj} +\gamma_{ijij}  & 0 & 0 & 0 \\
0 & \gamma_{jj} - \gamma_{ijij} & \gamma_{ji} & 0 \\
0 & \gamma_{ij} &\gamma_{ii} - \gamma_{ijij}  & 0 \\
0 & 0 & 0 &  \gamma_{ijij} 
\end{pmatrix}
\; , 
$%
}
\nonumber \\
\label{e:2RDM}
\end{eqnarray}
where $\gamma_{ij} = \braket{\Psi | a^\dagger_j a_i  |\Psi} $ denotes non-diagonal terms of the one-body density,  and $\gamma_{ijij} = \braket{ \Psi | a^\dagger_i a^\dagger_j a_j a_i| \Psi  } $ is an element of the two-nucleon density.
The two-orbital entanglement entropy becomes
\begin{eqnarray}
S_{ij}^{(2)} = - {\rm Tr}\left[ \rho^{(ij)} \ln (\rho^{(ij)}) \right] = - \sum_{k = 1}^4 \eta_k^{(ij)} \ln  \eta_k^{(ij)} \; ,
\end{eqnarray}
where $\eta_k^{(ij)}$ are the eigenvalues of $\rho^{(ij)}$.
The MI between these states becomes,
\begin{equation}
I(i:j)= \left( S_{i}^{(1)} + S_{j}^{(1)} - S_{ij}^{(2)} \right) \left( 1 - \delta_{ij} \right) \; .
\end{equation}
Consistent with works in quantum chemistry \cite{RISSLER2006519}, a factor of $\left( 1 - \delta_{ij} \right)$ has been introduced to ensure the vanishing of  the entanglement of a single-particle state with itself.
\\ \\
\begin{figure}[!ht]
\centering{
\includegraphics[width=0.85\columnwidth]{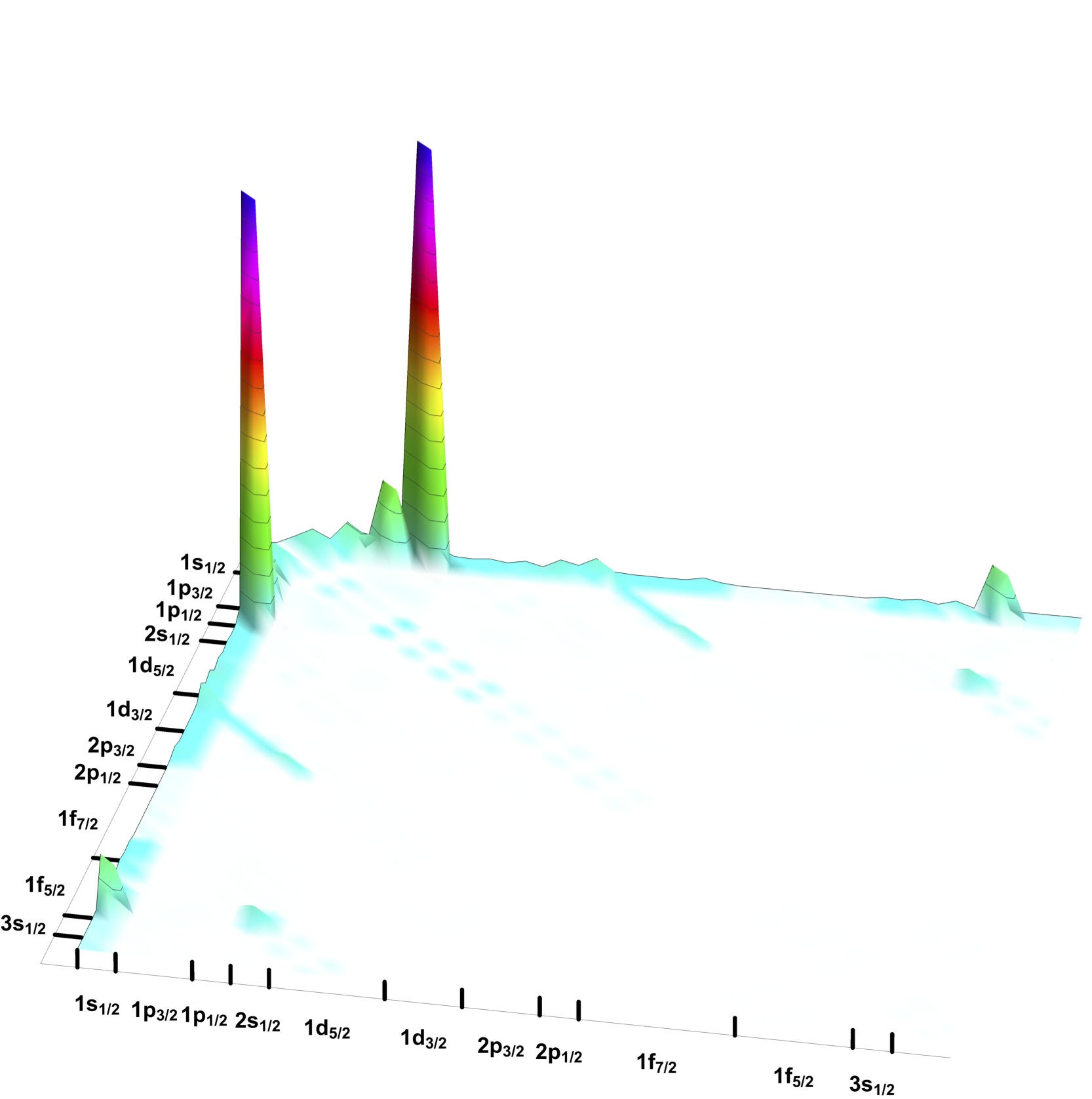}\\
\includegraphics[width=0.85\columnwidth]{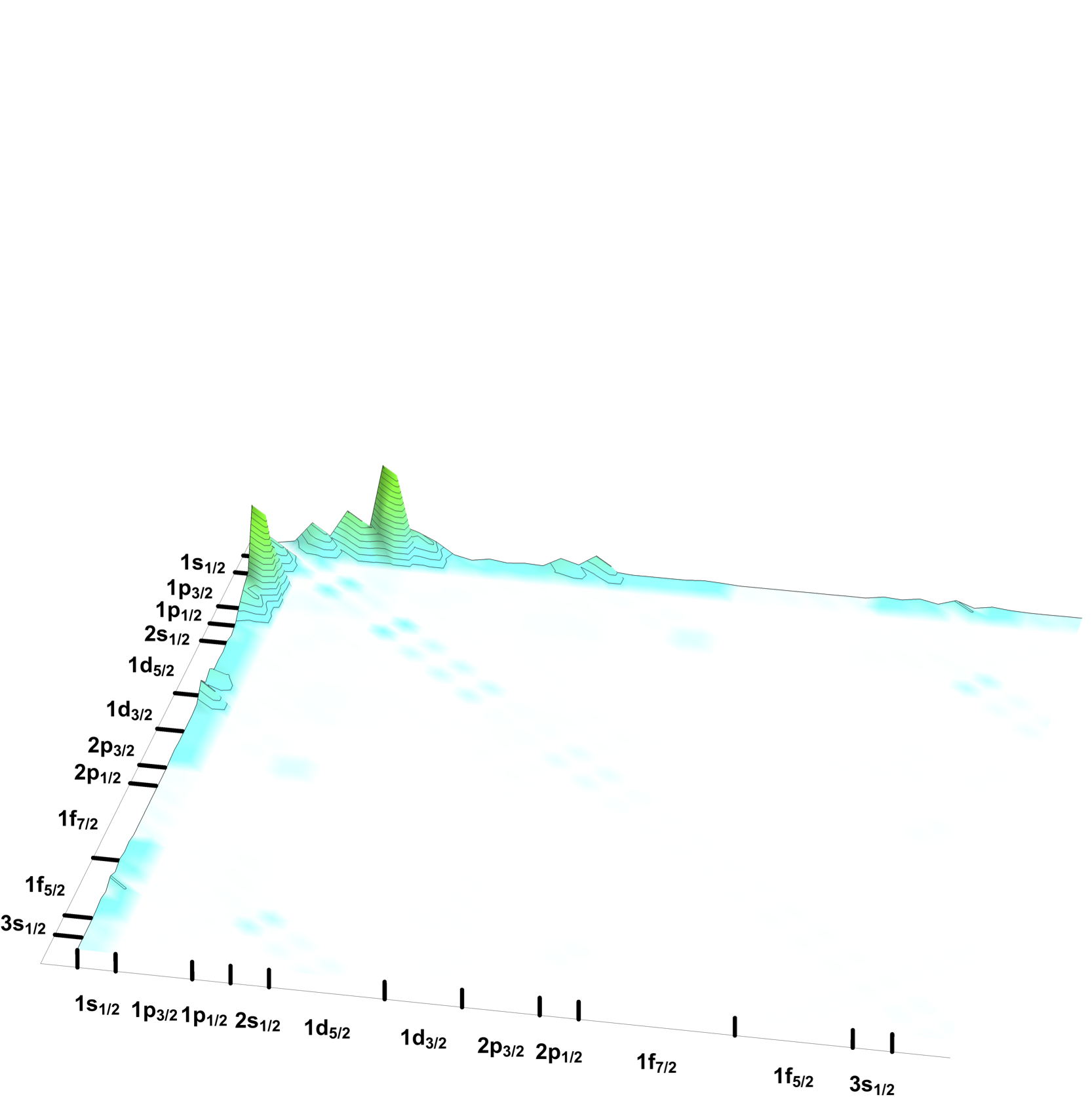}
}
\caption{(Color Online) The mutual information within the neutron-neutron orbitals of $^4$He using 5-shell active spaces in the HO (upper panel) and VNAT (lower panel) bases. The states are ordered by quantum numbers, and the vertical scales are the same.} 
\label{f:MI_nn_4He_3D}
\end{figure}

\begin{figure}[h]
\centering{\includegraphics[width=\columnwidth]{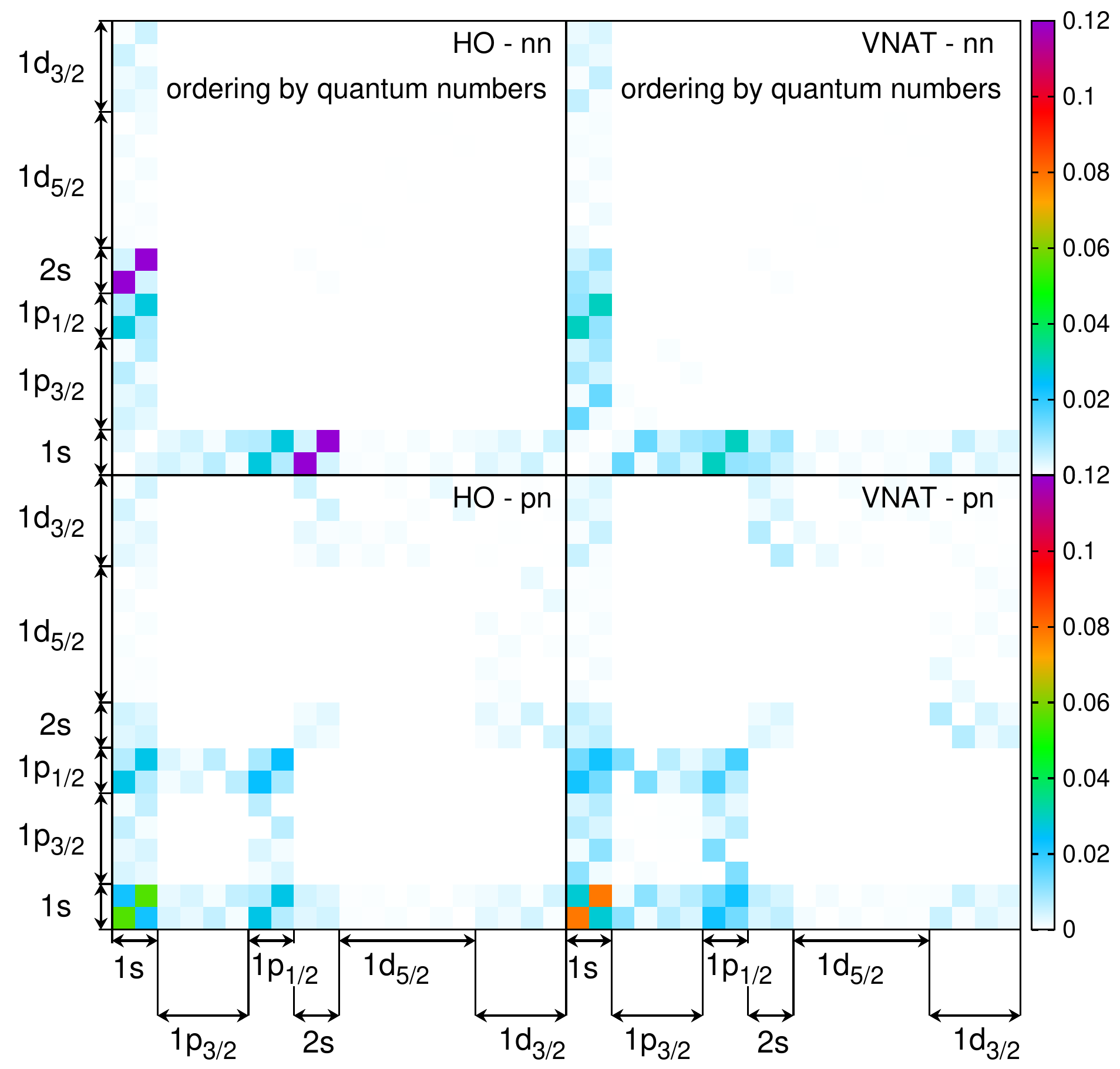}} 
\caption{(Color Online) Mutual information between two HO (left) and VNAT (right) states, obtained for a model space of $N_{tot}=5$ shells. Each pixel corresponds to the MI between states $i = (n_i,l_i,j_i,m_i,\tau_i)$. The states are ordered by quantum numbers $N_i=1n_i + l_i, j_i$ and AM projection $m_i=+1/2,-1/2,+3/2,-3/2$ etc. The top panels show neutron-neutron MI and the bottom panels show the proton-neutron MI. In the bottom panels, the proton (neutron) states are on the y (x) axis.} 
\label{f:MI_4He}
\end{figure}
Figure~\ref{f:MI_nn_4He_3D} shows the MI of two neutron orbitals (in 3D) computed from $^4$He ground state wavefunctions in the HO and VNAT bases using 5 active shells. In order to analyse the results more closely, Fig.~\ref{f:MI_4He} shows the neutron-neutron and proton-neutron MI (in 2D) for the lowest single-particle states. For direct comparison, the single-nucleon states are ordered by quantum numbers in both figures. 
\\
Let us first examine the neutron-neutron sector (Fig.~\ref{f:MI_nn_4He_3D} and top panels of Fig.~\ref{f:MI_4He}).  
In the HO basis, the most important correlations appear between states of the $1s$ shell and states of the $2s$ shell, with aligned AM projections. To a lesser extent MI between $1s$ and $3s$ shells, and between $1s$ and $1p_{1/2}$ orbitals are also important.
Remarkably, this large MI between the $s$ states is suppressed by 
approximately an order of magnitude in the VNAT basis, compared to the HO one. This is observed systematically when varying the size of the model space, as shown in Table~\ref{t:1s2sMInn}. 
Interestingly, the MI between time-reversed states (with same $nlj$ and opposite AM projection) of the $1s$, that could indicate isovector BCS-type neutron-neutron pairing, is weak in both bases.
\\

\begin{figure}[h]
\centering{\includegraphics[width=\columnwidth]{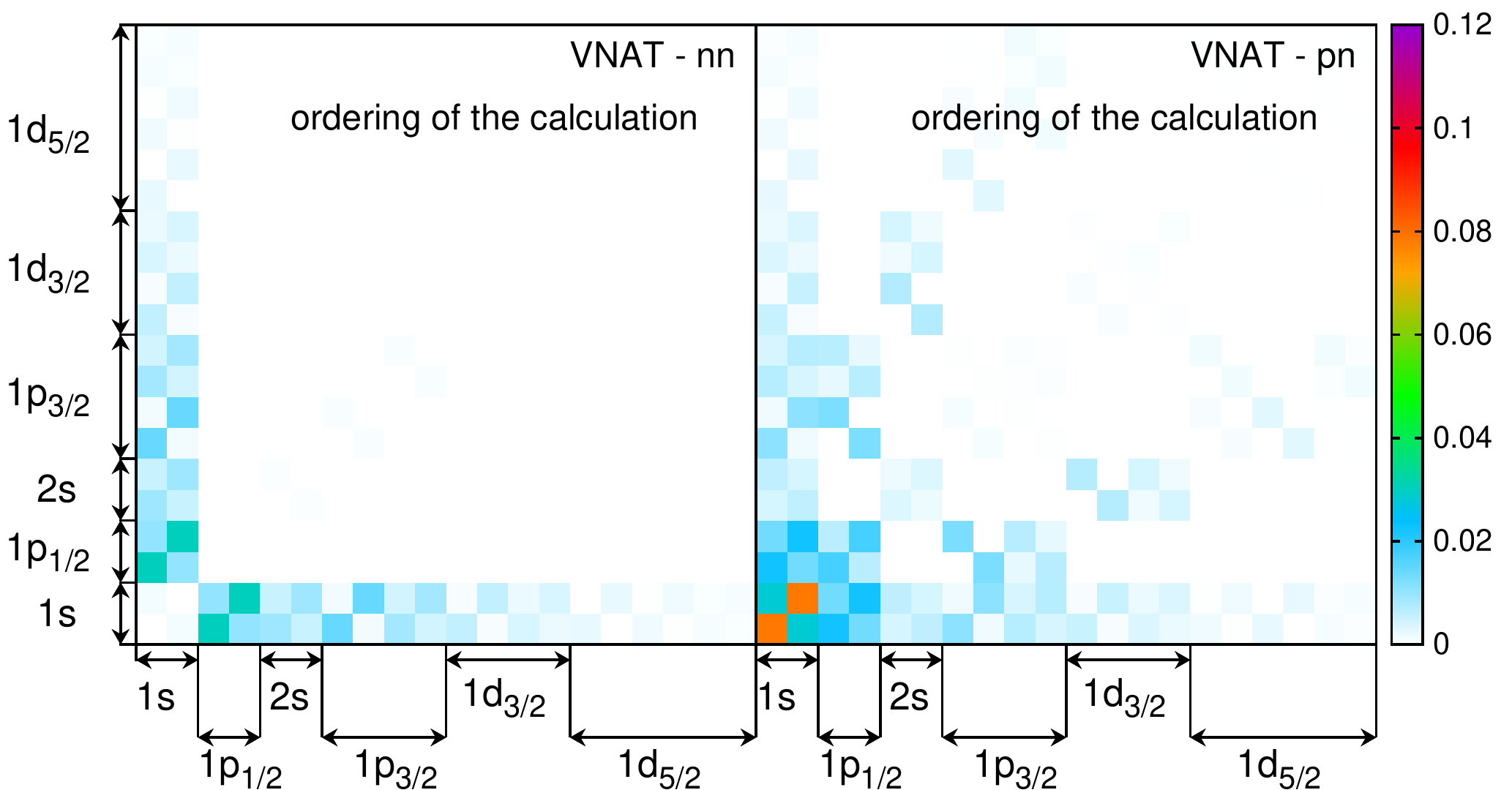}}
\caption{(Color Online) Neutron-neutron (left panel) and proton-neutron (right panel) MI obtained using VNAT orbitals ordered with occupation numbers, for $N_{tot}=5$ shells. In the right panel, the proton (neutron) states are shown on the y (x) axis.} 
\label{f:MI_VNAT_4He}
\end{figure}

\begin{table}[!ht]
\centering
\begin{tabular}{ccc} 
\hline   
\hline
$N_{tot}$ & HO           & VNAT\\
\hline
3 shells & 0.11      & 0.0040    \\ 
4 shells & 0.071     & 0.0047    \\ 
5 shells & 0.15      & 0.0091    \\
6 shells & 0.12      & 0.0075    \\
7 shells & 0.089     & 0.0083    \\ 
\hline
\hline
\end{tabular} 
\caption{The neutron-neutron mutual information of the $1s$-$2s$ orbitals.} 
\label{t:1s2sMInn}
\end{table}
In the proton-neutron sector (bottom panels of Fig.~\ref{f:MI_4He}), the strength of the MI remains similar in the HO and VNAT bases. This is  likely because the VNAT basis is obtained via a unitary transformation of the HO basis, which does not mix proton and neutron states, and thus, does not capture proton-neutron correlations. The most important couplings are of the type $1s$-$1s$, $1s$-$1p_{1/2}$ and $1p_{1/2}$-$1p_{1/2}$ with aligned AM projection. These are related to deuteron-type (J=1,T=0) correlations. \\
%
Again we remind the reader that
the orbitals resulting from the 
self-consistent VNAT calculation 
are in practice ordered by occupation numbers.
In that basis, the $1p_{1/2}$ orbitals 
are adjacent to the $1s_{1/2}$ orbitals,
and the 
neutron-neutron and proton-neutron MI,
shown in Fig.~\ref{f:MI_VNAT_4He} with the occupation number ordering, becomes more localized.
\\

\subsection{Negativity}
The negativity $\mathcal{N}(\rho^{(ij)})$ is a measure of entanglement that provides an upper bound to the amount of distillable entanglement. 
It and its variants such as log-negativity, play a central role in, for example, quantum communication.  In the context of nuclear physics, as will be considered in this section, the practical implications of negativity in a nucleus are not immediately obvious.
However, it is a distinct measure of entanglement beyond MI, and as such is expected to provide insight into nuclear structure and reactions.
Negativity is defined as the sum of the negative eigenvalues of the partially transposed two-orbital reduced density,
\begin{eqnarray}
\rho^{T(ij)}_{n_i n_j ,n_i' n_j'}  = \sum_{ C } \bra{C } \braket{n_j n_i' | \Psi}  \braket{\Psi | n_i n_j'} \ket{C} \; .
\end{eqnarray}
In the basis ${\ket{n_i n_j}}=\{\ket{00},\ket{01},\ket{10},\ket{11} \}$, this becomes
\begin{eqnarray}
 \resizebox{0.95\hsize}{!}{%
        $
\rho^{T(ij)} \ = \  
\begin{pmatrix}
1 - \gamma_{ii} - \gamma_{jj} + \gamma_{ijij}  & 0 & 0 & \gamma_{ji} \\
0 & \gamma_{jj} - \gamma_{ijij} & 0 & 0 \\
0 &0 &\gamma_{ii} - \gamma_{ijij}  & 0 \\
 \gamma_{ij} & 0 & 0 &  \gamma_{ijij} 
\end{pmatrix}
\; .
$%
}
        \nonumber \\
\label{e:T2RDM}
\end{eqnarray}
As $\rho^{T(ij)}$ differs from $\rho^{(ij)}$ only through the non-diagonal elements $\gamma_{ij}$, 
$\gamma_{ji}$, and since $ \gamma_{ij} =  \gamma_{ji} =0$ when $i$ and $j$ have different isospin projections, in the proton-neutron case  $\rho^{T(ij)} =\rho^{(ij)} $, which only has positive eigenvalues. Therefore the proton-neutron negativity vanishes. Moreover, due to other symmetries, $\gamma_{ij}$ can only be non-zero if $i$ and $j$ have same AM, AM projection and parity. Therefore distillable entanglement could only arise between two single-particle states that have these same quantum numbers, that is, between states that can mix though unitary transformations of the single-particle basis.
\begin{figure}[!ht]
\centering{
\includegraphics[width=0.85\columnwidth]{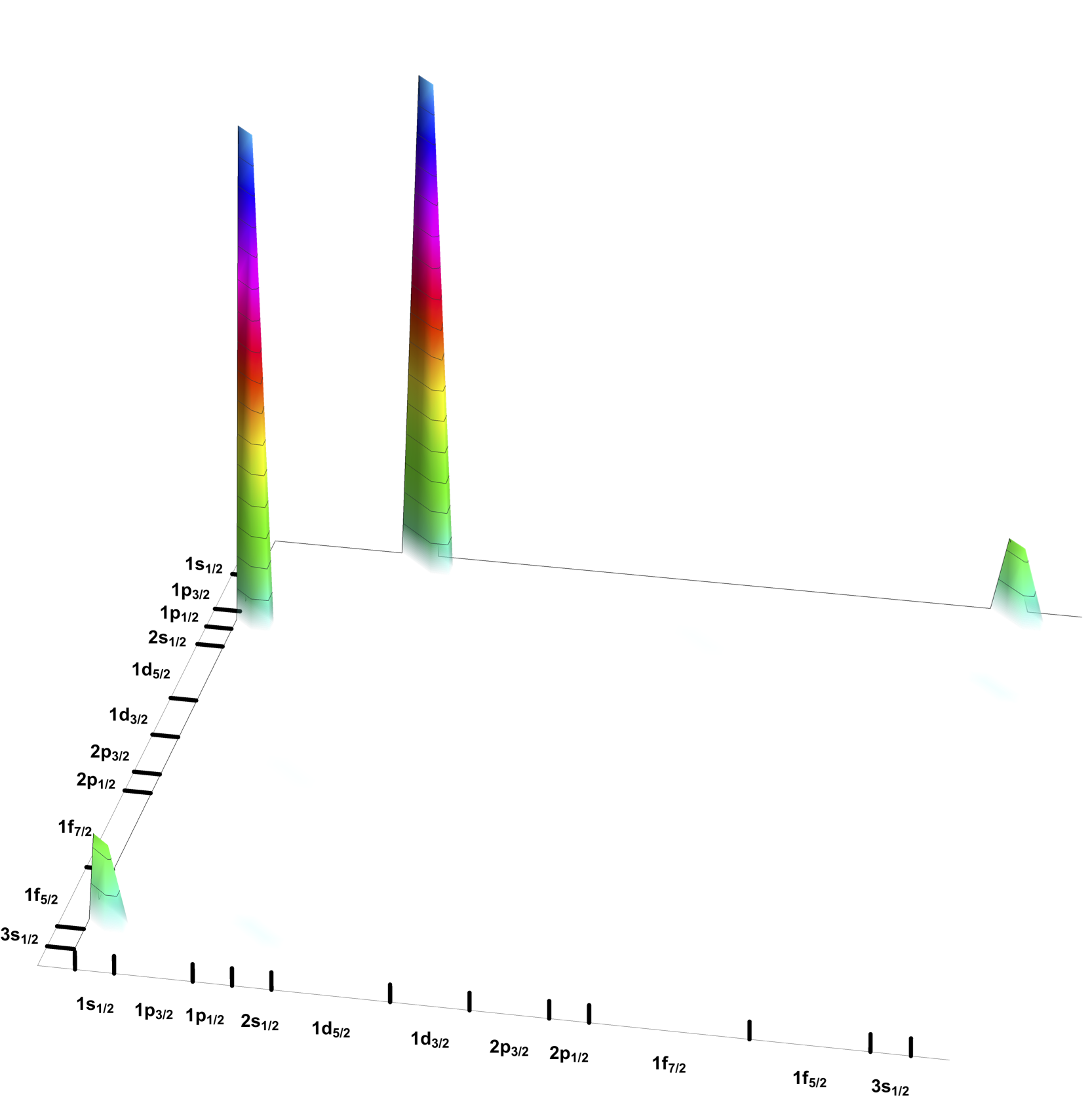}\\
\includegraphics[width=0.85\columnwidth]{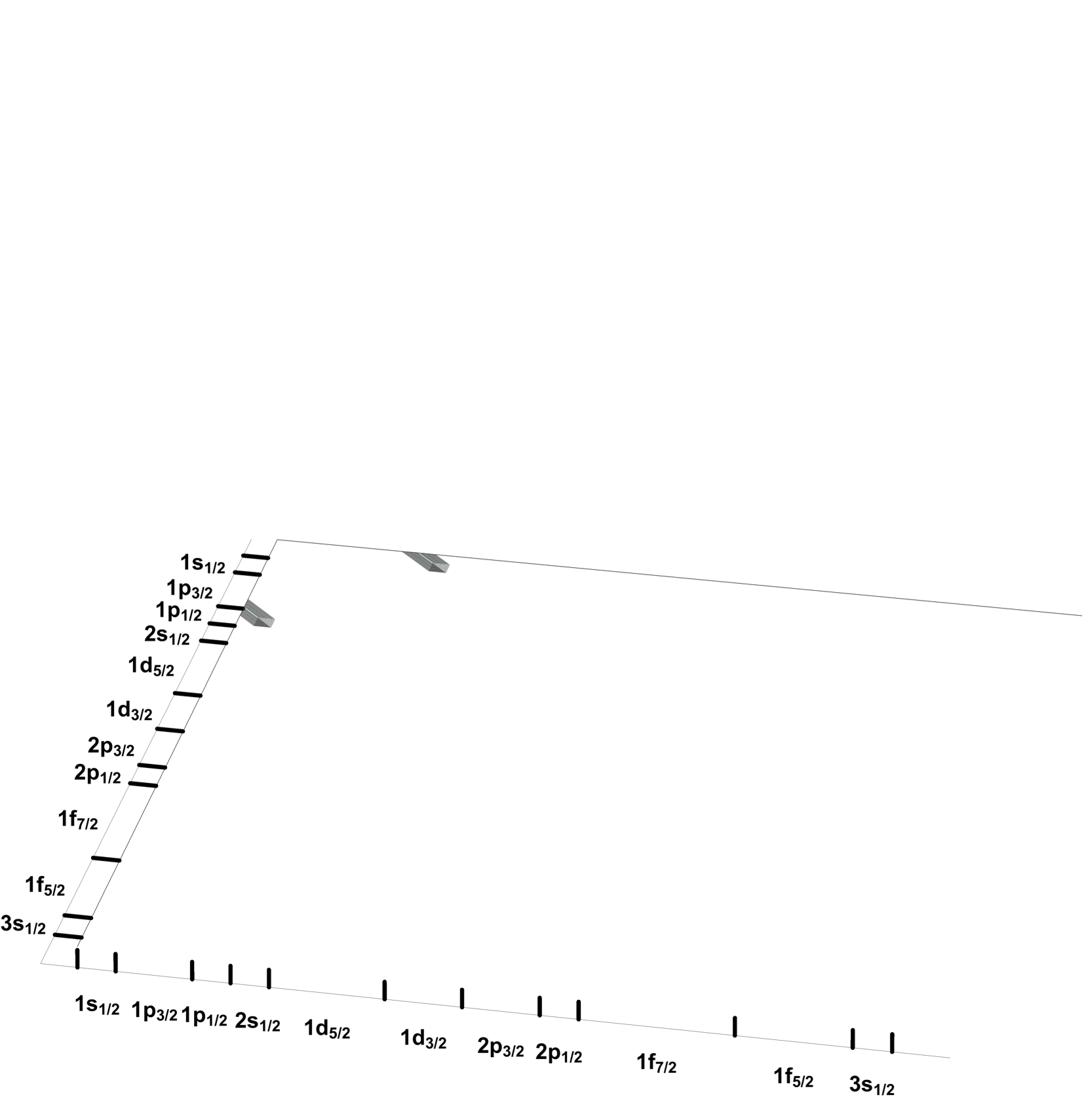}
}
\caption{(Color Online) 
The negativity within the neutron-neutron orbitals of $^4$He using 5-shell active spaces in the HO (upper panel) and VNAT (lower panel) bases. The vertical scales are the same in both panels.} 
\label{f:Neg_nn_4He_3D}
\end{figure}
The negativities within the neutron sector of $^4$He using the HO and VNAT bases with 5 active shells are shown in Fig.\ref{f:Neg_nn_4He_3D}.
A striking feature of the NAT and VNAT bases is that,
as the one-body density matrix $\gamma$ becomes diagonal, the negativity vanishes identically. Note that the small negativity found between the VNAT 1s and 2s shells of the order of $10^{-5}$ is due to the numerical precision of the self-consistent procedure.
\\
In the HO basis, the only non-negligible terms appear between s-shell orbitals for the case of $^4$He.
We present the corresponding values in Table~\ref{t:Neg} with increasing size of the model space. Generally,
we observe larger values of the negativity between the $1s$ orbitals and other $ns$ shells, with the negativity decreasing  with increasing 
$n$.
\\
Given the simple structure of the transposed two-orbital density, shown in Eq.~(\ref{e:T2RDM}), a condition for the appearance of non-zero negativity for the case of an arbitrary single-particle basis can be easily derived. This condition (detailed in Appendix~\ref{app3}) 
relates the non-diagonal elements $\gamma_{ij}$ to the occupation numbers and diagonal terms of the two-body density:
\begin{equation}
|\gamma_{ij}| \geq \sqrt{(1 - \gamma_{ii} - \gamma_{jj} ) \gamma_{ijij}  + (\gamma_{ijij} )^2 } \; .
\end{equation}

\begin{table*}
 \centering
\begin{tabular}{c|cccccc} 
\hline   
\hline
$N_{tot}$   & $1s$-$2s$ & $1s$-$3s$ &  $2s$-$3s$ & $1s$-$4s$ & $2s$-$4s$ & $3s$-$4s$ \\
\hline
2   & -      & -      & -      & -      & -      & -        \\
\hline
3   & 9.59 $\times 10^{-2}$ & -      & -      & -      & -      & -         \\
\hline
4   & 6.94 $\times 10^{-2}$ & -      & -      & -      & -      & -          \\
\hline
5   & 1.12$\times 10^{-1}$ & 1.67 $\times 10^{-2}$ & 2.50 $\times 10^{-5}$ & -      & -      & -  \\
\hline
6   & 1.01$\times 10^{-1}$ & 2.36$\times 10^{-2}$ &  3.10 $\times 10^{-5}$& 4.92 $\times 10^{-3}$ & 8.11 $\times 10^{-6}$ & 8.04$\times 10^{-7}$  \\
\hline
7   & 7.96 $\times 10^{-2}$   & 1.97 $\times 10^{-2}$ & 1.27$\times 10^{-5}$ & 4.28$\times 10^{-3}$ & 4.96 $\times 10^{-6}$ &
4.98 $\times 10^{-6}$ \\
\hline
\hline
\end{tabular}
\caption{The negativity between neutron states of the $s$ shells (with same AM projection) in $^4$He in the HO basis for different number of shells $N_{tot}$ in the active space. The negativity in the VNAT basis vanishes by definition.}
\label{t:Neg}
\end{table*}

\section{Entanglement in $^{6}$He}
\begin{figure}[h]
\centering{\includegraphics[width=\columnwidth]{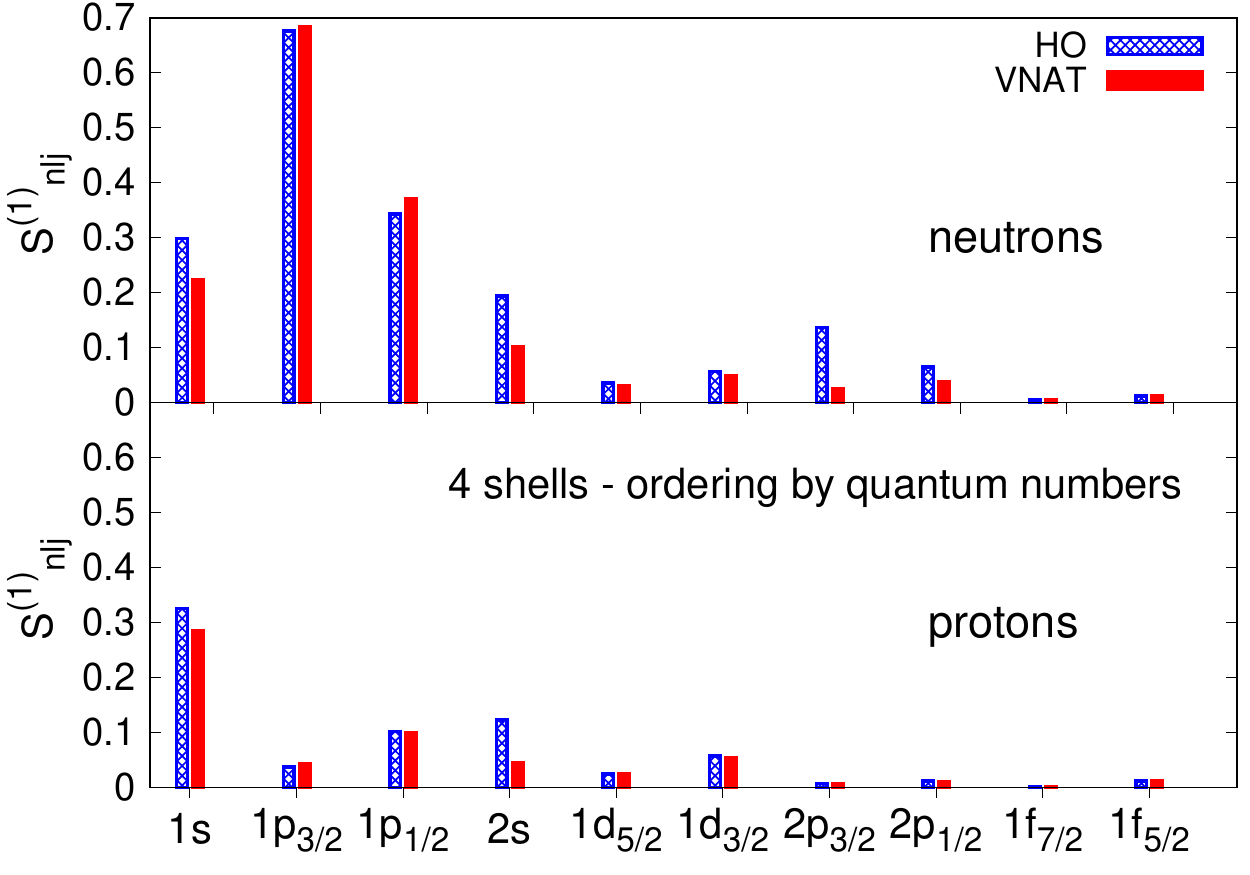}} 
\caption{(Color Online) The entanglement entropy of a single-neutron (top) and single-proton (bottom) HO and VNAT state,
in $^6$He obtained with $N_{tot}$=4 shells.} 
\label{f:S1_6He}
\end{figure}

\noindent
$^{6}$He is a halo nucleus consisting of two protons and four neutrons.
As such, it provides a "sandbox" in which to test basic aspects of entanglement in the context of the traditional nuclear shell model, where the naive neutron configuration is $(1s_{1/2})^2 (1p_{3/2})^2$ while the naive proton configuration is $(1s_{1/2})^2$.
The same numerical framework is used for $^{6}$He and $^4$He.  Specifically, all possible configurations (up to 6p-6h) in an active space comprising a given number of shells $N_{tot}$ are included.
\\
Figure~\ref{f:S1_6He} shows the single-orbital entanglement entropy in $^6$He
obtained with the HO and VNAT bases in a model space of $N_{tot}$=4 shells.
In the proton sector, the single-orbital entropy profile resembles that obtained in $^4$He. However, there is a small increase in the entropy of the states on the $1s$ shell, due to a decrease of their occupation number from $0.95$ to $0.92$ (in the VNAT basis) through proton-neutron interactions. 
In the neutron sector, the $1p_{3/2}$ and $1p_{1/2}$ sub-shells that the two extra neutrons are expected to primarily occupy appear as the most entangled orbitals.
In particular, $1p_{3/2}$ is almost maximally entangled, with occupation numbers of $~0.41$ and $~0.43$ in the HO and VNAT bases, respectively. 
%
\begin{table}
 \centering
\begin{tabular}{ccccc} 
\hline   
\hline
$N_{tot}$ & HO           & HF   & NAT    & VNAT\\
\hline
2 shells &  8.90      & -2.99  &  8.90  &  -6.32   \\
3 shells & -6.52      & -7.44  & -6.52  &  -13.91 \\
4 shells & -15.98     & -12.41 & -15.98 &  -19.41 \\
5 shells & -20.30     & -18.03 & -20.30 &  -22.50 \\
\hline 
\hline
\end{tabular}
\caption{The ground-state energy of $^6$He (in MeV) obtained in the different bases with different numbers of shells in the active space.}
\label{t:E_gs_6He}
\end{table} 
\\
%
%
\begin{figure}[h]
\centering{
\includegraphics[width=0.85\columnwidth]{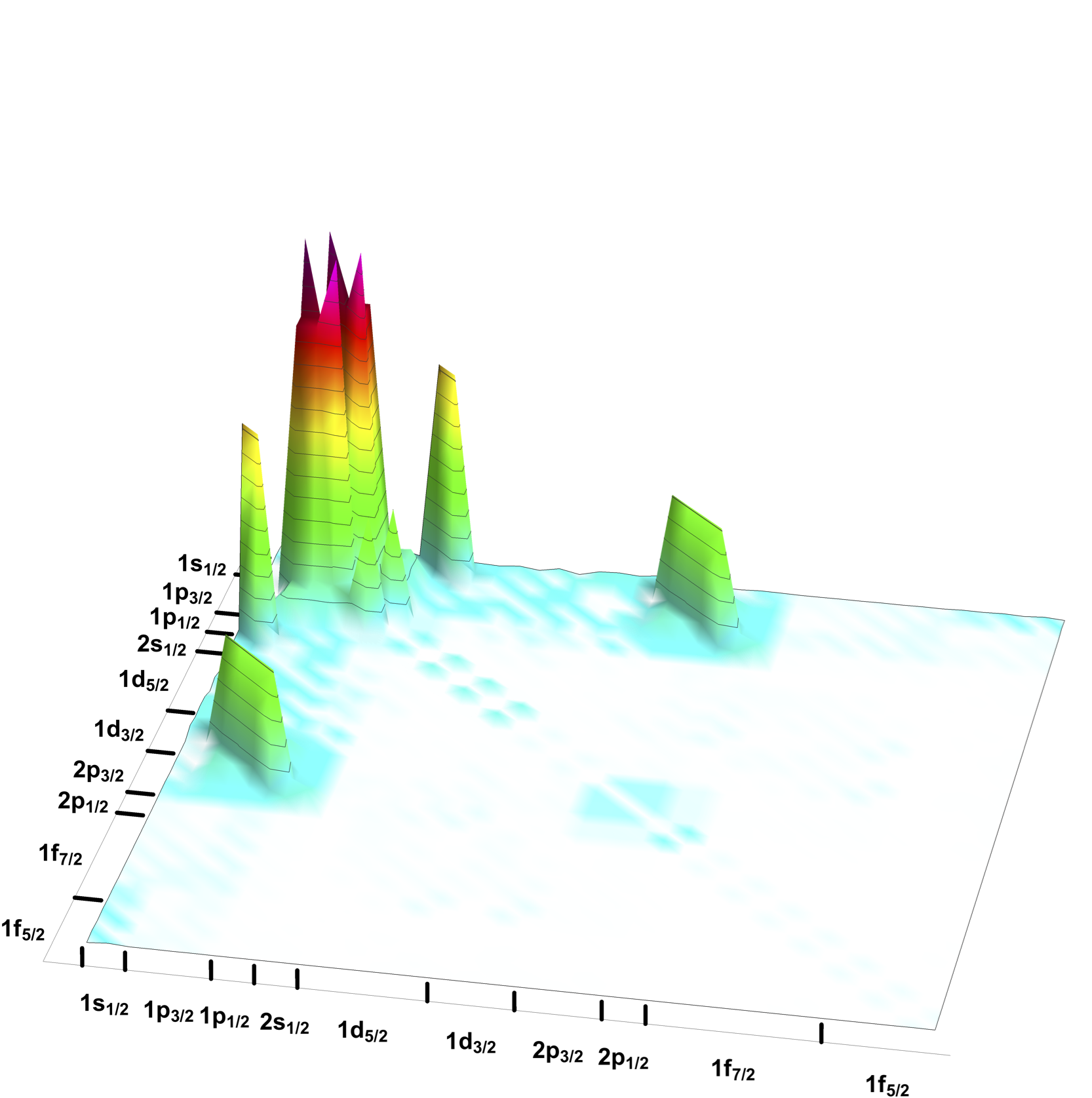}\\
\includegraphics[width=0.85\columnwidth]{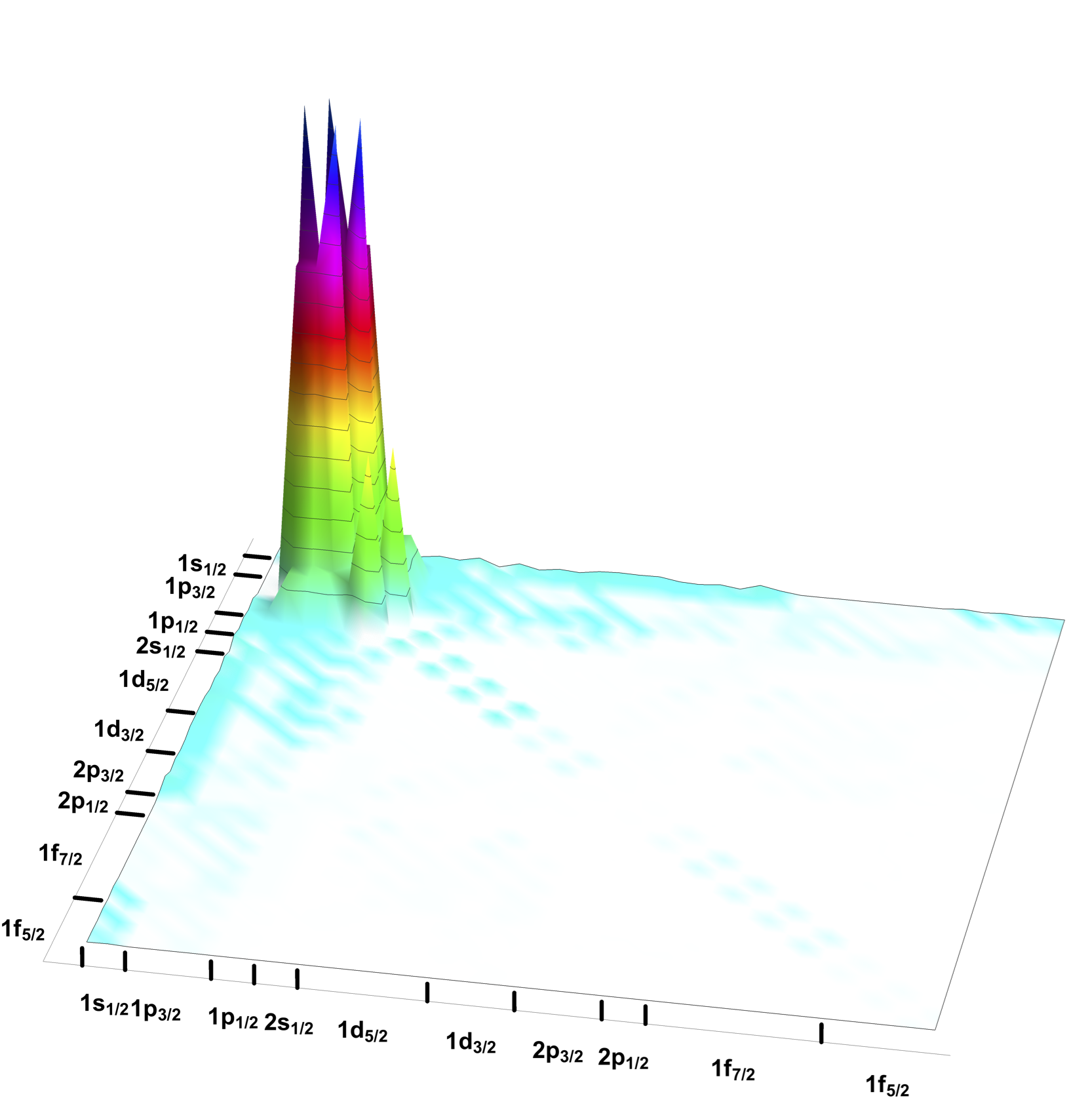}
}
\caption{ (Color Online) 
The mutual information within two neutron orbitals in $^6$He with the HO (upper panel) and VNAT (lower panel) bases using $N_{tot}$=4 active shells. } 
\label{f:MI_nn_6He_3D}
\end{figure}
\begin{figure}[h]
\centering{\includegraphics[width=\columnwidth]{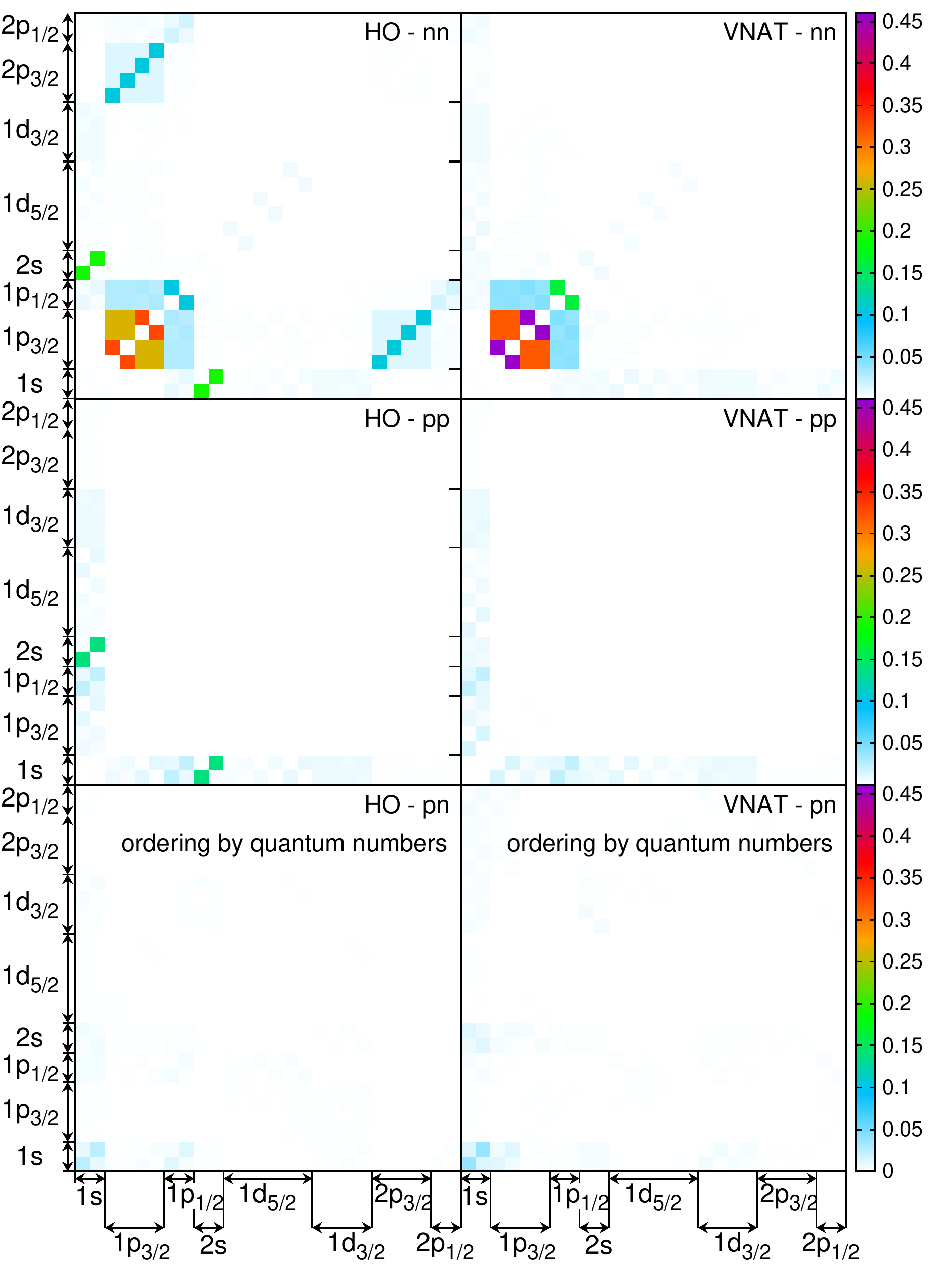}} 
\caption{(Color Online) Mutual information of HO (left) and VNAT (right) single-particle states in $^{6}$He obtained for an active space of $N_{tot}=$4 shells. The top panels show the MI between neutron orbitals, the middle panels show the MI between proton orbitals, and the bottom ones show the proton-neutron MI. (The $f$ shells are not shown as their MI is not visible). In the bottom panels the proton (neutron) states are on the y (x) axis.} 
\label{f:MI_6He}
\end{figure}
Figure~\ref{f:MI_nn_6He_3D} shows the MI (in 3D) within two HO and VNAT neutron orbitals.
In the HO basis, localized regions of MI are distributed between single-neutron states within the active model space.  In contrast, the two-neutron MI in the wavefunction in the VNAT basis is largely localized within the $1p$ shell, pointing to an emerging core-valence picture, where the two extra $p$-shell neutrons decouple from the $^{4}$He core.
To analyse the results in more details,
Fig.~\ref{f:MI_6He} shows the neutron-neutron (top), proton-proton (middle) and proton-neutron (bottom) MI in $^6$He.
Generally the MI in the proton-proton and proton-neutron sector is weak compared to the neutron-neutron sector, 
and this is particularly so in the VNAT basis.
Both bases show strong neutron-neutron MI within $1p_{3/2}$ states, and to a lesser extent within the $1p_{1/2}$, where the two halo neutrons primarily reside. This is a clear signature of the important isovector pairing correlations between these two neutrons, which is known to be responsible for the binding of $^{6}$He.
Within the $1p_{3/2}$, couplings between states with different AM projection e.g. with $|m_i|=1/2$ and $|m_j|=3/2$, are evident.
In the HO basis, we observe important MI between the $1s$ and $2s$ shells and also between the $1p$ and $2p$ shells. These couplings vanish in the VNAT basis. 
Thus, in that basis, $^{6}$He resembles much more a system of two neutrons orbiting in the $1p-$shell on top of a $^{4}$He core, as already seen in Fig.~\ref{f:MI_nn_6He_3D}.
Since the VNAT $1s$ shell does not couple to other neutron states, the single-orbital entropy $S^{(1)}_{1s} \approx 0.2$, shown in Fig.~\ref{f:S1_6He}, only results from interaction with proton states (mostly $1s$ and, to a lesser extent, $2s$ states).
This can be seen from the bottom panel of Fig.~\ref{f:MI_6He}.
\\
\\
Finally the ground state energy of $^6$He is given in Table~\ref{t:E_gs_6He} in the HO and VNAT bases as a function of the number of active shells. Because the present many-body scheme (all many-body configurations are included in the active space) leads to a fast growth of the size of the model space, we performed calculations up to $N_{tot}=5$. While the energies are not fully converged for this basis size, and thus suggest that the mutual information, and other entanglement measures, may not be fully converged either, we do observe that the patterns of entanglement emerge for small sizes of the model space. Increasing the number of active shells would modify only slightly the computed entanglement, but would not change the conclusion of this study.


\section{NAT versus VNAT basis}  \label{sec:NATvsVNAT}
\begin{figure}[!ht]
\centering{\includegraphics[width=\columnwidth]{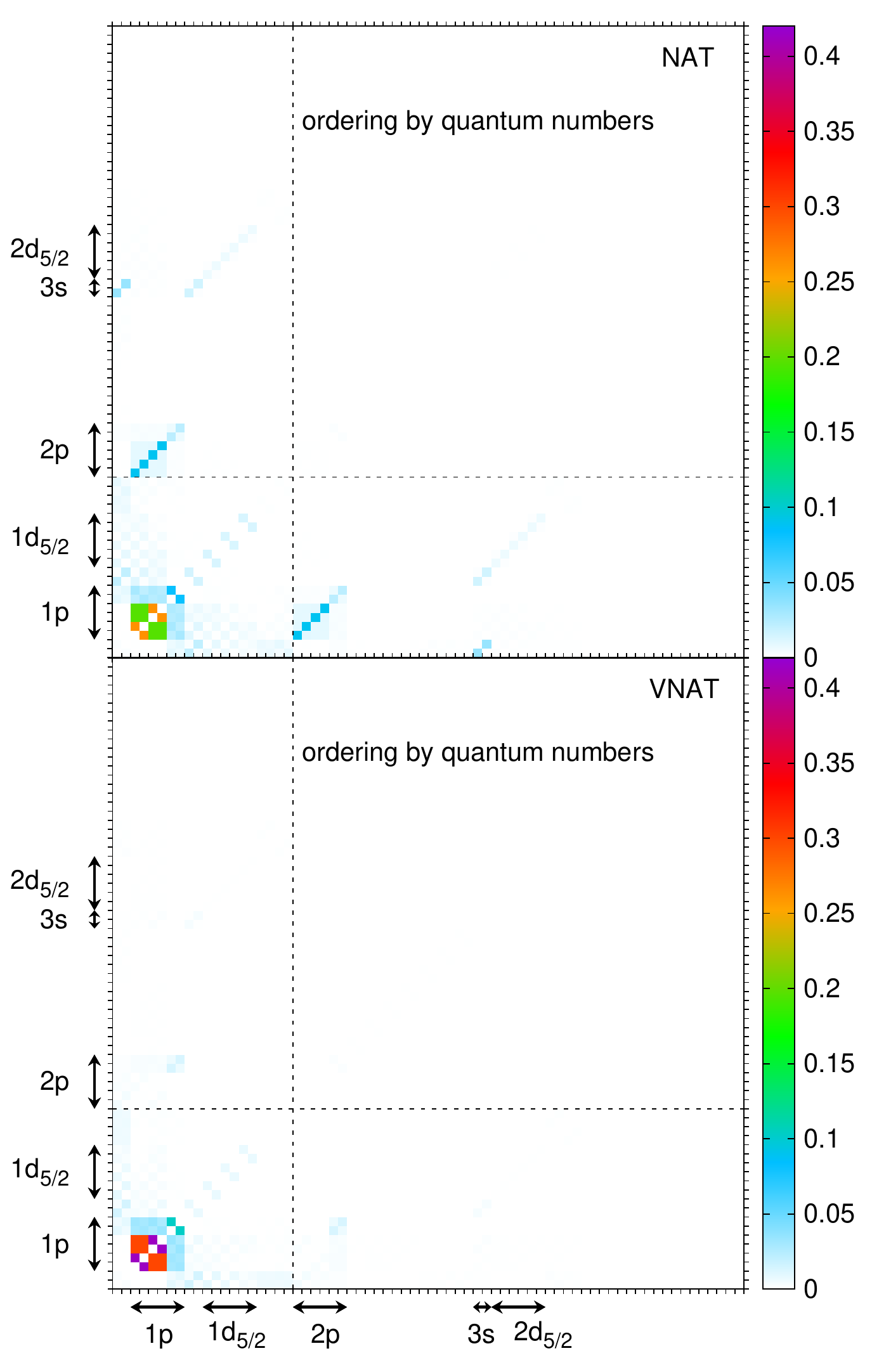}} 
\caption{(Color Online) Neutron-neutron MI for NAT (upper) and VNAT (lower) single-particle states in $^{6}$He obtained with $N_{tot}=3$ and $N'_{tot}=5$. The dashed lines show the initial active space of $N_{tot}=3$ shells.} 
\label{f:MI_nn_6He_NATvsVNAT}
\end{figure}

\noindent
In Section~\ref{sec:4He}, it was shown that the NAT and VNAT bases typically present similar entanglement profiles. This is understood as the entanglement measures are based on the computation of reduced density matrices, which cancel outside the active model spaces. However, the VNAT basis, which mixes both active and empty HO single-particle orbitals, leads to a faster convergence of the ground-state energy with respect to the size of the model space (see Table~\ref{t:E_gs_6He} and \cite{Robin:2020}).  This is to be compared with the NAT basis, which only mixes HO states in the active space, and thus does not improve the energy convergence compared to the HO basis. \\
In order to distinguish the NAT and VNAT bases,  the coupling between the active and inactive (empty) single-particle spaces needs to be quantified.
To do that, we perform initial calculations of the NAT and VNAT states in a model space of given $N_{tot}$ major shells, and, as a second step, use these bases to perform one diagonalization of the two-body Hamiltonian in a configuration space spanned by a larger single-particle basis, {\it i.e.} with $N'_{tot} > N_{tot}$.  One- and two-orbital entanglement measures can then be calculated.
Since the VNAT basis mixes HO states from both active and inactive spaces, entanglement measures 
are expected to be  weak between single-particle states below and above $N_{tot}$. \\
As an example, 
Fig.~\ref{f:MI_nn_6He_NATvsVNAT} shows the neutron-neutron MI in $^6$He obtained with $N_{tot}=3$ and $N'_{tot}=5$.
In the VNAT basis the couplings between the $1p$ and $2p$ shells are very small, even though the $2p$ shell was absent from the initial self-consistent calculation with $N_{tot}=3$. In the NAT basis however these couplings are sizeable. The same is true for the MI between the $1s$-$2s$ and $3s$ shells, and to a lesser extent for the MI between the $1d_{5/2}$-$2d_{5/2}$ states.
%

\section{Summary and conclusion}
\noindent
In this work we have explored measures of entanglement, entanglement entropy, mutual information and negativity in $^4$He and $^6$He in a selection of bases and a chiral interaction.  
The nuclear shell model, that successfully describes some features of nuclei, and has formed the basis for a large selection of increasingly sophisticated descriptions of nuclei and their interactions,
is in some ways pinned upon the fact that entanglement within a nucleus is somewhat localized. 
In that model, the localization of entanglement 
manifests itself in employing an active valence sector built upon an inert core, i.e. a tensor product system. While this is far from exact, residual interactions and sophisticated many-body techniques  build upon such tensor product starting points. In the present calculations of $^6$He, this core-valence structure emerges from the full 6-body computation. 
It is known that entanglement measures in systems with identical particles are basis dependent, and we have shown that commonly used bases employed for ab-initio nuclear structure calculations support quite different entanglement structures.  In particular, the widely used HO basis exhibits a somewhat distributed two-nucleon entanglement structure for both MI and negativity, and a somewhat larger single-orbital entanglement entropy.  In contrast, the VNAT basis exhibits a more compact two-nucleon MI and single-orbital entanglement entropy, and vanishing negativity by construction.  
The potential utility of MI, and more generally measures of entanglement, is made clearest in comparing $^4$He and $^6$He, where the additional two p-shell neutrons provide a substantial and structured MI within the p-shell.
The truncation scheme used in the present study limits the size of the active space. While the results for $^4$He are converged to better than $10^{-2}$, the $^6$He calculations are not fully converged at this point. Because of the exotic halo character of this nucleus, reaching an accurate value of the ground-state energy requires larger active spaces than we are presently able to employ, as well as a three-body force. Such improvements will be made in future works, and will also move us towards a complete quantification of uncertainties of entanglement measures.
\\
Studying the entanglement structure of nuclei may have future benefits when considering workflows for hybrid classical-quantum computations of nuclear structure and reactions. Elements of such computations with minimal or vanishing entanglement are amenable to classical computations, while those where entanglement is significant will be computed using a quantum device.  An optimal workflow would have the intrinsically quantum aspects of the computation performed using a quantum device, while those that are intrinsically classical would be best performed on a classical device.   The entanglement measures we have considered in this work could provide helpful diagnostics in designing workflows for such hybrid classical-quantum computations.
\\
While we have not provided evidence, it is possible that using a basis for nuclear many-body computations that has minimal support of entanglement entropy and two-nucleon MI and negativity may provide a better low-energy model to match to low-energy effective field theories and also the results of lattice QCD calculations.  At the physical point, the low-energy two-nucleon entanglement power is near minimal, and related to enhanced spin-flavor symmetries.  It seems natural to preserve this feature during matching to nuclear many-body systems in order to address more complex nuclear systems.  This point requires significantly more investigation before conclusive statements can be made.
\\
Our investigations suggest that there maybe utility in designing effective interactions that 
are organized by, to some extent,  entanglement.  We have not addressed this potential yet, but suggest that exploring the behavior of entanglement induced by forces as a function of evolution under SRG flow, 
or renormalization group flows, more generally,
has the potential to provide valuable insight.
\\
The results we have presented represent some of the first steps in an emerging line of investigation.
While this work focused on the link between entanglement and the convergence of binding energies, the relation to other observables, such as nuclear radii, which are directly related to the content of the wave function, will provide complementary insight. Such studies will be performed in forthcoming works.
Overall we are encouraged by the entanglement structures we have found, and intend to extend these studies to include multi-partite entanglement in nuclei and nuclear reactions, with a particular focus on extracting further insights into clustering, three-nucleon and four-nucleon forces. It is plausible, as demonstrated on the case of $^6$He, that nuclei near the drip line exhibit entanglement structures that differ from those enjoying the valley of stability. We will be pursuing such systems in upcoming research.
In the future these calculations of entanglement can also be used to improve our many-body scheme. In particular we plan to investigate a selection of orbitals based on one- or two-orbital entanglement measures ("\` a la DMRG") within the self-consistent procedure.

\section*{Acknowledgments}
\noindent
We thank Petr Navr\'atil for graciously providing matrix elements of the chiral interactions. 
We would also like to thank Natalie Klco, Alessandro Roggero and Guillaume Hupin for a number of fruitful discussions.
This work was supported, in part, by U.S. DOE Grant No. DE-FG02-00ER41132 and by JINA-CEE under U.S. NSF Grant No. PHY-1430152.
The computational work for this project was partly carried 
out on the Hyak
High Performance Computing and Data Ecosystem at
the University of Washington, supported, in part, by
the U.S. National Science Foundation Major Research
Instrumentation Award, Grant Number 0922770, the Institute for Nuclear Theory and the Physics Department at the University of Washington.
This research also used resources of the National Energy Research
Scientific Computing Center, a DOE Office of Science User Facility
supported by the Office of Science of the U.S. Department of Energy
under Contract No. DE-AC02-05CH11231.
\\
\\
{\it Note added}: After completion of this work we became aware of Ref.~\cite{Kruppa:2020rfa}, in which single-orbital entanglement entropy and two-orbital mutual information in two-nucleon systems are considered.

\onecolumngrid
\appendix
\section{Single-orbital reduced density} 
\label{app1}
The single-orbital reduced density matrix is
\begin{equation}
\rho^{(i)}_{n_i,n_i'}  = \sum_{ BC } \braket{ \Psi | BC } \ket{ n_i' }  \bra{n_i} \braket{ BC | \Psi } \; ,
\end{equation}
where $BC \equiv   (n_1  n_2 ... n_{i-1} n_{i+1} ... n_N )$.
Each fixed state $i$ can be occupied or empty so that we have a basis $\{ \ket{n_i} \}= \{ \ket{0}; \ket{1} = a^\dagger_i \ket{0} \} $. 
In this basis there are 4 matrix elements:
\begin{eqnarray}
\rho^{(i)}_{1,1}  &=& \sum_{ BC } \braket{ \Psi | BC} \ket{ 1 }  \bra{1} \braket{ BC | \Psi } \ =\  \sum_{ BC } \braket{ \Psi | BC } a^\dagger_i \ket{ 0 }  \bra{0} a_i \braket{ BC | \Psi } \nonumber \\
                         &=& \sum_{ BC }  \braket{ \Psi | BC } a^\dagger_i \underbrace{\sum_{n_i} \ket{ n_i}  \bra{n_i} }_{\hat 1} a_i \braket{ BC | \Psi }  - \sum_{ BC } \braket{ \Psi | BC } \underbrace{ a^\dagger_i \ket{ 1 }}_0  \bra{1} a_i \braket{ BC | \Psi } \nonumber \\
                         &=& \braket{ \Psi |a^\dagger_i a_i | \Psi}   
                        \ =\  \gamma_{ii} \; ,
\end{eqnarray}
\begin{eqnarray}
\rho^{(i)}_{0,0}  &=& \sum_{ BC } \braket{ \Psi | BC } \ket{ 0 }  \bra{0} \braket{ BC | \Psi } \nonumber \\
                         &=& \sum_{ BC } \sum_i \braket{ \Psi | BC } \ket{ n_i}  \bra{n_i} \braket{ BC | \Psi }  - \sum_{ BC } \braket{ \Psi | BC } \ket{ 1 }  \bra{1} \braket{ BC | \Psi } \nonumber \\
                         &=& \braket{\Psi | \Psi} - \rho^{(i)}_{1,1} 
                         \ =\  1 - \braket{ \Psi |a^\dagger_i a_i | \Psi} 
                         \ =\  1 - \gamma_{ii} \; ,
\end{eqnarray}
and
$\rho^{(i)}_{1,0}  = \rho^{(i)}_{0,1} = 0$
due to conservation of particle number.

\section{Two-orbital reduced density} 
\label{app2} 
The two-orbital reduced density matrix is
\begin{equation}
\rho^{(ij)}_{n_in_j,n_i'n_j'}  = \sum_{ C } \braket{ \Psi | C } \ket{ n_i' n_j'}  \bra{n_j n_i} \braket{ C | \Psi } \; ,
\end{equation}
where $\ket{C} = | n_1  n_2 ... n_{i-1} n_{i+1} ... n_{j-1} n_{j+1} ... n_N \rangle$. 
There are 4 states for the basis $\ket{n_i n_j}$ that we denote: 
\begin{eqnarray}
 \ket{\underbar 1} &\equiv& \ket{00} \ ,\ 
 \ket{\underbar 2} \ \equiv\ \ket{01} \ ,\ 
 \ket{\underbar 3}\ \equiv\  \ket{10} \ ,\ 
 \ket{\underbar 4}\ \equiv  \ket{11} \; .
 \end{eqnarray}
 The matrix elements of $\rho^{(ij)}$ are then
\begin{eqnarray}
\rho^{(ij)}_{\underbar 4,\underbar 4}  \equiv \rho^{(ij)}_{11,11} 
                         &=& \sum_{ C } \braket{ \Psi | C } \ket{ 11}  \bra{11} \braket{ C | \Psi } \ =\  \sum_{ C }   \braket{ \Psi | C } a^\dagger_i a^\dagger_j \ket{00} \bra{00} a_j a_i \braket{ C | \Psi } \nonumber \\
			&=& \sum_{ C }   \braket{ \Psi | C } a^\dagger_i a^\dagger_j \underbrace{ \sum_{n_i n_j}\ket{n_i n_j} \bra{n_j n_i} }_{\hat 1} a_j a_i \braket{ C | \Psi } 
			- \sum_{ C }   \braket{ \Psi | C } \underbrace {a^\dagger_i a^\dagger_j \ket{01}}_{0} \bra{10} a_j a_i \braket{ C | \Psi } \nonumber \\
			&&- \sum_{ C }   \braket{ \Psi | C } \underbrace {a^\dagger_i a^\dagger_j \ket{10}}_{0} \bra{01} a_j a_i \braket{ C | \Psi } 
			- \sum_{ C }   \braket{ \Psi | C } \underbrace {a^\dagger_i a^\dagger_j \ket{11}}_{0} \bra{11} a_j a_i \braket{ C | \Psi }  \nonumber \\
			&=& \braket{  \Psi | a^\dagger_i a^\dagger_j a_j a_i |\Psi} = \gamma_{ijij}\; ,
 \end{eqnarray}
\begin{eqnarray}
\rho^{(ij)}_{\underbar 3,\underbar 3}  \equiv \rho^{(ij)}_{10,10} 
                         &=& \sum_{ C } \braket{ \Psi | C } \ket{ 10}  \bra{01} \braket{ C | \Psi } \ =\  \sum_{ C }   \braket{ \Psi | C } a^\dagger_i  \ket{00} \bra{00} a_i \braket{ C | \Psi } \nonumber \\
			&=& \sum_{ C }   \braket{ \Psi | C } a^\dagger_i  \underbrace{ \sum_{n_i n_j}\ket{n_i n_j} \bra{n_j n_i} }_{\hat 1} a_i \braket{ C | \Psi } 
			- \sum_{ C }   \braket{ \Psi | C } a^\dagger_i \ket{01} \bra{10} a_i \braket{ C | \Psi } \nonumber \\
			&&- \sum_{ C }   \braket{ \Psi | C } \underbrace {a^\dagger_i \ket{10}}_{0} \bra{01}  a_i \braket{ C | \Psi } 
			- \sum_{ C }   \braket{ \Psi | C } \underbrace {a^\dagger_i \ket{11}}_{0} \bra{11} a_i \braket{ C | \Psi }  \nonumber \\
			&=& \braket{  \Psi | a^\dagger_i  a_i  |\Psi} - \sum_{ C }   \braket{ \Psi | C } a^\dagger_i a^\dagger_j \ket{00} \bra{00} a_j a_i \braket{ C | \Psi } \nonumber \\
			&=& \braket{  \Psi | a^\dagger_i  a_i  |\Psi} - \braket{  \Psi | a^\dagger_i a^\dagger_j a_j a_i |\Psi} = \gamma_{ii} - \gamma_{ijij} \; .
 \end{eqnarray}
 Similarly, 
\begin{eqnarray}
\rho^{(ij)}_{\underbar 2,\underbar 2}  \equiv \rho^{(ij)}_{01,01} = \braket{  \Psi | a^\dagger_j  a_j  |\Psi} - \braket{  \Psi | a^\dagger_i a^\dagger_j a_j a_i |\Psi} = \gamma_{jj} - \gamma_{ijij} \; ,
 \end{eqnarray}
and,
 \begin{eqnarray}
\rho^{(ij)}_{\underbar 1,\underbar 1}  \equiv \rho^{(ij)}_{00,00} 
			&=& \sum_{ C } \braket{ \Psi | C } \ket{ 00}  \bra{00} \braket{ C | \Psi } \nonumber \\
			&=& \sum_{ C }   \braket{ \Psi | C }  \underbrace{ \sum_{n_i n_j}\ket{n_i n_j} \bra{n_j n_i} }_{\hat 1} a_j a_i \braket{ C | \Psi } 
			- \sum_{ C }   \braket{ \Psi | C }  \ket{01}\bra{10}  \braket{ C | \Psi } \nonumber \\
			&&- \sum_{ C }   \braket{ \Psi | C }  \ket{10} \bra{01} \braket{ C | \Psi } 
			- \sum_{ C }   \braket{ \Psi | C } \ket{11} \bra{11} \braket{ C | \Psi }  \nonumber \\
			&=& \braket{\Psi |\Psi} - \rho^{(ij)}_{\underbar 2,\underbar 2} - \rho^{(ij)}_{\underbar 3,\underbar 3} - \rho^{(ij)}_{\underbar 4,\underbar 4} \nonumber \\
			&=& 1 - \braket{  \Psi | a^\dagger_j  a_j  |\Psi} - \braket{  \Psi | a^\dagger_i  a_i  |\Psi} + \braket{  \Psi | a^\dagger_i a^\dagger_j a_j a_i |\Psi} \nonumber \\
			&=& 1 - \gamma_{jj} -\gamma_{ii} + \gamma_{ijij} \; .
 \end{eqnarray}
Finally, the non-zero off-diagonal elements are
 \begin{eqnarray}
\rho^{(ij)}_{\underbar 2,\underbar 3}  \equiv \rho^{(ij)}_{01,10} 
		&=& \sum_{ C } \braket{ \Psi | C } \ket{ 10}  \bra{10} \braket{ C | \Psi } \ =\  \sum_{ C } \braket{ \Psi | C } a^\dagger_i \ket{ 00}  \bra{00} a_j \braket{ C | \Psi } \nonumber \\
		&=& \braket{  \Psi | a^\dagger_i  a_j |\Psi}  = \gamma_{ji}\; ,
 \end{eqnarray}
 and, 
  \begin{eqnarray}
\rho^{(ij)}_{\underbar 3,\underbar 2}  \equiv \rho^{(ij)}_{10,01}= \braket{  \Psi | a^\dagger_j  a_i |\Psi} = \gamma_{ij}  \; .
 \end{eqnarray}
 All other matrix elements cancel due to particle-number conservation.

\section{Condition for non-zero two-orbital negativity} 
\label{app3}
The negativity $N^{(ij)}$ is defined as the sum of the negative eigenvalues of the partially transposed two-orbital density, given in Eq.~(\ref{e:T2RDM}), which has the following structure
\begin{eqnarray}
\rho^{T(ij)}= 
\begin{pmatrix}
M_{11} & 0 & 0 &M_{14} \\
0 & M_{22} & 0 & 0 \\
0 &0 &M_{33}  & 0 \\
M_{41} & 0 & 0 &  M_{44}
\end{pmatrix}
\; ,
\end{eqnarray}
with 
\begin{eqnarray}
M_{11}  &=& 1 - \gamma_{ii} - \gamma_{jj} + \gamma_{ijij}  
\ ,\ 
M_{22} \ =\  \gamma_{jj} - \gamma_{ijij} \ ,\ 
M_{33} \ =\  \gamma_{ii} - \gamma_{ijij}  \\
M_{44} &=&  \gamma_{ijij} \ , \ 
M_{14} \ =\  M_{41} =  \gamma_{ji} =  \gamma_{ij} \; .
\end{eqnarray}

The eigenvalues of this matrix are
\begin{eqnarray}
\lambda_{1}  &=& M_{22} \ ,\ 
\lambda_{2} \ =\  M_{33} \\
\lambda_{3}  &=& \frac{1}{2} \left(  M_{11} + M_{44}  - \sqrt{M_{11}^2 + 4 M_{14}M_{41} - 2M_{11}M_{44} + M_{44}^2}  \right) \\
\lambda_{4}  &=& \frac{1}{2} \left(  M_{11} + M_{44}  + \sqrt{M_{11}^2 + 4 M_{14}M_{41} - 2M_{11}M_{44} + M_{44}^2}  \right) \; .
\end{eqnarray}

\begin{itemize}
\item $M_{11}$ and $M_{44}$ are eigenvalues of the two-orbital density in Eq.~(\ref{e:2RDM}), thus $M_{11}$ and $M_{44}$ are positive by definition, and thus $\lambda_{4} \ge 0$.
\item
$\lambda_{1}, \lambda_{2}\ge 0 $,
by considering the norms of 
$a_i|\Psi\rangle$ and 
$a_j a_i|\Psi\rangle$.
\item $\lambda_{3} \leq 0$  if 
\begin{equation}
M_{11} + M_{44}  \leq  \sqrt{M_{11}^2 + 4 M_{14}M_{41} - 2M_{11}M_{44} + M_{44}^2}
\end{equation}
which, after manipulations, gives
\begin{eqnarray}
1 - \gamma_{ii} - \gamma_{jj} + 2 \gamma_{ijij} \leq \sqrt{ (1 - \gamma_{ii} - \gamma_{jj} )^2 + 4 \gamma_{ij}^2} \\
\Leftrightarrow |\gamma_{ij}| \geq \sqrt{(1 - \gamma_{ii} - \gamma_{jj} ) \gamma_{ijij}  + (\gamma_{ijij} )^2 }
\ \ .
\end{eqnarray}

\end{itemize}

%
%
\newpage
\twocolumngrid
\bibliography{bib_HOVNATentangle}

\begin{thebibliography}{85}%
\makeatletter
\providecommand \@ifxundefined [1]{%
 \@ifx{#1\undefined}
}%
\providecommand \@ifnum [1]{%
 \ifnum #1\expandafter \@firstoftwo
 \else \expandafter \@secondoftwo
 \fi
}%
\providecommand \@ifx [1]{%
 \ifx #1\expandafter \@firstoftwo
 \else \expandafter \@secondoftwo
 \fi
}%
\providecommand \natexlab [1]{#1}%
\providecommand \enquote  [1]{``#1''}%
\providecommand \bibnamefont  [1]{#1}%
\providecommand \bibfnamefont [1]{#1}%
\providecommand \citenamefont [1]{#1}%
\providecommand \href@noop [0]{\@secondoftwo}%
\providecommand \href [0]{\begingroup \@sanitize@url \@href}%
\providecommand \@href[1]{\@@startlink{#1}\@@href}%
\providecommand \@@href[1]{\endgroup#1\@@endlink}%
\providecommand \@sanitize@url [0]{\catcode `\\12\catcode `\$12\catcode
  `\&12\catcode `\#12\catcode `\^12\catcode `\_12\catcode `\%12\relax}%
\providecommand \@@startlink[1]{}%
\providecommand \@@endlink[0]{}%
\providecommand \url  [0]{\begingroup\@sanitize@url \@url }%
\providecommand \@url [1]{\endgroup\@href {#1}{\urlprefix }}%
\providecommand \urlprefix  [0]{URL }%
\providecommand \Eprint [0]{\href }%
\providecommand \doibase [0]{http://dx.doi.org/}%
\providecommand \selectlanguage [0]{\@gobble}%
\providecommand \bibinfo  [0]{\@secondoftwo}%
\providecommand \bibfield  [0]{\@secondoftwo}%
\providecommand \translation [1]{[#1]}%
\providecommand \BibitemOpen [0]{}%
\providecommand \bibitemStop [0]{}%
\providecommand \bibitemNoStop [0]{.\EOS\space}%
\providecommand \EOS [0]{\spacefactor3000\relax}%
\providecommand \BibitemShut  [1]{\csname bibitem#1\endcsname}%
\let\auto@bib@innerbib\@empty
\bibitem [{\citenamefont {Politzer}(1973)}]{Politzer:1973fx}%
  \BibitemOpen
  \bibfield  {author} {\bibinfo {author} {\bibfnamefont {H.}~\bibnamefont
  {Politzer}},\ }\href {\doibase 10.1103/PhysRevLett.30.1346} {\bibfield
  {journal} {\bibinfo  {journal} {Phys. Rev. Lett.}\ }\textbf {\bibinfo
  {volume} {30}},\ \bibinfo {pages} {1346} (\bibinfo {year}
  {1973})}\BibitemShut {NoStop}%
\bibitem [{\citenamefont {Gross}\ and\ \citenamefont
  {Wilczek}(1973)}]{Gross:1973id}%
  \BibitemOpen
  \bibfield  {author} {\bibinfo {author} {\bibfnamefont {D.~J.}\ \bibnamefont
  {Gross}}\ and\ \bibinfo {author} {\bibfnamefont {F.}~\bibnamefont
  {Wilczek}},\ }\href {\doibase 10.1103/PhysRevLett.30.1343} {\bibfield
  {journal} {\bibinfo  {journal} {Phys. Rev. Lett.}\ }\textbf {\bibinfo
  {volume} {30}},\ \bibinfo {pages} {1343} (\bibinfo {year}
  {1973})}\BibitemShut {NoStop}%
\bibitem [{\citenamefont {Glashow}(1961)}]{Glashow:1961tr}%
  \BibitemOpen
  \bibfield  {author} {\bibinfo {author} {\bibfnamefont {S.}~\bibnamefont
  {Glashow}},\ }\href {\doibase 10.1016/0029-5582(61)90469-2} {\bibfield
  {journal} {\bibinfo  {journal} {Nucl. Phys.}\ }\textbf {\bibinfo {volume}
  {22}},\ \bibinfo {pages} {579} (\bibinfo {year} {1961})}\BibitemShut
  {NoStop}%
\bibitem [{\citenamefont {Weinberg}(1967)}]{Weinberg:1967tq}%
  \BibitemOpen
  \bibfield  {author} {\bibinfo {author} {\bibfnamefont {S.}~\bibnamefont
  {Weinberg}},\ }\href {\doibase 10.1103/PhysRevLett.19.1264} {\bibfield
  {journal} {\bibinfo  {journal} {Phys. Rev. Lett.}\ }\textbf {\bibinfo
  {volume} {19}},\ \bibinfo {pages} {1264} (\bibinfo {year}
  {1967})}\BibitemShut {NoStop}%
\bibitem [{\citenamefont {Salam}(1968)}]{Salam:1968rm}%
  \BibitemOpen
  \bibfield  {author} {\bibinfo {author} {\bibfnamefont {A.}~\bibnamefont
  {Salam}},\ }\href {\doibase 10.1142/9789812795915\_0034} {\bibfield
  {journal} {\bibinfo  {journal} {Conf. Proc. C}\ }\textbf {\bibinfo {volume}
  {680519}},\ \bibinfo {pages} {367} (\bibinfo {year} {1968})}\BibitemShut
  {NoStop}%
\bibitem [{\citenamefont {Wilson}(1974)}]{Wilson:1974sk}%
  \BibitemOpen
  \bibfield  {author} {\bibinfo {author} {\bibfnamefont {K.~G.}\ \bibnamefont
  {Wilson}},\ }\href {\doibase 10.1103/PhysRevD.10.2445} {\bibfield  {journal}
  {\bibinfo  {journal} {Phys. Rev. D}\ }\textbf {\bibinfo {volume} {10}},\
  \bibinfo {pages} {2445} (\bibinfo {year} {1974})}\BibitemShut {NoStop}%
\bibitem [{\citenamefont {Creutz}\ \emph {et~al.}(1979)\citenamefont {Creutz},
  \citenamefont {Jacobs},\ and\ \citenamefont {Rebbi}}]{Creutz:1979kf}%
  \BibitemOpen
  \bibfield  {author} {\bibinfo {author} {\bibfnamefont {M.}~\bibnamefont
  {Creutz}}, \bibinfo {author} {\bibfnamefont {L.}~\bibnamefont {Jacobs}}, \
  and\ \bibinfo {author} {\bibfnamefont {C.}~\bibnamefont {Rebbi}},\ }\href
  {\doibase 10.1103/PhysRevLett.42.1390} {\bibfield  {journal} {\bibinfo
  {journal} {Phys. Rev. Lett.}\ }\textbf {\bibinfo {volume} {42}},\ \bibinfo
  {pages} {1390} (\bibinfo {year} {1979})}\BibitemShut {NoStop}%
\bibitem [{\citenamefont {Balian}\ \emph {et~al.}(1974)\citenamefont {Balian},
  \citenamefont {Drouffe},\ and\ \citenamefont {Itzykson}}]{Balian:1974ts}%
  \BibitemOpen
  \bibfield  {author} {\bibinfo {author} {\bibfnamefont {R.}~\bibnamefont
  {Balian}}, \bibinfo {author} {\bibfnamefont {J.}~\bibnamefont {Drouffe}}, \
  and\ \bibinfo {author} {\bibfnamefont {C.}~\bibnamefont {Itzykson}},\ }\href
  {\doibase 10.1103/PhysRevD.10.3376} {\bibfield  {journal} {\bibinfo
  {journal} {Phys. Rev. D}\ ,\ \bibinfo {pages} {74}} (\bibinfo {year}
  {1974})}\BibitemShut {NoStop}%
\bibitem [{\citenamefont {{\bf [NPLQCD]}}\ \emph {et~al.}(2013)\citenamefont
  {{\bf [NPLQCD]}}, \citenamefont {Beane}, \citenamefont {Chang}, \citenamefont
  {Cohen}, \citenamefont {Detmold}, \citenamefont {Lin}, \citenamefont {Luu},
  \citenamefont {Orginos}, \citenamefont {Parreno}, \citenamefont {Savage},\
  and\ \citenamefont {Walker-Loud}}]{Beane:2012vq}%
  \BibitemOpen
  \bibfield  {author} {\bibinfo {author} {\bibnamefont {{\bf [NPLQCD]}}},
  \bibinfo {author} {\bibfnamefont {S.}~\bibnamefont {Beane}}, \bibinfo
  {author} {\bibfnamefont {E.}~\bibnamefont {Chang}}, \bibinfo {author}
  {\bibfnamefont {S.}~\bibnamefont {Cohen}}, \bibinfo {author} {\bibfnamefont
  {W.}~\bibnamefont {Detmold}}, \bibinfo {author} {\bibfnamefont
  {H.}~\bibnamefont {Lin}}, \bibinfo {author} {\bibfnamefont {T.}~\bibnamefont
  {Luu}}, \bibinfo {author} {\bibfnamefont {K.}~\bibnamefont {Orginos}},
  \bibinfo {author} {\bibfnamefont {A.}~\bibnamefont {Parreno}}, \bibinfo
  {author} {\bibfnamefont {M.}~\bibnamefont {Savage}}, \ and\ \bibinfo {author}
  {\bibfnamefont {A.}~\bibnamefont {Walker-Loud}},\ }\href {\doibase
  10.1103/PhysRevD.87.034506} {\bibfield  {journal} {\bibinfo  {journal} {Phys.
  Rev. D}\ }\textbf {\bibinfo {volume} {87}},\ \bibinfo {pages} {034506}
  (\bibinfo {year} {2013})},\ \Eprint {http://arxiv.org/abs/1206.5219}
  {arXiv:1206.5219 [hep-lat]} \BibitemShut {NoStop}%
\bibitem [{\citenamefont {Beane}\ \emph {et~al.}(2013)\citenamefont {Beane}
  \emph {et~al.}}]{Beane:2013br}%
  \BibitemOpen
  \bibfield  {author} {\bibinfo {author} {\bibfnamefont {S.}~\bibnamefont
  {Beane}} \emph {et~al.} (\bibinfo {collaboration} {NPLQCD}),\ }\href
  {\doibase 10.1103/PhysRevC.88.024003} {\bibfield  {journal} {\bibinfo
  {journal} {Phys. Rev. C}\ }\textbf {\bibinfo {volume} {88}},\ \bibinfo
  {pages} {024003} (\bibinfo {year} {2013})},\ \Eprint
  {http://arxiv.org/abs/1301.5790} {arXiv:1301.5790 [hep-lat]} \BibitemShut
  {NoStop}%
\bibitem [{\citenamefont {Yamazaki}\ \emph {et~al.}(2015)\citenamefont
  {Yamazaki}, \citenamefont {Ishikawa}, \citenamefont {Kuramashi},\ and\
  \citenamefont {Ukawa}}]{Yamazaki:2015asa}%
  \BibitemOpen
  \bibfield  {author} {\bibinfo {author} {\bibfnamefont {T.}~\bibnamefont
  {Yamazaki}}, \bibinfo {author} {\bibfnamefont {K.-i.}\ \bibnamefont
  {Ishikawa}}, \bibinfo {author} {\bibfnamefont {Y.}~\bibnamefont {Kuramashi}},
  \ and\ \bibinfo {author} {\bibfnamefont {A.}~\bibnamefont {Ukawa}},\ }\href
  {\doibase 10.1103/PhysRevD.92.014501} {\bibfield  {journal} {\bibinfo
  {journal} {Phys. Rev.}\ }\textbf {\bibinfo {volume} {D92}},\ \bibinfo {pages}
  {14501} (\bibinfo {year} {2015})},\ \Eprint {http://arxiv.org/abs/1502.04182}
  {arXiv:1502.04182 [hep-lat]} \BibitemShut {NoStop}%
\bibitem [{\citenamefont {Weinberg}(1979)}]{Weinberg:1978kz}%
  \BibitemOpen
  \bibfield  {author} {\bibinfo {author} {\bibfnamefont {S.}~\bibnamefont
  {Weinberg}},\ }\href@noop {} {\bibfield  {journal} {\bibinfo  {journal}
  {Physica}\ }\textbf {\bibinfo {volume} {A96}},\ \bibinfo {pages} {327}
  (\bibinfo {year} {1979})}\BibitemShut {NoStop}%
\bibitem [{\citenamefont {Weinberg}(1990)}]{Weinberg:1990rz}%
  \BibitemOpen
  \bibfield  {author} {\bibinfo {author} {\bibfnamefont {S.}~\bibnamefont
  {Weinberg}},\ }\href {\doibase 10.1016/0370-2693(90)90938-3} {\bibfield
  {journal} {\bibinfo  {journal} {Phys. Lett.}\ }\textbf {\bibinfo {volume}
  {B251}},\ \bibinfo {pages} {288} (\bibinfo {year} {1990})}\BibitemShut
  {NoStop}%
\bibitem [{\citenamefont {Weinberg}(1991)}]{Weinberg:1991um}%
  \BibitemOpen
  \bibfield  {author} {\bibinfo {author} {\bibfnamefont {S.}~\bibnamefont
  {Weinberg}},\ }\href {\doibase 10.1016/0550-3213(91)90231-l} {\bibfield
  {journal} {\bibinfo  {journal} {Nucl. Phys.}\ }\textbf {\bibinfo {volume}
  {B363}},\ \bibinfo {pages} {3} (\bibinfo {year} {1991})}\BibitemShut
  {NoStop}%
\bibitem [{\citenamefont {Kaplan}\ \emph
  {et~al.}(1998{\natexlab{a}})\citenamefont {Kaplan}, \citenamefont {Savage},\
  and\ \citenamefont {Wise}}]{Kaplan:1998tg}%
  \BibitemOpen
  \bibfield  {author} {\bibinfo {author} {\bibfnamefont {D.~B.}\ \bibnamefont
  {Kaplan}}, \bibinfo {author} {\bibfnamefont {M.~J.}\ \bibnamefont {Savage}},
  \ and\ \bibinfo {author} {\bibfnamefont {M.~B.}\ \bibnamefont {Wise}},\
  }\href {\doibase 10.1016/S0370-2693(98)00210-X} {\bibfield  {journal}
  {\bibinfo  {journal} {Phys. Lett. B}\ }\textbf {\bibinfo {volume} {424}},\
  \bibinfo {pages} {390} (\bibinfo {year} {1998}{\natexlab{a}})},\ \Eprint
  {http://arxiv.org/abs/nucl-th/9801034} {arXiv:nucl-th/9801034} \BibitemShut
  {NoStop}%
\bibitem [{\citenamefont {Kaplan}\ \emph
  {et~al.}(1998{\natexlab{b}})\citenamefont {Kaplan}, \citenamefont {Savage},\
  and\ \citenamefont {Wise}}]{Kaplan:1998we}%
  \BibitemOpen
  \bibfield  {author} {\bibinfo {author} {\bibfnamefont {D.~B.}\ \bibnamefont
  {Kaplan}}, \bibinfo {author} {\bibfnamefont {M.~J.}\ \bibnamefont {Savage}},
  \ and\ \bibinfo {author} {\bibfnamefont {M.~B.}\ \bibnamefont {Wise}},\
  }\href {\doibase 10.1016/S0550-3213(98)00440-4} {\bibfield  {journal}
  {\bibinfo  {journal} {Nucl. Phys. B}\ }\textbf {\bibinfo {volume} {534}},\
  \bibinfo {pages} {329} (\bibinfo {year} {1998}{\natexlab{b}})},\ \Eprint
  {http://arxiv.org/abs/nucl-th/9802075} {arXiv:nucl-th/9802075} \BibitemShut
  {NoStop}%
\bibitem [{\citenamefont {Szpigel}\ and\ \citenamefont
  {Perry}(1999)}]{Szpigel:1999gf}%
  \BibitemOpen
  \bibfield  {author} {\bibinfo {author} {\bibfnamefont {S.}~\bibnamefont
  {Szpigel}}\ and\ \bibinfo {author} {\bibfnamefont {R.~J.}\ \bibnamefont
  {Perry}},\ }\href@noop {} {\  (\bibinfo {year} {1999})},\ \Eprint
  {http://arxiv.org/abs/nucl-th/9906031} {arXiv:nucl-th/9906031} \BibitemShut
  {NoStop}%
\bibitem [{\citenamefont {Bogner}\ \emph {et~al.}(2003)\citenamefont {Bogner},
  \citenamefont {Kuo},\ and\ \citenamefont {Schwenk}}]{Bogner:2003wn}%
  \BibitemOpen
  \bibfield  {author} {\bibinfo {author} {\bibfnamefont {S.}~\bibnamefont
  {Bogner}}, \bibinfo {author} {\bibfnamefont {T.}~\bibnamefont {Kuo}}, \ and\
  \bibinfo {author} {\bibfnamefont {A.}~\bibnamefont {Schwenk}},\ }\href
  {\doibase 10.1016/j.physrep.2003.07.001} {\bibfield  {journal} {\bibinfo
  {journal} {Phys. Rept.}\ }\textbf {\bibinfo {volume} {386}},\ \bibinfo
  {pages} {1} (\bibinfo {year} {2003})},\ \Eprint
  {http://arxiv.org/abs/nucl-th/0305035} {arXiv:nucl-th/0305035} \BibitemShut
  {NoStop}%
\bibitem [{\citenamefont {Carlson}\ \emph {et~al.}(2015)\citenamefont
  {Carlson}, \citenamefont {Gandolfi}, \citenamefont {Pederiva}, \citenamefont
  {Pieper}, \citenamefont {Schiavilla}, \citenamefont {Schmidt},\ and\
  \citenamefont {Wiringa}}]{Carlson:2014vla}%
  \BibitemOpen
  \bibfield  {author} {\bibinfo {author} {\bibfnamefont {J.}~\bibnamefont
  {Carlson}}, \bibinfo {author} {\bibfnamefont {S.}~\bibnamefont {Gandolfi}},
  \bibinfo {author} {\bibfnamefont {F.}~\bibnamefont {Pederiva}}, \bibinfo
  {author} {\bibfnamefont {S.~C.}\ \bibnamefont {Pieper}}, \bibinfo {author}
  {\bibfnamefont {R.}~\bibnamefont {Schiavilla}}, \bibinfo {author}
  {\bibfnamefont {K.~E.}\ \bibnamefont {Schmidt}}, \ and\ \bibinfo {author}
  {\bibfnamefont {R.~B.}\ \bibnamefont {Wiringa}},\ }\href
  {https://doi.org/10.1103/RevModPhys.87.1067} {\bibfield  {journal} {\bibinfo
  {journal} {Rev. Mod. Phys.}\ }\textbf {\bibinfo {volume} {87}},\ \bibinfo
  {pages} {1067} (\bibinfo {year} {2015})},\ \Eprint
  {http://arxiv.org/abs/1412.3081} {arXiv:1412.3081 [nucl-th]} \BibitemShut
  {NoStop}%
\bibitem [{\citenamefont {Carlson}\ \emph {et~al.}(2017)\citenamefont {Carlson}
  \emph {et~al.}}]{Carlson:2017ebk}%
  \BibitemOpen
  \bibfield  {author} {\bibinfo {author} {\bibfnamefont {J.}~\bibnamefont
  {Carlson}} \emph {et~al.},\ }\href {\doibase 10.1016/j.ppnp.2016.11.002}
  {\bibfield  {journal} {\bibinfo  {journal} {Prog. Part. Nucl. Phys.}\
  }\textbf {\bibinfo {volume} {94}},\ \bibinfo {pages} {68} (\bibinfo {year}
  {2017})}\BibitemShut {NoStop}%
\bibitem [{\citenamefont {Gandolfi}\ \emph {et~al.}(2018)\citenamefont
  {Gandolfi}, \citenamefont {Carlson}, \citenamefont {Roggero}, \citenamefont
  {Lynn},\ and\ \citenamefont {Reddy}}]{Gandolfi:2017arm}%
  \BibitemOpen
  \bibfield  {author} {\bibinfo {author} {\bibfnamefont {S.}~\bibnamefont
  {Gandolfi}}, \bibinfo {author} {\bibfnamefont {J.}~\bibnamefont {Carlson}},
  \bibinfo {author} {\bibfnamefont {A.}~\bibnamefont {Roggero}}, \bibinfo
  {author} {\bibfnamefont {J.}~\bibnamefont {Lynn}}, \ and\ \bibinfo {author}
  {\bibfnamefont {S.}~\bibnamefont {Reddy}},\ }\href {\doibase
  10.1016/j.physletb.2018.07.073} {\bibfield  {journal} {\bibinfo  {journal}
  {Phys. Lett.}\ }\textbf {\bibinfo {volume} {B785}},\ \bibinfo {pages} {232}
  (\bibinfo {year} {2018})},\ \Eprint {http://arxiv.org/abs/1712.10236}
  {arXiv:1712.10236 [nucl-th]} \BibitemShut {NoStop}%
\bibitem [{\citenamefont {Contessi}\ \emph {et~al.}(2017)\citenamefont
  {Contessi}, \citenamefont {Lovato}, \citenamefont {Pederiva}, \citenamefont
  {Roggero}, \citenamefont {Kirscher},\ and\ \citenamefont {van
  Kolck}}]{Contessi:2017rww}%
  \BibitemOpen
  \bibfield  {author} {\bibinfo {author} {\bibfnamefont {L.}~\bibnamefont
  {Contessi}}, \bibinfo {author} {\bibfnamefont {A.}~\bibnamefont {Lovato}},
  \bibinfo {author} {\bibfnamefont {F.}~\bibnamefont {Pederiva}}, \bibinfo
  {author} {\bibfnamefont {A.}~\bibnamefont {Roggero}}, \bibinfo {author}
  {\bibfnamefont {J.}~\bibnamefont {Kirscher}}, \ and\ \bibinfo {author}
  {\bibfnamefont {U.}~\bibnamefont {van Kolck}},\ }\href {\doibase
  10.1016/j.physletb.2017.07.048} {\bibfield  {journal} {\bibinfo  {journal}
  {Phys. Lett. B}\ }\textbf {\bibinfo {volume} {772}},\ \bibinfo {pages} {839}
  (\bibinfo {year} {2017})},\ \Eprint {http://arxiv.org/abs/1701.06516}
  {arXiv:1701.06516 [nucl-th]} \BibitemShut {NoStop}%
\bibitem [{\citenamefont {Bansal}\ \emph {et~al.}(2018)\citenamefont {Bansal},
  \citenamefont {Binder}, \citenamefont {Ekstr{\"o}m}, \citenamefont {Hagen},
  \citenamefont {Jansen},\ and\ \citenamefont {Papenbrock}}]{Bansal:2017pwn}%
  \BibitemOpen
  \bibfield  {author} {\bibinfo {author} {\bibfnamefont {A.}~\bibnamefont
  {Bansal}}, \bibinfo {author} {\bibfnamefont {S.}~\bibnamefont {Binder}},
  \bibinfo {author} {\bibfnamefont {A.}~\bibnamefont {Ekstr{\"o}m}}, \bibinfo
  {author} {\bibfnamefont {G.}~\bibnamefont {Hagen}}, \bibinfo {author}
  {\bibfnamefont {G.}~\bibnamefont {Jansen}}, \ and\ \bibinfo {author}
  {\bibfnamefont {T.}~\bibnamefont {Papenbrock}},\ }\href {\doibase
  10.1103/PhysRevC.98.054301} {\bibfield  {journal} {\bibinfo  {journal} {Phys.
  Rev. C}\ }\textbf {\bibinfo {volume} {98}},\ \bibinfo {pages} {054301}
  (\bibinfo {year} {2018})},\ \Eprint {http://arxiv.org/abs/1712.10246}
  {arXiv:1712.10246 [nucl-th]} \BibitemShut {NoStop}%
\bibitem [{\citenamefont {Lonardoni}\ \emph {et~al.}(2018)\citenamefont
  {Lonardoni}, \citenamefont {Gandolfi}, \citenamefont {Lynn}, \citenamefont
  {Petrie}, \citenamefont {Carlson}, \citenamefont {Schmidt},\ and\
  \citenamefont {Schwenk}}]{Lonardoni:2018nob}%
  \BibitemOpen
  \bibfield  {author} {\bibinfo {author} {\bibfnamefont {D.}~\bibnamefont
  {Lonardoni}}, \bibinfo {author} {\bibfnamefont {S.}~\bibnamefont {Gandolfi}},
  \bibinfo {author} {\bibfnamefont {J.~E.}\ \bibnamefont {Lynn}}, \bibinfo
  {author} {\bibfnamefont {C.}~\bibnamefont {Petrie}}, \bibinfo {author}
  {\bibfnamefont {J.}~\bibnamefont {Carlson}}, \bibinfo {author} {\bibfnamefont
  {K.~E.}\ \bibnamefont {Schmidt}}, \ and\ \bibinfo {author} {\bibfnamefont
  {A.}~\bibnamefont {Schwenk}},\ }\href {\doibase 10.1103/PhysRevC.97.044318}
  {\bibfield  {journal} {\bibinfo  {journal} {Phys. Rev. C}\ }\textbf {\bibinfo
  {volume} {97}},\ \bibinfo {pages} {044318} (\bibinfo {year}
  {2018})}\BibitemShut {NoStop}%
\bibitem [{\citenamefont {Johnson}\ \emph {et~al.}(2018)\citenamefont
  {Johnson}, \citenamefont {Ormand}, \citenamefont {McElvain},\ and\
  \citenamefont {Shan}}]{Johnson:2018hrx}%
  \BibitemOpen
  \bibfield  {author} {\bibinfo {author} {\bibfnamefont {C.~W.}\ \bibnamefont
  {Johnson}}, \bibinfo {author} {\bibfnamefont {W.~E.}\ \bibnamefont {Ormand}},
  \bibinfo {author} {\bibfnamefont {K.~S.}\ \bibnamefont {McElvain}}, \ and\
  \bibinfo {author} {\bibfnamefont {H.}~\bibnamefont {Shan}},\ }\href@noop {}
  {\  (\bibinfo {year} {2018})},\ \Eprint {http://arxiv.org/abs/1801.08432}
  {arXiv:1801.08432 [physics.comp-ph]} \BibitemShut {NoStop}%
\bibitem [{\citenamefont {King}\ \emph {et~al.}(2020)\citenamefont {King},
  \citenamefont {Andreoli}, \citenamefont {Pastore}, \citenamefont {Piarulli},
  \citenamefont {Schiavilla}, \citenamefont {Wiringa}, \citenamefont
  {Carlson},\ and\ \citenamefont {Gandolfi}}]{King:2020wmp}%
  \BibitemOpen
  \bibfield  {author} {\bibinfo {author} {\bibfnamefont {G.~B.}\ \bibnamefont
  {King}}, \bibinfo {author} {\bibfnamefont {L.}~\bibnamefont {Andreoli}},
  \bibinfo {author} {\bibfnamefont {S.}~\bibnamefont {Pastore}}, \bibinfo
  {author} {\bibfnamefont {M.}~\bibnamefont {Piarulli}}, \bibinfo {author}
  {\bibfnamefont {R.}~\bibnamefont {Schiavilla}}, \bibinfo {author}
  {\bibfnamefont {R.~B.}\ \bibnamefont {Wiringa}}, \bibinfo {author}
  {\bibfnamefont {J.}~\bibnamefont {Carlson}}, \ and\ \bibinfo {author}
  {\bibfnamefont {S.}~\bibnamefont {Gandolfi}},\ }\href {\doibase
  10.1103/PhysRevC.102.025501} {\bibfield  {journal} {\bibinfo  {journal}
  {Phys. Rev. C}\ }\textbf {\bibinfo {volume} {102}},\ \bibinfo {pages}
  {025501} (\bibinfo {year} {2020})}\BibitemShut {NoStop}%
\bibitem [{\citenamefont {Barrett}\ \emph {et~al.}(2013)\citenamefont
  {Barrett}, \citenamefont {Navr{\'a}til},\ and\ \citenamefont
  {Vary}}]{BARRETT2013131}%
  \BibitemOpen
  \bibfield  {author} {\bibinfo {author} {\bibfnamefont {B.~R.}\ \bibnamefont
  {Barrett}}, \bibinfo {author} {\bibfnamefont {P.}~\bibnamefont
  {Navr{\'a}til}}, \ and\ \bibinfo {author} {\bibfnamefont {J.~P.}\
  \bibnamefont {Vary}},\ }\href {\doibase
  https://doi.org/10.1016/j.ppnp.2012.10.003} {\bibfield  {journal} {\bibinfo
  {journal} {Progress in Particle and Nuclear Physics}\ }\textbf {\bibinfo
  {volume} {69}},\ \bibinfo {pages} {131 } (\bibinfo {year}
  {2013})}\BibitemShut {NoStop}%
\bibitem [{\citenamefont {Roth}\ \emph {et~al.}(2011)\citenamefont {Roth},
  \citenamefont {Langhammer}, \citenamefont {Calci}, \citenamefont {Binder},\
  and\ \citenamefont {Navr\'atil}}]{PhysRevLett.107.072501}%
  \BibitemOpen
  \bibfield  {author} {\bibinfo {author} {\bibfnamefont {R.}~\bibnamefont
  {Roth}}, \bibinfo {author} {\bibfnamefont {J.}~\bibnamefont {Langhammer}},
  \bibinfo {author} {\bibfnamefont {A.}~\bibnamefont {Calci}}, \bibinfo
  {author} {\bibfnamefont {S.}~\bibnamefont {Binder}}, \ and\ \bibinfo {author}
  {\bibfnamefont {P.}~\bibnamefont {Navr\'atil}},\ }\href {\doibase
  10.1103/PhysRevLett.107.072501} {\bibfield  {journal} {\bibinfo  {journal}
  {Phys. Rev. Lett.}\ }\textbf {\bibinfo {volume} {107}},\ \bibinfo {pages}
  {072501} (\bibinfo {year} {2011})}\BibitemShut {NoStop}%
\bibitem [{\citenamefont {Vary}\ \emph {et~al.}(2015)\citenamefont {Vary} \emph
  {et~al.}}]{Vary:2015dda}%
  \BibitemOpen
  \bibfield  {author} {\bibinfo {author} {\bibfnamefont {J.~P.}\ \bibnamefont
  {Vary}} \emph {et~al.},\ }in\ \href@noop {} {\emph {\bibinfo {booktitle}
  {{International Conference Nuclear Theory in the Supercomputing Era}}}}\
  (\bibinfo {year} {2015})\ \Eprint {http://arxiv.org/abs/1507.04693}
  {arXiv:1507.04693 [nucl-th]} \BibitemShut {NoStop}%
\bibitem [{\citenamefont {Hergert}\ \emph {et~al.}(2016)\citenamefont
  {Hergert}, \citenamefont {Bogner}, \citenamefont {Morris}, \citenamefont
  {Schwenk},\ and\ \citenamefont {Tsukiyama}}]{HERGERT2016165}%
  \BibitemOpen
  \bibfield  {author} {\bibinfo {author} {\bibfnamefont {H.}~\bibnamefont
  {Hergert}}, \bibinfo {author} {\bibfnamefont {S.}~\bibnamefont {Bogner}},
  \bibinfo {author} {\bibfnamefont {T.}~\bibnamefont {Morris}}, \bibinfo
  {author} {\bibfnamefont {A.}~\bibnamefont {Schwenk}}, \ and\ \bibinfo
  {author} {\bibfnamefont {K.}~\bibnamefont {Tsukiyama}},\ }\href {\doibase
  https://doi.org/10.1016/j.physrep.2015.12.007} {\bibfield  {journal}
  {\bibinfo  {journal} {Physics Reports}\ }\textbf {\bibinfo {volume} {621}},\
  \bibinfo {pages} {165 } (\bibinfo {year} {2016})}\BibitemShut {NoStop}%
\bibitem [{\citenamefont {Stroberg}\ \emph {et~al.}(2017)\citenamefont
  {Stroberg}, \citenamefont {Calci}, \citenamefont {Hergert}, \citenamefont
  {Holt}, \citenamefont {Bogner}, \citenamefont {Roth},\ and\ \citenamefont
  {Schwenk}}]{PhysRevLett.118.032502}%
  \BibitemOpen
  \bibfield  {author} {\bibinfo {author} {\bibfnamefont {S.~R.}\ \bibnamefont
  {Stroberg}}, \bibinfo {author} {\bibfnamefont {A.}~\bibnamefont {Calci}},
  \bibinfo {author} {\bibfnamefont {H.}~\bibnamefont {Hergert}}, \bibinfo
  {author} {\bibfnamefont {J.~D.}\ \bibnamefont {Holt}}, \bibinfo {author}
  {\bibfnamefont {S.~K.}\ \bibnamefont {Bogner}}, \bibinfo {author}
  {\bibfnamefont {R.}~\bibnamefont {Roth}}, \ and\ \bibinfo {author}
  {\bibfnamefont {A.}~\bibnamefont {Schwenk}},\ }\href {\doibase
  10.1103/PhysRevLett.118.032502} {\bibfield  {journal} {\bibinfo  {journal}
  {Phys. Rev. Lett.}\ }\textbf {\bibinfo {volume} {118}},\ \bibinfo {pages}
  {032502} (\bibinfo {year} {2017})}\BibitemShut {NoStop}%
\bibitem [{\citenamefont {Hagen}\ \emph {et~al.}(2014)\citenamefont {Hagen},
  \citenamefont {Papenbrock}, \citenamefont {Hjorth-Jensen},\ and\
  \citenamefont {Dean}}]{Hagen:2013nca}%
  \BibitemOpen
  \bibfield  {author} {\bibinfo {author} {\bibfnamefont {G.}~\bibnamefont
  {Hagen}}, \bibinfo {author} {\bibfnamefont {T.}~\bibnamefont {Papenbrock}},
  \bibinfo {author} {\bibfnamefont {M.}~\bibnamefont {Hjorth-Jensen}}, \ and\
  \bibinfo {author} {\bibfnamefont {D.}~\bibnamefont {Dean}},\ }\href {\doibase
  10.1088/0034-4885/77/9/096302} {\bibfield  {journal} {\bibinfo  {journal}
  {Rept. Prog. Phys.}\ }\textbf {\bibinfo {volume} {77}},\ \bibinfo {pages}
  {096302} (\bibinfo {year} {2014})},\ \Eprint {http://arxiv.org/abs/1312.7872}
  {arXiv:1312.7872 [nucl-th]} \BibitemShut {NoStop}%
\bibitem [{\citenamefont {Lähde}\ \emph {et~al.}(2014)\citenamefont {Lähde},
  \citenamefont {Epelbaum}, \citenamefont {Krebs}, \citenamefont {Lee},
  \citenamefont {Meißner},\ and\ \citenamefont {Rupak}}]{Lahde:2013uqa}%
  \BibitemOpen
  \bibfield  {author} {\bibinfo {author} {\bibfnamefont {T.~A.}\ \bibnamefont
  {Lähde}}, \bibinfo {author} {\bibfnamefont {E.}~\bibnamefont {Epelbaum}},
  \bibinfo {author} {\bibfnamefont {H.}~\bibnamefont {Krebs}}, \bibinfo
  {author} {\bibfnamefont {D.}~\bibnamefont {Lee}}, \bibinfo {author}
  {\bibfnamefont {U.-G.}\ \bibnamefont {Meißner}}, \ and\ \bibinfo {author}
  {\bibfnamefont {G.}~\bibnamefont {Rupak}},\ }\href {\doibase
  10.1016/j.physletb.2014.03.023} {\bibfield  {journal} {\bibinfo  {journal}
  {Phys. Lett. B}\ }\textbf {\bibinfo {volume} {732}},\ \bibinfo {pages} {110}
  (\bibinfo {year} {2014})},\ \Eprint {http://arxiv.org/abs/1311.0477}
  {arXiv:1311.0477 [nucl-th]} \BibitemShut {NoStop}%
\bibitem [{\citenamefont {Som\`a}\ \emph {et~al.}(2014)\citenamefont {Som\`a},
  \citenamefont {Barbieri},\ and\ \citenamefont {Duguet}}]{PhysRevC.89.024323}%
  \BibitemOpen
  \bibfield  {author} {\bibinfo {author} {\bibfnamefont {V.}~\bibnamefont
  {Som\`a}}, \bibinfo {author} {\bibfnamefont {C.}~\bibnamefont {Barbieri}}, \
  and\ \bibinfo {author} {\bibfnamefont {T.}~\bibnamefont {Duguet}},\ }\href
  {\doibase 10.1103/PhysRevC.89.024323} {\bibfield  {journal} {\bibinfo
  {journal} {Phys. Rev. C}\ }\textbf {\bibinfo {volume} {89}},\ \bibinfo
  {pages} {024323} (\bibinfo {year} {2014})}\BibitemShut {NoStop}%
\bibitem [{\citenamefont {Cipollone}\ \emph {et~al.}(2013)\citenamefont
  {Cipollone}, \citenamefont {Barbieri},\ and\ \citenamefont
  {Navr\'atil}}]{PhysRevLett.111.062501}%
  \BibitemOpen
  \bibfield  {author} {\bibinfo {author} {\bibfnamefont {A.}~\bibnamefont
  {Cipollone}}, \bibinfo {author} {\bibfnamefont {C.}~\bibnamefont {Barbieri}},
  \ and\ \bibinfo {author} {\bibfnamefont {P.}~\bibnamefont {Navr\'atil}},\
  }\href {\doibase 10.1103/PhysRevLett.111.062501} {\bibfield  {journal}
  {\bibinfo  {journal} {Phys. Rev. Lett.}\ }\textbf {\bibinfo {volume} {111}},\
  \bibinfo {pages} {062501} (\bibinfo {year} {2013})}\BibitemShut {NoStop}%
\bibitem [{\citenamefont {Haxton}\ and\ \citenamefont
  {Luu}(2002)}]{Haxton:2002kb}%
  \BibitemOpen
  \bibfield  {author} {\bibinfo {author} {\bibfnamefont {W.}~\bibnamefont
  {Haxton}}\ and\ \bibinfo {author} {\bibfnamefont {T.}~\bibnamefont {Luu}},\
  }\href {\doibase 10.1103/PhysRevLett.89.182503} {\bibfield  {journal}
  {\bibinfo  {journal} {Phys. Rev. Lett.}\ }\textbf {\bibinfo {volume} {89}},\
  \bibinfo {pages} {182503} (\bibinfo {year} {2002})},\ \Eprint
  {http://arxiv.org/abs/nucl-th/0204072} {arXiv:nucl-th/0204072} \BibitemShut
  {NoStop}%
\bibitem [{\citenamefont {Signoracci}\ \emph {et~al.}(2011)\citenamefont
  {Signoracci}, \citenamefont {Brown},\ and\ \citenamefont
  {Hjorth-Jensen}}]{Signoracci:2010bz}%
  \BibitemOpen
  \bibfield  {author} {\bibinfo {author} {\bibfnamefont {A.}~\bibnamefont
  {Signoracci}}, \bibinfo {author} {\bibfnamefont {B.~A.}\ \bibnamefont
  {Brown}}, \ and\ \bibinfo {author} {\bibfnamefont {M.}~\bibnamefont
  {Hjorth-Jensen}},\ }\href {\doibase 10.1103/PhysRevC.83.024315} {\bibfield
  {journal} {\bibinfo  {journal} {Phys. Rev. C}\ }\textbf {\bibinfo {volume}
  {83}},\ \bibinfo {pages} {024315} (\bibinfo {year} {2011})},\ \Eprint
  {http://arxiv.org/abs/1009.4916} {arXiv:1009.4916 [nucl-th]} \BibitemShut
  {NoStop}%
\bibitem [{\citenamefont {Lietz}\ \emph {et~al.}(2017)\citenamefont {Lietz},
  \citenamefont {Novario}, \citenamefont {Jansen}, \citenamefont {Hagen},\ and\
  \citenamefont {Hjorth-Jensen}}]{Lietz:2016qfb}%
  \BibitemOpen
  \bibfield  {author} {\bibinfo {author} {\bibfnamefont {J.~G.}\ \bibnamefont
  {Lietz}}, \bibinfo {author} {\bibfnamefont {S.}~\bibnamefont {Novario}},
  \bibinfo {author} {\bibfnamefont {G.~R.}\ \bibnamefont {Jansen}}, \bibinfo
  {author} {\bibfnamefont {G.}~\bibnamefont {Hagen}}, \ and\ \bibinfo {author}
  {\bibfnamefont {M.}~\bibnamefont {Hjorth-Jensen}},\ }\enquote {\bibinfo
  {title} {Computational nuclear physics and post hartree-fock methods},}\ in\
  \href {\doibase 10.1007/978-3-319-53336-0_8} {\emph {\bibinfo {booktitle} {An
  Advanced Course in Computational Nuclear Physics: Bridging the Scales from
  Quarks to Neutron Stars}}},\ \bibinfo {editor} {edited by\ \bibinfo {editor}
  {\bibfnamefont {M.}~\bibnamefont {Hjorth-Jensen}}, \bibinfo {editor}
  {\bibfnamefont {M.~P.}\ \bibnamefont {Lombardo}}, \ and\ \bibinfo {editor}
  {\bibfnamefont {U.}~\bibnamefont {van Kolck}}}\ (\bibinfo  {publisher}
  {Springer International Publishing},\ \bibinfo {address} {Cham},\ \bibinfo
  {year} {2017})\ pp.\ \bibinfo {pages} {293--399}\BibitemShut {NoStop}%
\bibitem [{\citenamefont {Zelevinsky}\ \emph {et~al.}(1995)\citenamefont
  {Zelevinsky}, \citenamefont {Horoi},\ and\ \citenamefont
  {Brown}}]{ZELEVINSKY1995141}%
  \BibitemOpen
  \bibfield  {author} {\bibinfo {author} {\bibfnamefont {V.}~\bibnamefont
  {Zelevinsky}}, \bibinfo {author} {\bibfnamefont {M.}~\bibnamefont {Horoi}}, \
  and\ \bibinfo {author} {\bibfnamefont {B.~A.}\ \bibnamefont {Brown}},\ }\href
  {\doibase https://doi.org/10.1016/0370-2693(95)00324-E} {\bibfield  {journal}
  {\bibinfo  {journal} {Physics Letters B}\ }\textbf {\bibinfo {volume}
  {350}},\ \bibinfo {pages} {141 } (\bibinfo {year} {1995})}\BibitemShut
  {NoStop}%
\bibitem [{\citenamefont {Zelevinsky}\ \emph {et~al.}(1996)\citenamefont
  {Zelevinsky}, \citenamefont {Brown}, \citenamefont {Frazier},\ and\
  \citenamefont {Horoi}}]{ZELEVINSKY199685}%
  \BibitemOpen
  \bibfield  {author} {\bibinfo {author} {\bibfnamefont {V.}~\bibnamefont
  {Zelevinsky}}, \bibinfo {author} {\bibfnamefont {B.~A.}\ \bibnamefont
  {Brown}}, \bibinfo {author} {\bibfnamefont {N.}~\bibnamefont {Frazier}}, \
  and\ \bibinfo {author} {\bibfnamefont {M.}~\bibnamefont {Horoi}},\ }\href
  {\doibase https://doi.org/10.1016/S0370-1573(96)00007-5} {\bibfield
  {journal} {\bibinfo  {journal} {Physics Reports}\ }\textbf {\bibinfo {volume}
  {276}},\ \bibinfo {pages} {85 } (\bibinfo {year} {1996})}\BibitemShut
  {NoStop}%
\bibitem [{\citenamefont {Sokolov}\ \emph {et~al.}(1998)\citenamefont
  {Sokolov}, \citenamefont {Brown},\ and\ \citenamefont
  {Zelevinsky}}]{PhysRevE.58.56}%
  \BibitemOpen
  \bibfield  {author} {\bibinfo {author} {\bibfnamefont {V.~V.}\ \bibnamefont
  {Sokolov}}, \bibinfo {author} {\bibfnamefont {B.~A.}\ \bibnamefont {Brown}},
  \ and\ \bibinfo {author} {\bibfnamefont {V.}~\bibnamefont {Zelevinsky}},\
  }\href {\doibase 10.1103/PhysRevE.58.56} {\bibfield  {journal} {\bibinfo
  {journal} {Phys. Rev. E}\ }\textbf {\bibinfo {volume} {58}},\ \bibinfo
  {pages} {56} (\bibinfo {year} {1998})}\BibitemShut {NoStop}%
\bibitem [{\citenamefont {Volya}\ and\ \citenamefont
  {Zelevinsky}(2003)}]{VOLYA200327}%
  \BibitemOpen
  \bibfield  {author} {\bibinfo {author} {\bibfnamefont {A.}~\bibnamefont
  {Volya}}\ and\ \bibinfo {author} {\bibfnamefont {V.}~\bibnamefont
  {Zelevinsky}},\ }\href {\doibase
  https://doi.org/10.1016/j.physletb.2003.08.076} {\bibfield  {journal}
  {\bibinfo  {journal} {Physics Letters B}\ }\textbf {\bibinfo {volume}
  {574}},\ \bibinfo {pages} {27 } (\bibinfo {year} {2003})}\BibitemShut
  {NoStop}%
\bibitem [{\citenamefont {Guan}\ \emph {et~al.}(2013)\citenamefont {Guan},
  \citenamefont {Launey}, \citenamefont {Gu}, \citenamefont {Pan},\ and\
  \citenamefont {Draayer}}]{PhysRevC.88.044325}%
  \BibitemOpen
  \bibfield  {author} {\bibinfo {author} {\bibfnamefont {X.}~\bibnamefont
  {Guan}}, \bibinfo {author} {\bibfnamefont {K.~D.}\ \bibnamefont {Launey}},
  \bibinfo {author} {\bibfnamefont {J.}~\bibnamefont {Gu}}, \bibinfo {author}
  {\bibfnamefont {F.}~\bibnamefont {Pan}}, \ and\ \bibinfo {author}
  {\bibfnamefont {J.~P.}\ \bibnamefont {Draayer}},\ }\href {\doibase
  10.1103/PhysRevC.88.044325} {\bibfield  {journal} {\bibinfo  {journal} {Phys.
  Rev. C}\ }\textbf {\bibinfo {volume} {88}},\ \bibinfo {pages} {044325}
  (\bibinfo {year} {2013})}\BibitemShut {NoStop}%
\bibitem [{\citenamefont {Berges}\ \emph
  {et~al.}(2018{\natexlab{a}})\citenamefont {Berges}, \citenamefont
  {Floerchinger},\ and\ \citenamefont {Venugopalan}}]{Berges:2017zws}%
  \BibitemOpen
  \bibfield  {author} {\bibinfo {author} {\bibfnamefont {J.}~\bibnamefont
  {Berges}}, \bibinfo {author} {\bibfnamefont {S.}~\bibnamefont
  {Floerchinger}}, \ and\ \bibinfo {author} {\bibfnamefont {R.}~\bibnamefont
  {Venugopalan}},\ }\href {\doibase 10.1016/j.physletb.2018.01.068} {\bibfield
  {journal} {\bibinfo  {journal} {Phys. Lett.}\ }\textbf {\bibinfo {volume}
  {B778}},\ \bibinfo {pages} {442} (\bibinfo {year} {2018}{\natexlab{a}})},\
  \Eprint {http://arxiv.org/abs/1707.05338} {arXiv:1707.05338 [hep-ph]}
  \BibitemShut {NoStop}%
\bibitem [{\citenamefont {Berges}\ \emph
  {et~al.}(2018{\natexlab{b}})\citenamefont {Berges}, \citenamefont
  {Floerchinger},\ and\ \citenamefont {Venugopalan}}]{Berges:2017hne}%
  \BibitemOpen
  \bibfield  {author} {\bibinfo {author} {\bibfnamefont {J.}~\bibnamefont
  {Berges}}, \bibinfo {author} {\bibfnamefont {S.}~\bibnamefont
  {Floerchinger}}, \ and\ \bibinfo {author} {\bibfnamefont {R.}~\bibnamefont
  {Venugopalan}},\ }\href {\doibase 10.1007/JHEP04(2018)145} {\bibfield
  {journal} {\bibinfo  {journal} {JHEP}\ }\textbf {\bibinfo {volume} {04}},\
  \bibinfo {pages} {145} (\bibinfo {year} {2018}{\natexlab{b}})},\ \Eprint
  {http://arxiv.org/abs/1712.09362} {arXiv:1712.09362 [hep-th]} \BibitemShut
  {NoStop}%
\bibitem [{\citenamefont {Berges}\ \emph {et~al.}(2019)\citenamefont {Berges},
  \citenamefont {Floerchinger},\ and\ \citenamefont
  {Venugopalan}}]{Berges:2018cny}%
  \BibitemOpen
  \bibfield  {author} {\bibinfo {author} {\bibfnamefont {J.}~\bibnamefont
  {Berges}}, \bibinfo {author} {\bibfnamefont {S.}~\bibnamefont
  {Floerchinger}}, \ and\ \bibinfo {author} {\bibfnamefont {R.}~\bibnamefont
  {Venugopalan}},\ }\bibfield  {booktitle} {\emph {\bibinfo {booktitle}
  {{Proceedings, 27th International Conference on Ultrarelativistic
  Nucleus-Nucleus Collisions (Quark Matter 2018): Venice, Italy, May 14-19,
  2018}}},\ }\href {\doibase 10.1016/j.nuclphysa.2018.12.008} {\bibfield
  {journal} {\bibinfo  {journal} {Nucl. Phys.}\ }\textbf {\bibinfo {volume}
  {A982}},\ \bibinfo {pages} {819} (\bibinfo {year} {2019})},\ \Eprint
  {http://arxiv.org/abs/1812.08120} {arXiv:1812.08120 [hep-th]} \BibitemShut
  {NoStop}%
\bibitem [{\citenamefont {Ho}\ and\ \citenamefont {Hsu}(2016)}]{Ho:2015rga}%
  \BibitemOpen
  \bibfield  {author} {\bibinfo {author} {\bibfnamefont {C.~M.}\ \bibnamefont
  {Ho}}\ and\ \bibinfo {author} {\bibfnamefont {S.~D.~H.}\ \bibnamefont
  {Hsu}},\ }\href {\doibase 10.1142/S0217732316501108} {\bibfield  {journal}
  {\bibinfo  {journal} {Mod. Phys. Lett.}\ }\textbf {\bibinfo {volume} {A31}},\
  \bibinfo {pages} {1650110} (\bibinfo {year} {2016})},\ \Eprint
  {http://arxiv.org/abs/1506.03696} {arXiv:1506.03696 [hep-th]} \BibitemShut
  {NoStop}%
\bibitem [{\citenamefont {Kovner}\ and\ \citenamefont
  {Lublinsky}(2015)}]{Kovner:2015hga}%
  \BibitemOpen
  \bibfield  {author} {\bibinfo {author} {\bibfnamefont {A.}~\bibnamefont
  {Kovner}}\ and\ \bibinfo {author} {\bibfnamefont {M.}~\bibnamefont
  {Lublinsky}},\ }\href {\doibase 10.1103/PhysRevD.92.034016} {\bibfield
  {journal} {\bibinfo  {journal} {Phys. Rev.}\ }\textbf {\bibinfo {volume}
  {D92}},\ \bibinfo {pages} {034016} (\bibinfo {year} {2015})},\ \Eprint
  {http://arxiv.org/abs/1506.05394} {arXiv:1506.05394 [hep-ph]} \BibitemShut
  {NoStop}%
\bibitem [{\citenamefont {Kovner}\ \emph {et~al.}(2019)\citenamefont {Kovner},
  \citenamefont {Lublinsky},\ and\ \citenamefont {Serino}}]{Kovner:2018rbf}%
  \BibitemOpen
  \bibfield  {author} {\bibinfo {author} {\bibfnamefont {A.}~\bibnamefont
  {Kovner}}, \bibinfo {author} {\bibfnamefont {M.}~\bibnamefont {Lublinsky}}, \
  and\ \bibinfo {author} {\bibfnamefont {M.}~\bibnamefont {Serino}},\ }\href
  {\doibase 10.1016/j.physletb.2018.10.043} {\bibfield  {journal} {\bibinfo
  {journal} {Phys. Lett.}\ }\textbf {\bibinfo {volume} {B792}},\ \bibinfo
  {pages} {4} (\bibinfo {year} {2019})},\ \Eprint
  {http://arxiv.org/abs/1806.01089} {arXiv:1806.01089 [hep-ph]} \BibitemShut
  {NoStop}%
\bibitem [{\citenamefont {Armesto}\ \emph {et~al.}(2019)\citenamefont
  {Armesto}, \citenamefont {Dominguez}, \citenamefont {Kovner}, \citenamefont
  {Lublinsky},\ and\ \citenamefont {Skokov}}]{Armesto:2019mna}%
  \BibitemOpen
  \bibfield  {author} {\bibinfo {author} {\bibfnamefont {N.}~\bibnamefont
  {Armesto}}, \bibinfo {author} {\bibfnamefont {F.}~\bibnamefont {Dominguez}},
  \bibinfo {author} {\bibfnamefont {A.}~\bibnamefont {Kovner}}, \bibinfo
  {author} {\bibfnamefont {M.}~\bibnamefont {Lublinsky}}, \ and\ \bibinfo
  {author} {\bibfnamefont {V.}~\bibnamefont {Skokov}},\ }\href {\doibase
  10.1007/JHEP05(2019)025} {\bibfield  {journal} {\bibinfo  {journal} {JHEP}\
  }\textbf {\bibinfo {volume} {05}},\ \bibinfo {pages} {025} (\bibinfo {year}
  {2019})},\ \Eprint {http://arxiv.org/abs/1901.08080} {arXiv:1901.08080
  [hep-ph]} \BibitemShut {NoStop}%
\bibitem [{\citenamefont {Kharzeev}\ and\ \citenamefont
  {Levin}(2017)}]{Kharzeev:2017qzs}%
  \BibitemOpen
  \bibfield  {author} {\bibinfo {author} {\bibfnamefont {D.~E.}\ \bibnamefont
  {Kharzeev}}\ and\ \bibinfo {author} {\bibfnamefont {E.~M.}\ \bibnamefont
  {Levin}},\ }\href {\doibase 10.1103/PhysRevD.95.114008} {\bibfield  {journal}
  {\bibinfo  {journal} {Phys. Rev.}\ }\textbf {\bibinfo {volume} {D95}},\
  \bibinfo {pages} {114008} (\bibinfo {year} {2017})},\ \Eprint
  {http://arxiv.org/abs/1702.03489} {arXiv:1702.03489 [hep-ph]} \BibitemShut
  {NoStop}%
\bibitem [{\citenamefont {Tu}\ \emph {et~al.}(2019)\citenamefont {Tu},
  \citenamefont {Kharzeev},\ and\ \citenamefont {Ullrich}}]{Tu:2019ouv}%
  \BibitemOpen
  \bibfield  {author} {\bibinfo {author} {\bibfnamefont {Z.}~\bibnamefont
  {Tu}}, \bibinfo {author} {\bibfnamefont {D.~E.}\ \bibnamefont {Kharzeev}}, \
  and\ \bibinfo {author} {\bibfnamefont {T.}~\bibnamefont {Ullrich}},\
  }\href@noop {} {\  (\bibinfo {year} {2019})},\ \Eprint
  {http://arxiv.org/abs/1904.11974} {arXiv:1904.11974 [hep-ph]} \BibitemShut
  {NoStop}%
\bibitem [{\citenamefont {Tu}\ \emph {et~al.}(2020)\citenamefont {Tu},
  \citenamefont {Kharzeev},\ and\ \citenamefont {Ullrich}}]{Baker:2017wtt}%
  \BibitemOpen
  \bibfield  {author} {\bibinfo {author} {\bibfnamefont {Z.}~\bibnamefont
  {Tu}}, \bibinfo {author} {\bibfnamefont {D.~E.}\ \bibnamefont {Kharzeev}}, \
  and\ \bibinfo {author} {\bibfnamefont {T.}~\bibnamefont {Ullrich}},\ }\href
  {\doibase 10.1103/PhysRevLett.124.062001} {\bibfield  {journal} {\bibinfo
  {journal} {Phys. Rev. Lett.}\ }\textbf {\bibinfo {volume} {124}},\ \bibinfo
  {pages} {062001} (\bibinfo {year} {2020})},\ \Eprint
  {http://arxiv.org/abs/1712.04558} {arXiv:1712.04558 [hep-ph]} \BibitemShut
  {NoStop}%
\bibitem [{\citenamefont {Beane}\ and\ \citenamefont
  {Ehlers}(2019)}]{Beane:2019loz}%
  \BibitemOpen
  \bibfield  {author} {\bibinfo {author} {\bibfnamefont {S.~R.}\ \bibnamefont
  {Beane}}\ and\ \bibinfo {author} {\bibfnamefont {P.}~\bibnamefont {Ehlers}},\
  }\href {\doibase 10.1142/S0217732320500480} {\bibfield  {journal} {\bibinfo
  {journal} {Mod. Phys. Lett. A}\ }\textbf {\bibinfo {volume} {35}},\ \bibinfo
  {pages} {2050048} (\bibinfo {year} {2019})},\ \Eprint
  {http://arxiv.org/abs/1905.03295} {arXiv:1905.03295 [hep-ph]} \BibitemShut
  {NoStop}%
\bibitem [{\citenamefont {Beane}\ \emph {et~al.}(2019)\citenamefont {Beane},
  \citenamefont {Kaplan}, \citenamefont {Klco},\ and\ \citenamefont
  {Savage}}]{Beane:2018oxh}%
  \BibitemOpen
  \bibfield  {author} {\bibinfo {author} {\bibfnamefont {S.~R.}\ \bibnamefont
  {Beane}}, \bibinfo {author} {\bibfnamefont {D.~B.}\ \bibnamefont {Kaplan}},
  \bibinfo {author} {\bibfnamefont {N.}~\bibnamefont {Klco}}, \ and\ \bibinfo
  {author} {\bibfnamefont {M.~J.}\ \bibnamefont {Savage}},\ }\href {\doibase
  10.1103/PhysRevLett.122.102001} {\bibfield  {journal} {\bibinfo  {journal}
  {Phys. Rev. Lett.}\ }\textbf {\bibinfo {volume} {122}},\ \bibinfo {pages}
  {102001} (\bibinfo {year} {2019})},\ \Eprint
  {http://arxiv.org/abs/1812.03138} {arXiv:1812.03138 [nucl-th]} \BibitemShut
  {NoStop}%
\bibitem [{\citenamefont {Wigner}(1937)}]{Wigner:1936dx}%
  \BibitemOpen
  \bibfield  {author} {\bibinfo {author} {\bibfnamefont {E.}~\bibnamefont
  {Wigner}},\ }\href {\doibase 10.1103/PhysRev.51.106} {\bibfield  {journal}
  {\bibinfo  {journal} {Phys. Rev.}\ }\textbf {\bibinfo {volume} {51}},\
  \bibinfo {pages} {106} (\bibinfo {year} {1937})}\BibitemShut {NoStop}%
\bibitem [{\citenamefont {{\bf [NPLQCD]}}\ \emph {et~al.}(2017)\citenamefont
  {{\bf [NPLQCD]}}, \citenamefont {Wagman}, \citenamefont {Winter},
  \citenamefont {Chang}, \citenamefont {Davoudi}, \citenamefont {Detmold},
  \citenamefont {Orginos}, \citenamefont {Savage},\ and\ \citenamefont
  {Shanahan}}]{Wagman:2017tmp}%
  \BibitemOpen
  \bibfield  {author} {\bibinfo {author} {\bibnamefont {{\bf [NPLQCD]}}},
  \bibinfo {author} {\bibfnamefont {M.~L.}\ \bibnamefont {Wagman}}, \bibinfo
  {author} {\bibfnamefont {F.}~\bibnamefont {Winter}}, \bibinfo {author}
  {\bibfnamefont {E.}~\bibnamefont {Chang}}, \bibinfo {author} {\bibfnamefont
  {Z.}~\bibnamefont {Davoudi}}, \bibinfo {author} {\bibfnamefont
  {W.}~\bibnamefont {Detmold}}, \bibinfo {author} {\bibfnamefont
  {K.}~\bibnamefont {Orginos}}, \bibinfo {author} {\bibfnamefont {M.~J.}\
  \bibnamefont {Savage}}, \ and\ \bibinfo {author} {\bibfnamefont {P.~E.}\
  \bibnamefont {Shanahan}},\ }\href {\doibase 10.1103/PhysRevD.96.114510}
  {\bibfield  {journal} {\bibinfo  {journal} {Phys. Rev. D}\ }\textbf {\bibinfo
  {volume} {96}},\ \bibinfo {pages} {114510} (\bibinfo {year} {2017})},\
  \Eprint {http://arxiv.org/abs/1706.06550} {arXiv:1706.06550 [hep-lat]}
  \BibitemShut {NoStop}%
\bibitem [{\citenamefont {Kaplan}\ and\ \citenamefont
  {Savage}(1996)}]{Kaplan:1995yg}%
  \BibitemOpen
  \bibfield  {author} {\bibinfo {author} {\bibfnamefont {D.~B.}\ \bibnamefont
  {Kaplan}}\ and\ \bibinfo {author} {\bibfnamefont {M.~J.}\ \bibnamefont
  {Savage}},\ }\href {\doibase 10.1016/0370-2693(95)01277-X} {\bibfield
  {journal} {\bibinfo  {journal} {Phys. Lett. B}\ }\textbf {\bibinfo {volume}
  {365}},\ \bibinfo {pages} {244} (\bibinfo {year} {1996})},\ \Eprint
  {http://arxiv.org/abs/hep-ph/9509371} {arXiv:hep-ph/9509371} \BibitemShut
  {NoStop}%
\bibitem [{\citenamefont {Kaplan}\ and\ \citenamefont
  {Manohar}(1997)}]{Kaplan:1996rk}%
  \BibitemOpen
  \bibfield  {author} {\bibinfo {author} {\bibfnamefont {D.~B.}\ \bibnamefont
  {Kaplan}}\ and\ \bibinfo {author} {\bibfnamefont {A.~V.}\ \bibnamefont
  {Manohar}},\ }\href {\doibase 10.1103/PhysRevC.56.76} {\bibfield  {journal}
  {\bibinfo  {journal} {Phys. Rev. C}\ }\textbf {\bibinfo {volume} {56}},\
  \bibinfo {pages} {76} (\bibinfo {year} {1997})},\ \Eprint
  {http://arxiv.org/abs/nucl-th/9612021} {arXiv:nucl-th/9612021} \BibitemShut
  {NoStop}%
\bibitem [{\citenamefont {Detmold}\ \emph {et~al.}(2014)\citenamefont
  {Detmold}, \citenamefont {McCullough},\ and\ \citenamefont
  {Pochinsky}}]{Detmold:2014kba}%
  \BibitemOpen
  \bibfield  {author} {\bibinfo {author} {\bibfnamefont {W.}~\bibnamefont
  {Detmold}}, \bibinfo {author} {\bibfnamefont {M.}~\bibnamefont {McCullough}},
  \ and\ \bibinfo {author} {\bibfnamefont {A.}~\bibnamefont {Pochinsky}},\
  }\href {\doibase 10.1103/PhysRevD.90.114506} {\bibfield  {journal} {\bibinfo
  {journal} {Phys. Rev. D}\ }\textbf {\bibinfo {volume} {90}},\ \bibinfo
  {pages} {114506} (\bibinfo {year} {2014})},\ \Eprint
  {http://arxiv.org/abs/1406.4116} {arXiv:1406.4116 [hep-lat]} \BibitemShut
  {NoStop}%
\bibitem [{\citenamefont {Pine}\ \emph {et~al.}(2013)\citenamefont {Pine},
  \citenamefont {Lee},\ and\ \citenamefont {Rupak}}]{Pine:2013zja}%
  \BibitemOpen
  \bibfield  {author} {\bibinfo {author} {\bibfnamefont {M.}~\bibnamefont
  {Pine}}, \bibinfo {author} {\bibfnamefont {D.}~\bibnamefont {Lee}}, \ and\
  \bibinfo {author} {\bibfnamefont {G.}~\bibnamefont {Rupak}},\ }\href
  {\doibase 10.1140/epja/i2013-13151-3} {\bibfield  {journal} {\bibinfo
  {journal} {Eur. Phys. J. A}\ }\textbf {\bibinfo {volume} {49}},\ \bibinfo
  {pages} {151} (\bibinfo {year} {2013})},\ \Eprint
  {http://arxiv.org/abs/1309.2616} {arXiv:1309.2616 [nucl-th]} \BibitemShut
  {NoStop}%
\bibitem [{\citenamefont {Elhatisari}\ \emph {et~al.}(2016)\citenamefont
  {Elhatisari}, \citenamefont {Lee}, \citenamefont {Meißner},\ and\
  \citenamefont {Rupak}}]{Elhatisari:2016hby}%
  \BibitemOpen
  \bibfield  {author} {\bibinfo {author} {\bibfnamefont {S.}~\bibnamefont
  {Elhatisari}}, \bibinfo {author} {\bibfnamefont {D.}~\bibnamefont {Lee}},
  \bibinfo {author} {\bibfnamefont {U.-G.}\ \bibnamefont {Meißner}}, \ and\
  \bibinfo {author} {\bibfnamefont {G.}~\bibnamefont {Rupak}},\ }\href
  {\doibase 10.1140/epja/i2016-16174-2} {\bibfield  {journal} {\bibinfo
  {journal} {Eur. Phys. J. A}\ }\textbf {\bibinfo {volume} {52}},\ \bibinfo
  {pages} {174} (\bibinfo {year} {2016})},\ \Eprint
  {http://arxiv.org/abs/1603.02333} {arXiv:1603.02333 [nucl-th]} \BibitemShut
  {NoStop}%
\bibitem [{\citenamefont {Hill}\ and\ \citenamefont
  {Wootters}(1997)}]{Hill_1997}%
  \BibitemOpen
  \bibfield  {author} {\bibinfo {author} {\bibfnamefont {S.}~\bibnamefont
  {Hill}}\ and\ \bibinfo {author} {\bibfnamefont {W.~K.}\ \bibnamefont
  {Wootters}},\ }\href {\doibase 10.1103/physrevlett.78.5022} {\bibfield
  {journal} {\bibinfo  {journal} {Physical Review Letters}\ }\textbf {\bibinfo
  {volume} {78}},\ \bibinfo {pages} {5022–5025} (\bibinfo {year}
  {1997})}\BibitemShut {NoStop}%
\bibitem [{\citenamefont {Życzkowski}\ \emph {et~al.}(1998)\citenamefont
  {Życzkowski}, \citenamefont {Horodecki}, \citenamefont {Sanpera},\ and\
  \citenamefont {Lewenstein}}]{yczkowski_1998}%
  \BibitemOpen
  \bibfield  {author} {\bibinfo {author} {\bibfnamefont {K.}~\bibnamefont
  {Życzkowski}}, \bibinfo {author} {\bibfnamefont {P.}~\bibnamefont
  {Horodecki}}, \bibinfo {author} {\bibfnamefont {A.}~\bibnamefont {Sanpera}},
  \ and\ \bibinfo {author} {\bibfnamefont {M.}~\bibnamefont {Lewenstein}},\
  }\href {\doibase 10.1103/physreva.58.883} {\bibfield  {journal} {\bibinfo
  {journal} {Physical Review A}\ }\textbf {\bibinfo {volume} {58}},\ \bibinfo
  {pages} {883–892} (\bibinfo {year} {1998})}\BibitemShut {NoStop}%
\bibitem [{\citenamefont {Belavkin}\ and\ \citenamefont
  {Ohya}(2002)}]{Belavkin_2002}%
  \BibitemOpen
  \bibfield  {author} {\bibinfo {author} {\bibfnamefont {V.~P.}\ \bibnamefont
  {Belavkin}}\ and\ \bibinfo {author} {\bibfnamefont {M.}~\bibnamefont
  {Ohya}},\ }\href {\doibase 10.1098/rspa.2001.0867} {\bibfield  {journal}
  {\bibinfo  {journal} {Proceedings of the Royal Society of London. Series A:
  Mathematical, Physical and Engineering Sciences}\ }\textbf {\bibinfo {volume}
  {458}},\ \bibinfo {pages} {209–231} (\bibinfo {year} {2002})}\BibitemShut
  {NoStop}%
\bibitem [{\citenamefont {Eisert}(2006)}]{eisert2006entanglement}%
  \BibitemOpen
  \bibfield  {author} {\bibinfo {author} {\bibfnamefont {J.}~\bibnamefont
  {Eisert}},\ }\href@noop {} {\enquote {\bibinfo {title} {Entanglement in
  quantum information theory},}\ } (\bibinfo {year} {2006}),\ \Eprint
  {http://arxiv.org/abs/quant-ph/0610253} {arXiv:quant-ph/0610253 [quant-ph]}
  \BibitemShut {NoStop}%
\bibitem [{\citenamefont {Horodecki}\ \emph {et~al.}(2009)\citenamefont
  {Horodecki}, \citenamefont {Horodecki}, \citenamefont {Horodecki},\ and\
  \citenamefont {Horodecki}}]{RevModPhys.81.865}%
  \BibitemOpen
  \bibfield  {author} {\bibinfo {author} {\bibfnamefont {R.}~\bibnamefont
  {Horodecki}}, \bibinfo {author} {\bibfnamefont {P.}~\bibnamefont
  {Horodecki}}, \bibinfo {author} {\bibfnamefont {M.}~\bibnamefont
  {Horodecki}}, \ and\ \bibinfo {author} {\bibfnamefont {K.}~\bibnamefont
  {Horodecki}},\ }\href {\doibase 10.1103/RevModPhys.81.865} {\bibfield
  {journal} {\bibinfo  {journal} {Rev. Mod. Phys.}\ }\textbf {\bibinfo {volume}
  {81}},\ \bibinfo {pages} {865} (\bibinfo {year} {2009})}\BibitemShut
  {NoStop}%
\bibitem [{\citenamefont {Vidal}\ and\ \citenamefont
  {Werner}(2002)}]{PhysRevA.65.032314}%
  \BibitemOpen
  \bibfield  {author} {\bibinfo {author} {\bibfnamefont {G.}~\bibnamefont
  {Vidal}}\ and\ \bibinfo {author} {\bibfnamefont {R.~F.}\ \bibnamefont
  {Werner}},\ }\href {\doibase 10.1103/PhysRevA.65.032314} {\bibfield
  {journal} {\bibinfo  {journal} {Phys. Rev. A}\ }\textbf {\bibinfo {volume}
  {65}},\ \bibinfo {pages} {032314} (\bibinfo {year} {2002})}\BibitemShut
  {NoStop}%
\bibitem [{\citenamefont {Plenio}(2005)}]{Plenio_2005}%
  \BibitemOpen
  \bibfield  {author} {\bibinfo {author} {\bibfnamefont {M.~B.}\ \bibnamefont
  {Plenio}},\ }\href {\doibase 10.1103/physrevlett.95.090503} {\bibfield
  {journal} {\bibinfo  {journal} {Physical Review Letters}\ }\textbf {\bibinfo
  {volume} {95}} (\bibinfo {year} {2005}),\
  10.1103/physrevlett.95.090503}\BibitemShut {NoStop}%
\bibitem [{\citenamefont {Gorton}(2018)}]{GortonThesis}%
  \BibitemOpen
  \bibfield  {author} {\bibinfo {author} {\bibfnamefont {O.~C.}\ \bibnamefont
  {Gorton}},\ }\emph {\bibinfo {title} {Efficient modeling of nuclei through
  coupling of proton and neutron wavefunctions}},\ \href@noop {} {Master's
  thesis},\ \bibinfo  {school} {San Diego State University} (\bibinfo {year}
  {2018})\BibitemShut {NoStop}%
\bibitem [{\citenamefont {Gorton}\ and\ \citenamefont
  {Johnson}(2019)}]{GortonJohnson2019a}%
  \BibitemOpen
  \bibfield  {author} {\bibinfo {author} {\bibfnamefont {O.}~\bibnamefont
  {Gorton}}\ and\ \bibinfo {author} {\bibfnamefont {C.~W.}\ \bibnamefont
  {Johnson}},\ }\href {http://esnt.cea.fr/Phocea/Page/index.php?id=84}
  {\enquote {\bibinfo {title} {Entanglement entropy and proton-neutron
  interactions},}\ } (\bibinfo {year} {2019}),\ \bibinfo {note} {{ESNT}
  workshop on proton-neutron pairing,
  http://esnt.cea.fr/Phocea/Page/index.php?id=84}\BibitemShut {NoStop}%
\bibitem [{\citenamefont {Shi}(2003)}]{Shi2003}%
  \BibitemOpen
  \bibfield  {author} {\bibinfo {author} {\bibfnamefont {Y.}~\bibnamefont
  {Shi}},\ }\href {\doibase 10.1103/PhysRevA.67.024301} {\bibfield  {journal}
  {\bibinfo  {journal} {Phys. Rev. A}\ }\textbf {\bibinfo {volume} {67}},\
  \bibinfo {pages} {024301} (\bibinfo {year} {2003})}\BibitemShut {NoStop}%
\bibitem [{\citenamefont {Benatti}\ \emph {et~al.}(2014)\citenamefont
  {Benatti}, \citenamefont {Floreanini},\ and\ \citenamefont
  {Titimbo}}]{Benatti:2014gaa}%
  \BibitemOpen
  \bibfield  {author} {\bibinfo {author} {\bibfnamefont {F.}~\bibnamefont
  {Benatti}}, \bibinfo {author} {\bibfnamefont {R.}~\bibnamefont {Floreanini}},
  \ and\ \bibinfo {author} {\bibfnamefont {K.}~\bibnamefont {Titimbo}},\ }\href
  {\doibase 10.1142/S1230161214400034} {\bibfield  {journal} {\bibinfo
  {journal} {Open Syst. Info. Dyn.}\ }\textbf {\bibinfo {volume} {21}},\
  \bibinfo {pages} {1440003} (\bibinfo {year} {2014})},\ \Eprint
  {http://arxiv.org/abs/1403.3178} {arXiv:1403.3178 [quant-ph]} \BibitemShut
  {NoStop}%
\bibitem [{\citenamefont {Lo~Franco}\ and\ \citenamefont
  {Compagno}(2016)}]{LoFranco2016}%
  \BibitemOpen
  \bibfield  {author} {\bibinfo {author} {\bibfnamefont {R.}~\bibnamefont
  {Lo~Franco}}\ and\ \bibinfo {author} {\bibfnamefont {G.}~\bibnamefont
  {Compagno}},\ }\href {\doibase 10.1038/srep20603} {\bibfield  {journal}
  {\bibinfo  {journal} {Scientific Reports}\ }\textbf {\bibinfo {volume} {6}},\
  \bibinfo {pages} {20603} (\bibinfo {year} {2016})}\BibitemShut {NoStop}%
\bibitem [{\citenamefont {Rissler}\ \emph {et~al.}(2006)\citenamefont
  {Rissler}, \citenamefont {Noack},\ and\ \citenamefont
  {White}}]{RISSLER2006519}%
  \BibitemOpen
  \bibfield  {author} {\bibinfo {author} {\bibfnamefont {J.}~\bibnamefont
  {Rissler}}, \bibinfo {author} {\bibfnamefont {R.~M.}\ \bibnamefont {Noack}},
  \ and\ \bibinfo {author} {\bibfnamefont {S.~R.}\ \bibnamefont {White}},\
  }\href {\doibase https://doi.org/10.1016/j.chemphys.2005.10.018} {\bibfield
  {journal} {\bibinfo  {journal} {Chemical Physics}\ }\textbf {\bibinfo
  {volume} {323}},\ \bibinfo {pages} {519 } (\bibinfo {year}
  {2006})}\BibitemShut {NoStop}%
\bibitem [{\citenamefont {Boguslawski}\ and\ \citenamefont
  {Tecmer}(2015)}]{Boguslawski2015}%
  \BibitemOpen
  \bibfield  {author} {\bibinfo {author} {\bibfnamefont {K.}~\bibnamefont
  {Boguslawski}}\ and\ \bibinfo {author} {\bibfnamefont {P.}~\bibnamefont
  {Tecmer}},\ }\href {\doibase 10.1002/qua.24832} {\bibfield  {journal}
  {\bibinfo  {journal} {Int. J. Quantum Chem.}\ }\textbf {\bibinfo {volume}
  {115}},\ \bibinfo {pages} {1289–1295} (\bibinfo {year} {2015})}\BibitemShut
  {NoStop}%
\bibitem [{\citenamefont {Legeza}\ \emph {et~al.}(2015)\citenamefont {Legeza},
  \citenamefont {Veis}, \citenamefont {Poves},\ and\ \citenamefont
  {Dukelsky}}]{PhysRevC.92.051303}%
  \BibitemOpen
  \bibfield  {author} {\bibinfo {author} {\bibfnamefont {{\"O}.}~\bibnamefont
  {Legeza}}, \bibinfo {author} {\bibfnamefont {L.}~\bibnamefont {Veis}},
  \bibinfo {author} {\bibfnamefont {A.}~\bibnamefont {Poves}}, \ and\ \bibinfo
  {author} {\bibfnamefont {J.}~\bibnamefont {Dukelsky}},\ }\href {\doibase
  10.1103/PhysRevC.92.051303} {\bibfield  {journal} {\bibinfo  {journal} {Phys.
  Rev. C}\ }\textbf {\bibinfo {volume} {92}},\ \bibinfo {pages} {051303}
  (\bibinfo {year} {2015})}\BibitemShut {NoStop}%
\bibitem [{\citenamefont {Robin}\ \emph {et~al.}(2016)\citenamefont {Robin},
  \citenamefont {Pillet}, \citenamefont {Peña~Arteaga},\ and\ \citenamefont
  {Berger}}]{Robin:2015}%
  \BibitemOpen
  \bibfield  {author} {\bibinfo {author} {\bibfnamefont {C.}~\bibnamefont
  {Robin}}, \bibinfo {author} {\bibfnamefont {N.}~\bibnamefont {Pillet}},
  \bibinfo {author} {\bibfnamefont {D.}~\bibnamefont {Peña~Arteaga}}, \ and\
  \bibinfo {author} {\bibfnamefont {J.}~\bibnamefont {Berger}},\ }\href
  {\doibase 10.1103/PhysRevC.93.024302} {\bibfield  {journal} {\bibinfo
  {journal} {Phys. Rev. C}\ }\textbf {\bibinfo {volume} {93}},\ \bibinfo
  {pages} {024302} (\bibinfo {year} {2016})},\ \Eprint
  {http://arxiv.org/abs/1509.08694} {arXiv:1509.08694 [nucl-th]} \BibitemShut
  {NoStop}%
\bibitem [{\citenamefont {Robin}\ \emph {et~al.}(2017)\citenamefont {Robin},
  \citenamefont {Pillet}, \citenamefont {Dupuis}, \citenamefont {Le~Bloas},
  \citenamefont {Arteaga},\ and\ \citenamefont {Berger}}]{Robin:2016}%
  \BibitemOpen
  \bibfield  {author} {\bibinfo {author} {\bibfnamefont {C.}~\bibnamefont
  {Robin}}, \bibinfo {author} {\bibfnamefont {N.}~\bibnamefont {Pillet}},
  \bibinfo {author} {\bibfnamefont {M.}~\bibnamefont {Dupuis}}, \bibinfo
  {author} {\bibfnamefont {J.}~\bibnamefont {Le~Bloas}}, \bibinfo {author}
  {\bibfnamefont {D.~P.}\ \bibnamefont {Arteaga}}, \ and\ \bibinfo {author}
  {\bibfnamefont {J.}~\bibnamefont {Berger}},\ }\href {\doibase
  10.1103/PhysRevC.95.044315} {\bibfield  {journal} {\bibinfo  {journal} {Phys.
  Rev. C}\ }\textbf {\bibinfo {volume} {95}},\ \bibinfo {pages} {044315}
  (\bibinfo {year} {2017})},\ \Eprint {http://arxiv.org/abs/1611.03445}
  {arXiv:1611.03445 [nucl-th]} \BibitemShut {NoStop}%
\bibitem [{\citenamefont {Robin}\ \emph {et~al.}()\citenamefont {Robin},
  \citenamefont {Hupin}, \citenamefont {Pillet},\ and\ \citenamefont
  {Navr\'atil}}]{Robin:2020}%
  \BibitemOpen
  \bibfield  {author} {\bibinfo {author} {\bibfnamefont {C.}~\bibnamefont
  {Robin}}, \bibinfo {author} {\bibfnamefont {G.}~\bibnamefont {Hupin}},
  \bibinfo {author} {\bibfnamefont {N.}~\bibnamefont {Pillet}}, \ and\ \bibinfo
  {author} {\bibfnamefont {P.}~\bibnamefont {Navr\'atil}},\ }\href@noop {} {\
  }\bibinfo {note} {In preparation}\BibitemShut {NoStop}%
\bibitem [{\citenamefont {Constantinou}\ \emph {et~al.}(2017)\citenamefont
  {Constantinou}, \citenamefont {Caprio}, \citenamefont {Vary},\ and\
  \citenamefont {Maris}}]{Constantinou:2016}%
  \BibitemOpen
  \bibfield  {author} {\bibinfo {author} {\bibfnamefont {C.}~\bibnamefont
  {Constantinou}}, \bibinfo {author} {\bibfnamefont {M.}~\bibnamefont
  {Caprio}}, \bibinfo {author} {\bibfnamefont {J.}~\bibnamefont {Vary}}, \ and\
  \bibinfo {author} {\bibfnamefont {P.}~\bibnamefont {Maris}},\ }\href
  {\doibase 10.1007/s41365-017-0332-6} {\bibfield  {journal} {\bibinfo
  {journal} {Nucl. Sci. Tech.}\ }\textbf {\bibinfo {volume} {28}},\ \bibinfo
  {pages} {179} (\bibinfo {year} {2017})},\ \Eprint
  {http://arxiv.org/abs/1605.04976} {arXiv:1605.04976 [nucl-th]} \BibitemShut
  {NoStop}%
\bibitem [{\citenamefont {Tichai}\ \emph {et~al.}(2019)\citenamefont {Tichai},
  \citenamefont {Müller}, \citenamefont {Vobig},\ and\ \citenamefont
  {Roth}}]{Tichai:2018}%
  \BibitemOpen
  \bibfield  {author} {\bibinfo {author} {\bibfnamefont {A.}~\bibnamefont
  {Tichai}}, \bibinfo {author} {\bibfnamefont {J.}~\bibnamefont {Müller}},
  \bibinfo {author} {\bibfnamefont {K.}~\bibnamefont {Vobig}}, \ and\ \bibinfo
  {author} {\bibfnamefont {R.}~\bibnamefont {Roth}},\ }\href {\doibase
  10.1103/PhysRevC.99.034321} {\bibfield  {journal} {\bibinfo  {journal} {Phys.
  Rev. C}\ }\textbf {\bibinfo {volume} {99}},\ \bibinfo {pages} {034321}
  (\bibinfo {year} {2019})},\ \Eprint {http://arxiv.org/abs/1809.07571}
  {arXiv:1809.07571 [nucl-th]} \BibitemShut {NoStop}%
\bibitem [{\citenamefont {Ekstr\"om}\ \emph {et~al.}(2013)\citenamefont
  {Ekstr\"om}, \citenamefont {Baardsen}, \citenamefont {Forss\'en},
  \citenamefont {Hagen}, \citenamefont {Hjorth-Jensen}, \citenamefont {Jansen},
  \citenamefont {Machleidt}, \citenamefont {Nazarewicz}, \citenamefont
  {Papenbrock}, \citenamefont {Sarich},\ and\ \citenamefont
  {Wild}}]{NNLO_OPT_2013}%
  \BibitemOpen
  \bibfield  {author} {\bibinfo {author} {\bibfnamefont {A.}~\bibnamefont
  {Ekstr\"om}}, \bibinfo {author} {\bibfnamefont {G.}~\bibnamefont {Baardsen}},
  \bibinfo {author} {\bibfnamefont {C.}~\bibnamefont {Forss\'en}}, \bibinfo
  {author} {\bibfnamefont {G.}~\bibnamefont {Hagen}}, \bibinfo {author}
  {\bibfnamefont {M.}~\bibnamefont {Hjorth-Jensen}}, \bibinfo {author}
  {\bibfnamefont {G.~R.}\ \bibnamefont {Jansen}}, \bibinfo {author}
  {\bibfnamefont {R.}~\bibnamefont {Machleidt}}, \bibinfo {author}
  {\bibfnamefont {W.}~\bibnamefont {Nazarewicz}}, \bibinfo {author}
  {\bibfnamefont {T.}~\bibnamefont {Papenbrock}}, \bibinfo {author}
  {\bibfnamefont {J.}~\bibnamefont {Sarich}}, \ and\ \bibinfo {author}
  {\bibfnamefont {S.~M.}\ \bibnamefont {Wild}},\ }\href {\doibase
  10.1103/PhysRevLett.110.192502} {\bibfield  {journal} {\bibinfo  {journal}
  {Phys. Rev. Lett.}\ }\textbf {\bibinfo {volume} {110}},\ \bibinfo {pages}
  {192502} (\bibinfo {year} {2013})}\BibitemShut {NoStop}%
\bibitem [{\citenamefont {Gigena}\ and\ \citenamefont
  {Rossignoli}(2015)}]{PhysRevA.92.042326}%
  \BibitemOpen
  \bibfield  {author} {\bibinfo {author} {\bibfnamefont {N.}~\bibnamefont
  {Gigena}}\ and\ \bibinfo {author} {\bibfnamefont {R.}~\bibnamefont
  {Rossignoli}},\ }\href {\doibase 10.1103/PhysRevA.92.042326} {\bibfield
  {journal} {\bibinfo  {journal} {Phys. Rev. A}\ }\textbf {\bibinfo {volume}
  {92}},\ \bibinfo {pages} {042326} (\bibinfo {year} {2015})}\BibitemShut
  {NoStop}%
\bibitem [{\citenamefont {Kruppa}\ \emph {et~al.}(2020)\citenamefont {Kruppa},
  \citenamefont {Kov{\'a}cs}, \citenamefont {Salamon},\ and\ \citenamefont
  {Legeza}}]{Kruppa:2020rfa}%
  \BibitemOpen
  \bibfield  {author} {\bibinfo {author} {\bibfnamefont {A.}~\bibnamefont
  {Kruppa}}, \bibinfo {author} {\bibfnamefont {J.}~\bibnamefont {Kov{\'a}cs}},
  \bibinfo {author} {\bibfnamefont {P.}~\bibnamefont {Salamon}}, \ and\
  \bibinfo {author} {\bibfnamefont {{\"O}.}~\bibnamefont {Legeza}},\ }\href
  {\doibase 10.1088/1361-6471/abc2dd} {\bibfield  {journal} {\bibinfo
  {journal} {J. Phys. G: Nucl. Part. Phys.}\ }\textbf {\bibinfo {volume}
  {48}},\ \bibinfo {pages} {025107} (\bibinfo {year} {2020})},\ \Eprint
  {http://arxiv.org/abs/2006.07448} {arXiv:2006.07448 [nucl-th]} \BibitemShut
  {NoStop}%
\end{thebibliography}%

\end{document}